\documentclass[fleqn,usenatbib]{mnras}
\usepackage{graphicx}
\usepackage{amsmath}
\usepackage{amssymb}
\usepackage{bm}
\usepackage{multirow}
\usepackage{newtxtext,newtxmath}
\usepackage[T1]{fontenc}
\usepackage{subcaption}
\DeclareRobustCommand{\VAN}[3]{#2}
\let\VANthebibliography\thebibliography
\def\thebibliography{\DeclareRobustCommand{\VAN}[3]{##3}\VANthebibliography}
\graphicspath{{./}{figs/}}
\usepackage{scalerel,tikz}
\usetikzlibrary{svg.path}
\definecolor{orcidlogocol}{HTML}{A6CE39}
\tikzset{orcidlogo/.pic={
 \fill[orcidlogocol] svg{M256,128c0,70.7-57.3,128-128,128C57.3,256,0,198.7,0,128C0,57.3,57.3,0,128,0C198.7,0,256,57.3,256,128z};
 \fill[white] svg{M86.3,186.2H70.9V79.1h15.4v48.4V186.2z}
 svg{M108.9,79.1h41.6c39.6,0,57,28.3,57,53.6c0,27.5-21.5,53.6-56.8,53.6h-41.8V79.1z M124.3,172.4h24.5c34.9,0,42.9-26.5,42.9-39.7c0-21.5-13.7-39.7-43.7-39.7h-23.7V172.4z}
 svg{M88.7,56.8c0,5.5-4.5,10.1-10.1,10.1c-5.6,0-10.1-4.6-10.1-10.1c0-5.6,4.5-10.1,10.1-10.1C84.2,46.7,88.7,51.3,88.7,56.8z};
}}
\newcommand\orcidicon[1]{\href{https://orcid.org/#1}{\mbox{\scalerel*{
\begin{tikzpicture}[yscale=-1,transform shape]
\pic{orcidlogo};
\end{tikzpicture}
}{|}}}}
\title[\ahkash \, Hybrid particle-in-cell code]{\ahkash: a new Hybrid particle-in-cell code for simulations of astrophysical collisionless plasma}
\author[Achikanath Chirakkara et al.]{
Radhika Achikanath Chirakkara,$^{\orcidicon{0000-0001-5583-5038}\,1}$\thanks{E-mail: \href{mailto:radhika.achikanathchirakkara@anu.edu.au}{radhika.achikanathchirakkara@anu.edu.au}}
Christoph Federrath,$^{\orcidicon{0000-0002-0706-2306}\,1,2}$
and Amit Seta$^{\orcidicon{0000-0001-9708-0286}\,1}$
\\
$^{1}$Research School of Astronomy and Astrophysics, Australian National University, Canberra, ACT 2611, Australia\\
$^{2}$Australian Research Council Centre of Excellence in All Sky Astrophysics (ASTRO3D), Canberra, ACT 2611, Australia
}
\date{Accepted XXX. Received YYY; in original form ZZZ}
\pubyear{2024}
\newcommand\Eq[1]{Eq.\,\ref{#1}}
\newcommand\Fig[1]{Fig.~\ref{#1}}
\newcommand\Sec[1]{Sec.~\ref{#1}}
\newcommand\Tab[1]{Tab.~\ref{#1}}
\newcommand\App[1]{Appendix~\ref{#1}}

\newcommand{\Mach}{{\mathcal{M}}}      

\newcommand{\Rm}{{\rm Rm}}
\renewcommand{\vec}[1]{\boldsymbol{#1}}	
\newcommand{\dd}{\mathrm{d}}        

\newcommand{\m}{{\rm m}}      
\newcommand{\kpc}{{\rm kpc}}  
\newcommand{\s}{{\rm s}}      

\newcommand{\ratio}{E_{\rm{mag}}/E_{\rm{kin}}}

\newcommand{\Rmhyper}{{\rm Rm_{\rm h}}}

\newcommand{\initmagnetisation}{(r_{\rm Larmor}/L)_{\rm init}}

\newcommand{\ngrid}{N_{\rm{grid}}}
\newcommand{\nppc}{N_{\rm{ppc}}}
\newcommand{\nsmooth}{N_{\rm{smooth}}}
\newcommand{\mfp}{\lambda_{\rm mfp}}
\newcommand{\vel}{\vec{v}}
\renewcommand{\vec}[1]{\mathbf{#1}}
\newcommand{\Vturb}{v_{\rm turb}}
\newcommand{\Vtherm}{v_{\rm th}}
\newcommand{\Vthtarget}{(v_{\rm th})_{\rm target}}
\newcommand{\Lturb}{L_{\rm turb}}
\newcommand{\tth}{t_{\rm th}}
\newcommand{\tcool}{t_{\rm cool}}
\newcommand{\ted}{t_{0}}

\newcommand{\E}{\vec{E}}
\newcommand{\B}{\vec{B}}
\newcommand{\J}{\vec{J}}
\newcommand{\rL}{r_{\rm Larmor}}
\newcommand{\tL}{t_{\rm Larmor}}
\newcommand{\dx}{\Delta x}
\newcommand{\dy}{\Delta y}
\newcommand{\dz}{\Delta z}
\newcommand{\dt}{\Delta t}
\newcommand{\dl}{\Delta l}
\newcommand{\ex}{E_{x}}

\newcommand{\bx}{B_{x}}
\newcommand{\by}{B_{y}}
\newcommand{\bz}{B_{z}}
\newcommand{\vpavg}{\langle v_{\rm p} \rangle}
\newcommand{\ahkash}{\texttt{AHKASH}}
\begin{document}
\label{firstpage}
\pagerange{\pageref{firstpage}--\pageref{lastpage}}
\maketitle
\begin{abstract}
We introduce \texttt{A}strophysical \texttt{H}ybrid-\texttt{K}inetic simulations with the \texttt{flASH} code ($\ahkash$) -- a new Hybrid particle-in-cell (PIC) code developed within the framework of the multi-physics code \texttt{FLASH}. The new code uses a second-order accurate Boris integrator and a predictor-predictor-corrector algorithm for advancing the Hybrid-kinetic equations, using the constraint transport method to ensure that magnetic fields are divergence-free. The code supports various interpolation schemes between the particles and grid cells, with post-interpolation smoothing to reduce finite particle noise. We further implement a $\delta f$ method to study instabilities in weakly collisional plasmas. The new code is tested on standard physical problems such as the motion of charged particles in uniform and spatially varying magnetic fields, the propagation of Alfv\'en and whistler waves, and Landau damping of ion acoustic waves. We test different interpolation kernels and demonstrate the necessity of performing post-interpolation smoothing. We couple the \texttt{TurbGen} turbulence driving module to the new Hybrid PIC code, allowing us to test the code on the highly complex physical problem of the turbulent dynamo. To investigate steady-state turbulence with a fixed sonic Mach number, it is important to maintain isothermal plasma conditions. Therefore, we introduce a novel cooling method for Hybrid PIC codes and provide tests and calibrations of this method to keep the plasma isothermal. We describe and test the `hybrid precision' method, which significantly reduces (by a factor $\sim1.5$) the computational cost, without compromising the accuracy of the numerical solutions. Finally, we test the parallel scalability of the new code, showing excellent scaling up to 10,000~cores.
\end{abstract}

\section{Introduction}
Many astrophysical plasmas, such as the solar wind \citep{Verscharen+2019}, hot accretion flows in the galactic center \citep{EHTCollab2019}, and the hot intracluster medium of galaxy clusters \citep{Schekochihin&Cowley2006, Kunz+2022} are weakly collisional or collisionless plasmas, and their evolution is coupled with magnetic fields of dynamically important strength. The collisionality of a plasma can be determined from the ratio of the Coulomb mean free path ($\mfp$) to the characteristic length scale of the system ($L$). For example, the hot interstellar medium is characterised by $\mfp/L \sim 10^{-3} - 10^{-2}$, which can be suitably modelled as a collisional plasma \citep{Ferriere2020}, where the magneto-hydrodynamical (MHD) approximation holds reasonably well. For the solar wind, $\mfp/L \sim 1$, for hot accretion flows, $\mfp/L \sim 1$, and the hot intracluster medium, $\mfp/L \sim 0.1$, making these plasmas weakly collisional or collisionless. 
Such astrophysical systems are interesting laboratories to study the nature of magnetized collisionless and weakly collisional plasma and it is important to develop numerical schemes that can suitably model such systems, which is the aim of this work. 

Magnetohydrodynamic (MHD) codes have been used extensively to model magnetized astrophysical gas \citep{Fryxelletal2000,Dubey+2008a,Pluto2007, Athena2008,Pencil, Athena++2020,Pencil2021}.
Assuming gas as a conducting fluid, these codes solve the continuity equation, Navier Stokes equation, induction equation for magnetic field evolution and energy equation for a given equation of state. MHD numerical simulations have become an established method to model astrophysical plasma in the last few decades. However, when the plasma is weakly collisional or collisionless, the continuum limit of the fluid equations breaks down and MHD is no longer valid for such plasmas. In such situations, we resort to a kinetic treatment of the plasma \citep{Kulsrud2005}, which is our approach for this work.

Kinetic codes solve the Vlasov or the Vlasov-Landau equation \citep{Birdsall&Langdon1991}, to study the evolution of the ion distribution functions, along with Maxwell's equations. Numerically this can be achieved in many ways, including solving the six-dimensional (three spatial + three velocities) equations for the distribution function \citep{Vlasiator2023} or using a particle-in-cell (PIC) approach to sample the moments of the distribution function while using a three-dimensional spatial computational grid \citep{TRISTAN-MP2005, Zeltron2013,SHARP2017,Smilei2018,Runko2022}. Here we focus on the PIC method. The PIC method can simulate the physics on the ion and electron time and length scales, but doing so makes this approach computationally expensive and challenging. Moreover, it can be very difficult to run these simulations for larger physically interesting time and length scales, as the scale separation between physical processes is often large. For example, in the intracluster medium, the dynamic length scale $\sim 10 \, \kpc$, the ion Larmor scale $\sim 10^{-12} \, \kpc$ and the electron Larmor scale $\sim 10^{-14} \, \kpc$, making PIC codes infeasible for certain physical problems.

We adopt the middle ground approach of solving the Hybrid-kinetic equations using so-called \emph{Hybrid PIC} in our new code $\ahkash$ \footnote{$\ahkash$ is an acronym for Astrophysical Hybrid-Kinetic simulations with the \texttt{flASH} code. $\ahkash$ (pronounced as ākāś) means `the sky' in many Indian languages such as Hindi, Malayalam, and Gujarati.}, which builds on the architecture of the $\texttt{FLASH}$ code \citep{Fryxelletal2000, Dubey+2008a, Dubey+2008b}. In Hybrid PIC, positively charged ions are modelled as particles, which sample the moments of the ion distribution function, while the electrons are modelled as a mass-less fluid \footnote{We neglect the inertia of electrons in comparison to the protons as the ratio of electron to proton mass is small ($\approx 1/1836$) and thus model electrons as a mass-less fluid.} \citep{Horowitz+1989, Lipatov2002, Bagdonat&Motschmann2002, Gargat+2007, Muller+2011, Kunz+2014a, Le+2023}. We note that we solve the Hybrid-kinetic equations in the non-relativistic limit (the ratio of typical speeds to the speed of light in the hot ICM $\lesssim 10^{-3}$). With this approach, we capture the physics at the ion scales, like the Larmor gyration of ions and the damping of ion acoustic waves, without having to encapsulate electron scale physics, such as Debye oscillations. This also significantly reduces the computational cost of performing these simulations relative to the full PIC approach, while still being more computationally expensive than MHD simulations, but allowing us to model the physics of weakly collisional plasmas.

Another numerical complication is modelling turbulence which is prevalent in most astrophysics systems. Turbulence is also present in weakly collisional astrophysical environments like the hot intracluster medium (ICM) of galaxy clusters and solar wind. Turbulence in such environments may be driven by merger events, feedback from active galactic nuclei, or by in-falling galaxies stirring the ICM \citep{Simionescu+2019}. Observational signatures of turbulence in such environments have been studied by \cite{HitomiCollaboration2016, Gatuzz+2022a, Gatuzz+2022b, Gatuzz+2023}. Regions of the ICM can also host subsonic or supersonic turbulence of rotational or irrotational nature \citep{Zinger+2018}. This is also true of collisionless astrophysical plasma like the solar wind \citep{Howes+2014, Bruno&Carbone2013, Klein&Howes2015, Verscharen+2019}.

The viscous heating of plasma due to turbulence, and the thermodynamics of weakly collisional plasma are different from collisional MHD plasma \citep{Howes2010, Kunz+2011, Squire+2023a}. It is therefore important to study plasma turbulence and heating of collisionless plasmas like the solar wind, by using PIC methods. For example, recent studies use Hybrid PIC to study the heating of ions and electrons in the solar wind \citep{Arzamasskiy+2019, Squire+2023b}. The Hybrid PIC approach also enables the study of magnetic fields in a turbulent, magnetized, weakly collisional plasma \citep{St-Onge&Kunz2018, AchikanathChirakkara+2023}. While \cite{Rinconetal2016} solve the Vlasov equation in their numerical experiments, \cite{St-Onge&Kunz2018} and \cite{AchikanathChirakkara+2023} use Hybrid PIC simulations. We note that the numerical experiments described in \cite{AchikanathChirakkara+2023} have been performed using \ahkash.

Turbulence affects the thermodynamics of the plasma and controls the evolution of magnetic fields. Therefore, to properly understand the evolution of weakly collisional astrophysical systems, it is important to numerically model turbulence. Several kinetic codes have been used to study turbulence in weakly collisional or collisionless plasma \citep{TRISTAN-MP2005, Kunz+2014a}. Turbulence dissipates energy and heats the plasma which makes it difficult to maintain isothermal conditions, especially for controlled numerical experiments. Here we aim to accurately model the effects of turbulence in collisionless plasma with our code using an Ornstein-Uhlenbeck process \citep{Federrath+2010,FederrathEtAl2022ascl}, especially to maintain the isothermal nature of plasma via the novel cooling in the presence of plasma heating due to turbulence.

We use the second-order accurate predictor-predictor-corrector integration scheme to update the particles and electromagnetic fields in our code, similar to \citet{Kunz+2014a}. We use the Boris integrator \citep{Boris1970} and the constrained transport method to update the magnetic field while keeping it divergence-free \citep{Yee1966}. We also implemented a $\delta f$ method to enable the study of instabilities in collisonless plasma. We study the effect of using different interpolation kernels, performing smoothing operations and varying the number of particles per grid cell in our numerical simulations. Finally, we have implemented a new plasma cooling method, which can be used to maintain a steady sonic Mach number in driven fully-developed turbulence in weakly collisional plasmas, even when such plasmas experience strong local compression and/or rarefaction. This method is important for studying turbulence in collisionless plasma, particularly in the supersonic regime.

The rest of this paper is organised as follows. We describe the Hybrid-kinetic equations and the related numerical methods we use in \Sec{sec:methods}. We test the accuracy of our new Hybrid PIC code with various standard test problems in \Sec{sec:tests}. We discuss the interpolation tests in \Sec{sec:interpolation_tests} and describe the ion cooling method we developed alongside numerical tests in \Sec{sec:cooling_tests}. We discuss the `hybrid precision' method, performance of the code and present scaling tests in \Sec{sec:code_performance}. Finally, we summarize our study in \Sec{sec:summary}.

\section{Methods}
\label{sec:methods}
In this section, we present the equations and methods used in $\ahkash$. \Sec{sec:hybridkinetic_eqns} introduces the equations that govern the motion and coupling of positively-charged ions (particles) with an electron fluid in Hybrid-kinetic plasmas. \Sec{sec:hybridpic_algorithm} describes the operations that the code performs in each time step. To track ion trajectories, we use the Boris integrator, described in \Sec{sec:Boris_push}. The particle-grid interpolation and smoothing operations are explained in \Sec{sec:interpolation&filtering}. The $\delta f$ method is introduced in \Sec{sec:deltaf}, and the magnetic field update is described in \Sec{sec:CT}. We discuss the predictor-predictor-corrector integration scheme in \Sec{sec:integration_scheme}, with the corrections to the interpolated electromagnetic fields explained in \Sec{sec:E&B_corrections}, and higher-order numerical hyper-resistivity discussed in \Sec{sec:hyperres}. Finally, we describe the turbulent driving in \Sec{sec:driving}, and the novel cooling method, introduced to overcome plasma heating, in \Sec{sec:cooling}. Time-step constraints are discussed in \Sec{sec:timesteps}.

\subsection{Hybrid-kinetic equations}
\label{sec:hybridkinetic_eqns}
The evolution of a weakly collisional plasma is governed by the Vlasov equation, which ensures the conservation of phase-space density, 
\begin{align}
    &\frac{\dd f_{\rm i} (t, \vec{r}, \vel)}{\dd t} = 0, \nonumber \\
    & \frac{\partial f_{\rm i}}{\partial t} + \frac{\partial \vec{r}}{\partial t} \cdot \frac{\partial f_{\rm i}}{\partial \vec{r}}  + \frac{\partial \vel}{\partial t} \cdot \frac{\partial f_{\rm i}}{\partial \vel}   = 0, \nonumber \\ 
    &\frac{\partial f_{\rm i}}{\partial t}  + \vel \cdot \frac{\partial f_{\rm i}}{\partial \vec{r}} + \frac{q_{\rm i}}{m_{\rm i}} (\E + \vel \times \B) \cdot 
    \frac{\partial f_{\rm i}}{\partial \vel} = 0,
    \label{eqn:vlasoveqn}
\end{align}
which describes the evolution of the ion distribution function, $f_{\rm i} (t, \vec{r}, \vel)$, where $t$ is time, $\vec{r}$ is the position vector, and $\vel$ is the velocity vector\footnote{Note that $\frac{\partial f}{\partial \vec{r}} = \left(\frac{\partial f}{\partial x}, \frac{\partial f}{\partial y}, \frac{\partial f}{\partial z}\right)$, and $\frac{\partial f}{\partial \vel} = \left(\frac{\partial f}{\partial v_x}, \frac{\partial f}{\partial v_y}, \frac{\partial f}{\partial v_z}\right)$.}. The electromagnetic force on the ions is described by the Lorentz force, $(q_{\rm i}/m_{\rm i}) (\E + \vel \times \B)$, where $q_{\rm i}$ and $m_{\rm i}$ are the charge and mass of the ions, respectively. $\E$ and $\B$ are the electric and magnetic fields experienced by the ions. We solve \Eq{eqn:vlasoveqn} using the PIC approach, for which we can write the equations of motion for the ions,
\begin{align}
        & \frac{\dd \mathbf{r}_{p}} {\dd t}  = \vel_{p},
        \label{eqn:part_evolution2} \\
        & \frac{\dd \vel_{p} }{\dd t} = \frac{q_{\rm i}}{m_{\rm i}}\left(\E + \vel_{p} \times \B\right),  \label{eqn:part_evolution1} 
\end{align}
where $\vec{r}_{p}$ and $\vel_{p}$ are the position and velocity of the ${p}^{\rm th}$ particle, respectively, and ${p} = 1, 2, \dots, P$, where $P$ is the total number of particles. We note that this model can be easily expanded for multiple positively-charged ion species, however, here we only consider protons.

The evolution of the electrons is described by the Vlasov-Landau equation (similar to \Eq{eqn:vlasoveqn}, but now with a collision operator on the right-hand side),
\begin{equation}
    \frac{\partial f_{\rm e}}{\partial t} + \vel \cdot \frac{\partial f_{\rm e}}{\partial \vec{r}} + \frac{q_{\rm e}}{m_{\rm e}} (\E + \vel \times \B) \cdot 
    \frac{\partial f_{\rm e}}{\partial \vel} = C[f_{\rm e}],
    \label{eqn:vlasovlandaueqn}
\end{equation}
where $f_{\rm e} (t, \vec{r}, \vel)$, $q_{\rm e}$, and $m_{\rm e}$, are the electron distribution function, charge, and mass of the electrons, respectively, and $C [f_{\rm e}]$ is the electron collision operator.

Performing an expansion of the electron distribution function in powers of the electron-to-ion mass ratio $(m_{\rm e}/m_{\rm i})^{1/2}$, we can simplify the Vlasov-Landau equation (Appendix~A1.1 of \citet{Rosin+2011}). Up to the lowest orders in this expansion, one finds that the zeroth-order distribution function is Maxwellian as shown in Appendix~A1.3 of \citet{Rosin+2011}. Multiplying the expression obtained from \Eq{eqn:vlasovlandaueqn} up to order $(m_{\rm e}/m_{\rm i})^{1/2}$ by $m_{\rm e} v$, integrating over velocity space and using that the zeroth-order electron distribution function is Maxwellian, as detailed in Appendix~A1.5 of \citet{Rosin+2011}, we obtain the following the Ohm's law for the electric field with resistivity (index `nores'),
\begin{equation}
    \E_{\rm nores} = -\frac{\left(\J_{\rm e} \times \B \right)}{\rho_{\rm e}} + \frac{\nabla p_{\rm e}}{\rho_{\rm e}},
\label{eqn:electron_momentum_eqn}
\end{equation}
where $\J_{\rm e}$, $\rho_{\rm e}$, and $p_{\rm e}$, are the electron current, electron charge density, and electron pressure, respectively. The electron collision operator is modelled using Ohmic resistivity, which gives us
\begin{equation}
    \E = -\frac{\left(\J_{\rm e} \times \B \right)}{\rho_{\rm e}} + k_{\rm B} T_{\rm e} \frac{\nabla n_{\rm e}}{\rho_{\rm e}} + \eta \mu_{0} \J, 
\label{eqn:electron_momentum_eqn2}
\end{equation}
where $\J$ is the total current and $\eta$ is the magnetic diffusivity. This additional term acts as a sink for magnetic energy.  $\mu_0$ is the magnetic permeability constant. 
We have also used the ideal gas equation of state in \Eq{eqn:electron_momentum_eqn2} with a constant temperature for the electrons, $p_{\rm e} = n_{\rm e} k_{\rm B} T_{\rm e}$, where $n_{\rm e}$, $T_{\rm e}$, and $k_{\rm B}$, are the electron number density, constant electron temperature, and the Boltzmann constant, respectively. The total current is the sum of the ion current ($\J_{\rm i}$) and the electron current ($\J_{\rm e}$),
\begin{equation}
    \J = \J_{\rm i} + \J_{\rm e}. 
\end{equation}
The charge density of the ions, $\rho_{\rm i}$, can be obtained from the zeroth moment of the  distribution function and can be written as 
\begin{equation}\label{eqn:rhoI_def}
    \rho_{\rm i} = q_{\rm i} \int f_{\rm i} \, d^{3} \vel,
\end{equation}
and assuming quasi-neutrality of the plasma, we have
\begin{equation}
    \rho_{\rm i} + \rho_{\rm e} = 0,
\end{equation}
which implies that the charge density of the ions and electrons is the same, $\rho_{\rm i} = -{\rho_{\rm e}}$.

The first moment of the distribution function corresponds to the mean (or bulk) velocity of the ions, which defines the ion current,
\begin{equation}\label{eqn:JI_def}
    \J_{\rm i} = q_{\rm i} \int \vel f_{\rm i} \, d^{3} \vel .
\end{equation}
Using this and $\rho_{\rm e} = q_{\rm e} n_{\rm e}$ in \Eq{eqn:electron_momentum_eqn2}, the Ohm's law can be re-written as 
\begin{equation}
    \E = \frac{\left( \J - \J_{\rm i} \right) \times \B}{\rho_{\rm i}} - \frac{k_{\rm B} T_{\rm e}}{q_{\rm e}} \frac{\nabla \rho_{\rm i}}{\rho_{\rm i}} + \mu_{0} \eta \J.
\label{eqn:ohms_law_hp}
\end{equation}
Finally, the evolution of the positively-charged ions and electrons and the Ohm's law (\Eq{eqn:ohms_law_hp}) are coupled to the following Maxwell's equations,
\begin{equation}
\J = \frac{\nabla \times \B}{\mu_{0}},
\label{eqn:amperes_law}
\end{equation}
with the magnetic field calculated from Faraday's law, 
\begin{equation}
    \frac{\partial \B}{\partial t} = - \nabla \times \E,
    \label{eqn:faradays_law}  
\end{equation}
and the solenoidal condition for the magnetic field,
\begin{equation}
\nabla \cdot \B = 0.
\label{eqn:divB}
\end{equation}

\subsection{Hybrid PIC algorithm}
\label{sec:hybridpic_algorithm}
\begin{figure*}
    \centering
    \includegraphics[width=0.7\linewidth]{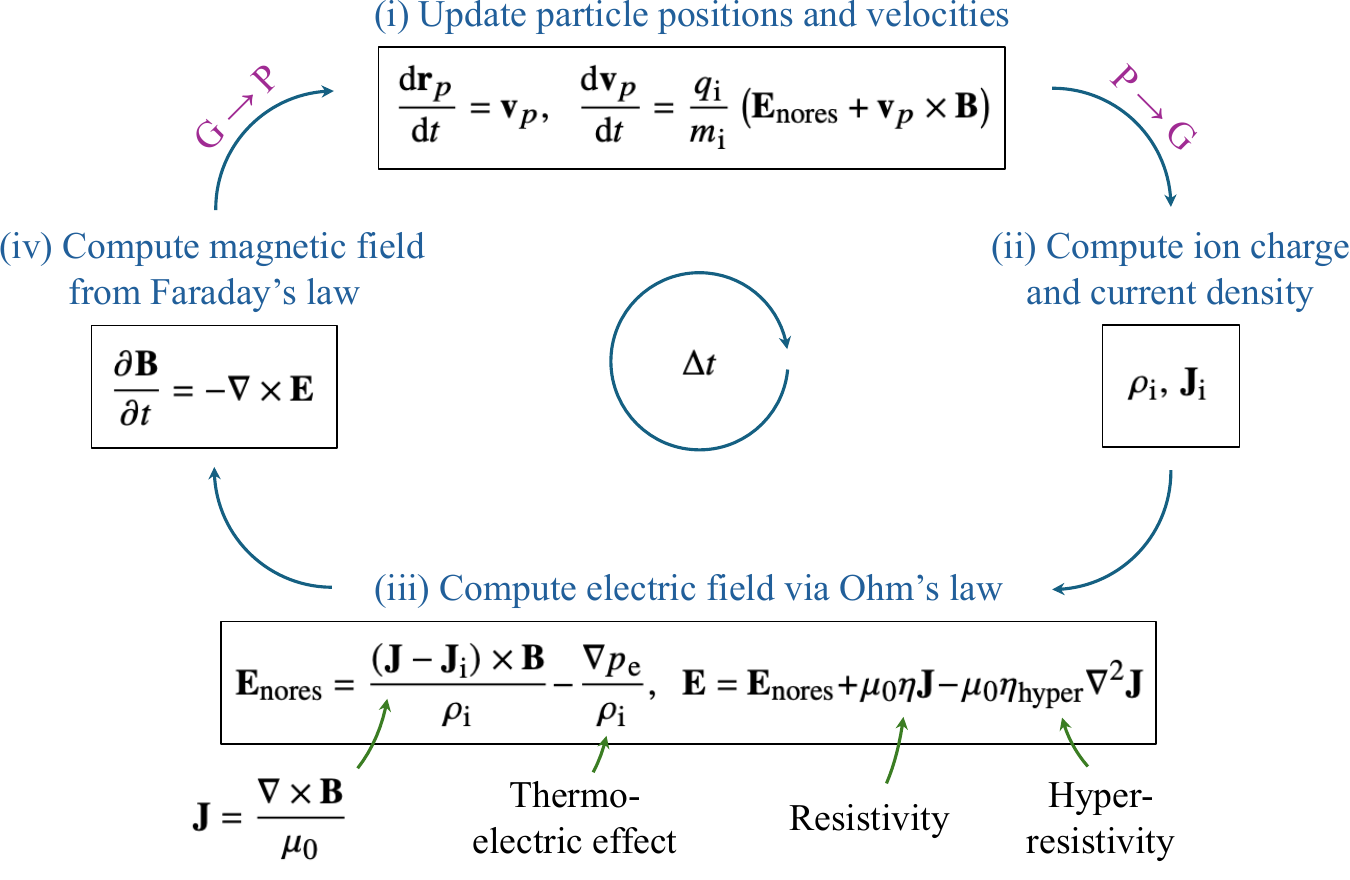}
    \caption{Simplified schematic of the Hybrid particle-in-cell (Hybrid PIC) algorithm for one time step $\Delta t$ (see \Sec{sec:hybridpic_algorithm}). (i) The particle positions and velocities are updated based on the velocity and acceleration from the Lorentz force, respectively (see \Sec{sec:hybridkinetic_eqns} and \Sec{sec:Boris_push}). (ii) Following this, the updated charge density, $\rho_{\rm i}$, and the ion current, ${\J}_{\rm i}$, are deposited onto the computational grid using an interpolation operation (see \Sec{sec:interpolation&filtering} and \Sec{sec:deltaf}). (iii) The charge density and ion current are source terms in the Ohm's law, \Eq{eqn:ohms_law_hp_hyperres}, which is used to compute the electric field, $\E$, on the grid. (iv) The magnetic field, $\B$, is then calculated using Faraday's law, \Eq{eqn:faradays_law}, on the grid (see \Sec{sec:CT}). Finally, the updated electric and magnetic fields are interpolated from the grid to the particles for the next particle evolution step.}
    \label{fig:hybidPIC_loop}
\end{figure*}

We solve the Hybrid-kinetic equations described in \Sec{sec:hybridkinetic_eqns} using the PIC method. Particles are used to sample the moments of the ion distribution function, $f_{\rm i}$. In this section, we go through a simplified description of what the code does in one computational time step, $\dt$. A schematic diagram of this simplified algorithm is shown in \Fig{fig:hybidPIC_loop} and the computational loop is described below.

\begin{enumerate}
    \item Particle evolution: In the presence of electric and magnetic fields, the ions are accelerated by the Lorentz force, $q_{\rm i}/m_{\rm i} (\E_{\rm nores} + \vel\times\B)$, where $\E_{\rm nores}$ is the electric field experienced by the ions, excluding resistive effects. The resistive term is excluded here to ensure the conservation of momentum of the overall system (see Sec.~4.5.2 of \citet{Lipatov2002} and \citet{Bagdonat2004}). The particle velocities and positions are updated based on this acceleration and velocity, respectively. More details on the particle evolution are provided in \Sec{sec:Boris_push}.
    \item Depositing moments of the distribution function onto the computational grid: The ion charge density and current are obtained from the zeroth and first moments of the distribution function, using \Eq{eqn:rhoI_def} and \Eq{eqn:JI_def}. These moments are sampled by the particles and deposited onto the grid, using an interpolation method (see \Sec{sec:interpolation&filtering} and \Sec{sec:deltaf}).
    \item Computing the electric field (Ohm's law): The charge density and current on the grid are used in the Ohm's law, \Eq{eqn:ohms_law_hp}. The total current is calculated at this stage using Amp\`ere's law, \Eq{eqn:amperes_law} (see \Sec{sec:hybridkinetic_eqns}).
    \item Computing the magnetic field (Faraday's law): The magnetic field update on the grid is described in more detail in \Sec{sec:CT}. The $\E$ and $\B$ fields are now interpolated from the grid to the particles for the next cycle. 
\end{enumerate}

Our code uses the predictor-predictor-corrector integration algorithm \citep{Kunz+2014a}, which is more sophisticated and involved than the simplified description above, and is second-order accurate, capable of stably propagating Alfv\'{e}n and whistler waves. We provide details on the full scheme in \Sec{sec:integration_scheme}. The time step, $\dt$, is computed as the minimum of several particle and wave time steps, as discussed in detail in \Sec{sec:timesteps}.

\subsection{Particle integrator}
\label{sec:Boris_push}
To update the particle positions and velocities, we solve the equations of motion, \Eq{eqn:part_evolution2} and~\ref{eqn:part_evolution1}, using the Boris integrator. The Boris integrator is commonly used in PIC codes for its ability to accurately evolve particle dynamics for a large number of time steps \citep{Boris1970,Qin+2013}. The velocity update in one time step, $\dt$, can be denoted as $\vel^{t} \rightarrow \vel^{t+\dt}$, where $\vel^{t}$ is the velocity at time $t$ and $\vel^{t+\dt}$ is the velocity at time $t + \dt$. Then, $\vel^{t+\dt}$ is calculated via
\begin{align}
        & \vel^{-} = \vel^{t} + a{\widetilde{\E}}^{t+\dt/2}_{\rm nores} \label{eqn:boris_push1} \\ 
        & \mathbf{v}' = \vel^{-} + a \vel^{-} \times \widetilde{\B}^{t + \dt/2} \label{eqn:boris_push2} \\ 
        & \vel^{+} = \vel^{-} + ab \mathbf{v}' \times \widetilde{\B}^{t+ \dt/2} \label{eqn:boris_push3} \\ 
        & \vel^{t+\dt} = \vel^{+} + a{\widetilde{\E}}^{t+\dt/2}_{\rm nores} \label{eqn:boris_push4} , 
\end{align}
where $\vel^{-}$, $\mathbf{v}'$ and $\vel^{+}$ are intermediate time-step solutions for the velocity. The coefficients $a$ and $b$ are calculated each time-step and defined as $a = q_{\rm i}/(2m_{\rm i}) \dt$ and $b = {2}/({1 + a^{2}\widetilde{\B}^{2}})$. $\widetilde{\E}^{t+\dt/2}_{\rm nores}$ and $\widetilde{\B}^{t + \dt/2}$ are the electric and magnetic fields interpolated to the particles at the intermediate time-step, $t+\dt/2$ (how these are obtained is described in \Sec{sec:integration_scheme}). As shown above, the scheme splits the electric field evolution into two steps. \Eq{eqn:boris_push1} describes the first electric field acceleration step at the beginning of the scheme. The magnetic field evolution is done in between the electric field updates, in the form of two rotations as shown by \Eq{eqn:boris_push2} and~\ref{eqn:boris_push3} \citep[see fig.~4 in][]{Boris1970}. The final electric field advancement is done in \Eq{eqn:boris_push4}, and the velocity update is complete.

The update of the particle positions, $\vec{r}^{t} \rightarrow \vec{r}^{t + \dt}$, is carried out as
\begin{equation}
    \vec{r}^{t + \dt} = \vec{r}^{t} + \frac{\dt}{2} (\vel^t + \vel^{t + \dt}).
\end{equation}

\subsection{Depositing moments from particles}
\label{sec:interpolation&filtering}
\subsubsection{Interpolation}
\label{sec:interpolation}
The PIC method requires particle-to-grid and grid-to-particle interpolations (see Fig.~\ref{fig:hybidPIC_loop}). The grid is used to evolve the electromagnetic fields, while the particles represent the ions, which exhibit Lagrangian motion governed by the electric and magnetic fields through the equations of motion, \Eq{eqn:part_evolution2} and~\ref{eqn:part_evolution1}.

In general, the interpolation operation provides us with the interpolated scalar quantity $\mathcal{Q}$ at position $\vec{r}$, based on a quantity $Q$ at positions $\vec{r}_l$, via
\begin{equation} \label{eq:interp}
    \mathcal{Q} (\vec{r}) = \sum_{l=1}^{L} Q(\vec{r}_l)\,W(\vec{r} - \vec{r}_{l}),
\end{equation}
where $W(\vec{r} - \vec{r}_{l})$ is the weight (interpolation kernel) function, and $l=1\dots L$ denotes the index of all $L$ particles (cells) within the kernel, for interpolations from the particles (grid cells) to the grid (particles). We describe the nearest-grid-point (NGP), cloud-in-cell (CIC), and the triangular-shaped-cloud (TSC) weight functions in \App{app:weight_function}. For vector quantities, Eq.~(\ref{eq:interp}) is performed component-wise.

\begin{figure*}
\centering
   \begin{subfigure}{0.49\linewidth}
   \centering
   \includegraphics[scale=1.4, trim={0.02cm 0.02cm 0.05cm 0.02cm},clip]{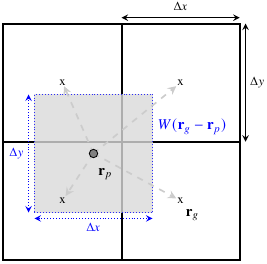}
   \caption{Particle-to-grid (\mbox{P $\rightarrow$ G}) interpolation}
   \label{fig:p-g} 
\end{subfigure}
\hfill
\begin{subfigure}{0.49\linewidth}
   \centering
   \includegraphics[scale=1.4, trim={0.02cm 0.02cm 0.02cm 0.02cm},clip]{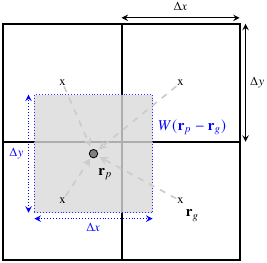}
   \caption{Grid-to-particle (\mbox{G $\rightarrow$ P}) interpolation}
   \label{fig:g-p}
\end{subfigure}
\centering
\caption{Schematic diagram describing (a) particle-to-grid (\mbox{P $\rightarrow$ G}) and (b) grid-to-particle (\mbox{G $\rightarrow$ P}) interpolations with the cloud-in-cell weight function in two-dimensions. The shaded region shows the cloud-in-cell interpolation kernel ($W$), which has the same dimensions as the grid cell, centred on the particle position, $\vec{r}_{p}$. The cell centres, $\vec{r}_{g}$, are depicted by the cross symbols. The arrows show the direction of the interpolation (a) from the particles to the grid cell centres and (b) from the grid cells to the particle (see \Sec{sec:interpolation} for further details).}
\label{fig:weight_function}
\end{figure*}

For particle-to-grid (\mbox{P $\rightarrow$ G}) interpolations, $\vec{r}_{l} = \vec{r}_{p}$ are the particle positions and $\vec{r} = \vec{r}_{g}$ is the centre of each computational grid cell in \Eq{eq:interp}. This is shown by \Fig{fig:p-g}. The quantity $Q(\vec{r}_l) = Q (\vec{r}_{p})$ may be further expressed as a product, $Q (\vec{r}_{p}) = q_p \vel_p^m$, where $q_p$ and $\vel_{p}$ are the charge and the velocity of the ${p}^{\rm th}$ particle, respectively, and $m = 0, 1, 2, ...$ corresponds to the $m^{\rm th}$ moment of the ion distribution function. The most common application of \mbox{P $\rightarrow$ G} interpolations is to compute the charge density, which is the zeroth moment ($m=0$), $\rho_{\rm i}=\mathcal{Q} (\vec{r})/\Delta V$, where $\Delta V=\Delta x \Delta y \Delta z$, based on the charge of each particle, $Q(\vec{r}_{p})=q_p$. Likewise, the first moment ($m=1$) represents the ion current, $\J_{\rm i} = \mathcal{Q} (\vec{r})/\Delta V$, by setting $Q(\vec{r}_{p})=q_p \vel_p$. 

For grid-to-particle (\mbox{G $\rightarrow$ P}) interpolations, $\vec{r}_{l} = \vec{r}_{g}$ and $\vec{r} = \vec{r}_{p}$ in \Eq{eq:interp}. In this case, $Q(\vec{r}_l) = Q(\vec{r}_{g})$ is the grid quantity to be interpolated to the particles. For example, the electromagnetic fields are interpolated from the centre of the computational grid cells to each particle's position using the same weight function used to perform the \mbox{P $\rightarrow$ G} interpolations. This is described by \Fig{fig:g-p}. We note that the default interpolation kernel we use is the CIC weight function.

The influence of using different kernels (NGP, CIC, TSC) is quantified in Sec.~\ref{sec:interpolation_tests}.

\subsubsection{Post-interpolation smoothing to reduce particle noise}
\label{sec:smoothing}
As described above, we sample the moments of the ion distribution function using averages over a population of particles. However, since one can only use a finite number of particles, the distribution function can only be sampled to a certain accuracy, and therefore finite particle counts always introduce noise in any of the particle-to-grid interpolated quantities. This can be particularly problematic for the charge density, as the particle noise may introduce spurious electric fields, due to the fact that the thermo-electric term in Eq.~(\ref{eqn:ohms_law_hp}) generates electric fields based on gradients in the electron pressure (which is proportional to the charge density). Thus, noise in the charge density may lead to large spurious gradients. To address this issue, we perform a smoothing operation on the grid for the ion charge density and current, every time these moments are deposited from the particles to the grid.

The smoothing (filtering) operation is defined such that a \mbox{P $\to$ G} interpolated quantity, $D$, at cell index $(i,j,k)$ is transformed as
\begin{equation}
    D (i,j,k) \to \frac{D(i-1,j,k)}{4} + \frac{D (i,j,k)}{2} + \frac{D(i+1,j,k)}{4},
    \label{eq:smoothing}
\end{equation}
in the $i^{\rm th}$ spatial direction. In case of a 3D simulation, this operation is performed in the $i^{\rm th}$, $j^{\rm th}$ and $k^{\rm th}$ direction, consecutively. This means that during a single smoothing operation, cell $(i,j,k)$ passes 1/2 of its value to its left and right neighbours, and receives 1/4 of the value of its left and right neighbour, respectively, therefore conserving the overall sum.

The smoothing operation (Eq.~\ref{eq:smoothing}) can be repeated  $N_\mathrm{smooth}$ times to achieve a smoother and smoother representation of the interpolated quantity, $D(i,j,k)$. We investigate the appropriateness of the smoothing operation in Sec.~\ref{sec:interpolation_tests} below. We find that using 2~smoothing passes ($N_\mathrm{smooth}=2$) avoids excessive smoothing, retaining small-scale detail, and at the same time provides an appropriate reduction of particle noise, at least in applications of turbulent dynamo amplification. We note that the details of required smoothing passes may depend on the specific problem and on the affordable number of particles per cell (with tests provided in Sec.~\ref{sec:interpolation_tests}).

\subsection{Studying perturbations with Hybrid particle-in-cell simulations}
\label{sec:deltaf}
The $\delta f$ method can be used to study perturbations and instabilities in collisionless plasmas \citep{Kunz+2014b, Kunz+2014a}. In this method, the particles are used to sample the difference between the full distribution function, $f$, and the zeroth-order equilibrium distribution function, $f_{0}$,
\begin{equation}
    \delta f = f - f_{0}.
\end{equation}
First, the ion charge density is deposited onto the grid as 
\begin{equation}
    \rho_{\rm i} (\vec{r}_g) = \rho_{0} (\vec{r}_g) + \Tilde{\rho}_{\rm i} (\vec{r}_g),
\end{equation}
where $\rho_{0} (\vec{r}_g)$ is the equilibrium charge density, and $\Tilde{\rho_{\rm i}} (\vec{r}_g)$ is obtained by a \mbox{P $\rightarrow$ G} interpolation with $Q(\vec{r}_{p})=q_p w_{p}$ in \Eq{eq:interp}, where $w_{p}$ is the factor the particle moments are weighted with in the $\delta f$ method \citep{Parker&Lee1993,Denton&Kotschenreuther1995}. The ion current is deposited as 
\begin{equation}
    \J_{\rm i} (\vec{r}_g) = \rho_{0}(\vec{r}_g) v_{0}(\vec{r}_g) + \Tilde{\J}_{\rm i}(\vec{r}_g),
\end{equation}
where $v_{0}(\vec{r}_g)$ is the equilibrium mean velocity, obtained from $f_{0}$. $\Tilde{\J}_{\rm i} (\vec{r}_g)$ is obtained by a \mbox{P $\rightarrow$ G} interpolation with $Q(\vec{r}_{p})=q_p\vel_p w_p$ in \Eq{eq:interp}. A common choice for the equilibrium distribution function is the Maxwell distribution function, for which $\rho_{0} (\vec{r}_g) = \rho_{0} = q_{\rm i} n_{0}$, where $n_{0}$ is the mean number density, and $v_{0}(\vec{r}_g) = 0$. The $\delta f$ method is useful to study the plasma when $\delta f \ll f$. However, as the perturbations grow and $\delta f \sim f$, the $\delta f$ method is no longer suitable to model the plasma, which is usually the case when turbulence is present.  
\subsection{Constrained transport and solenoidality of magnetic fields}
\label{sec:CT}
\begin{figure*}
    \centering
    \includegraphics[scale=1.0, trim={0cm 0cm 0cm 0cm},clip]{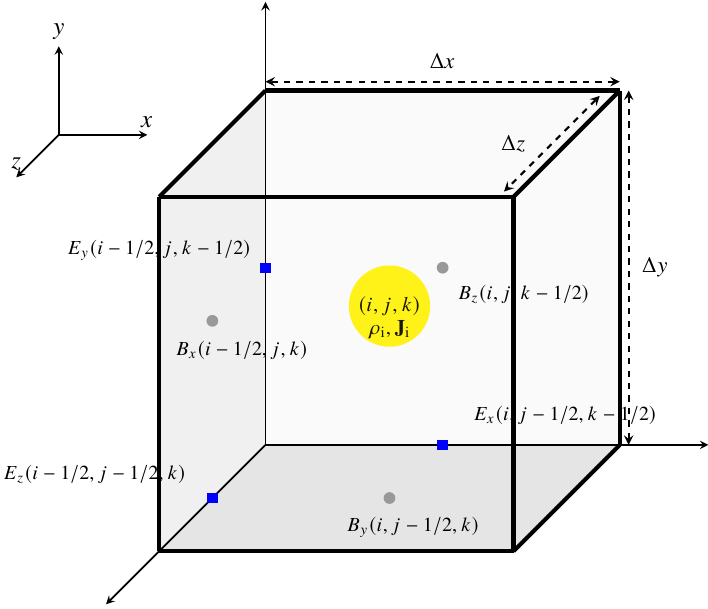}
    \caption{Staggered construction of the electric and magnetic fields for the constrained transport method to solve Faraday's law, \Eq{eqn:faradays_law}. The electric fields are depicted by the blue squares on the edge centres and the magnetic fields are shown by the grey circles on the face centres of the grid cell. The charge density and ion current are always deposited on the grid-cell centre, shown by the yellow circle.}
    \label{fig:constrainted_transport}
\end{figure*}

To update the magnetic fields we use the constrained transport (CT) method on the three-dimensional computational grid to solve Faraday's law, \Eq{eqn:faradays_law} \citep{Yee1966}. \Fig{fig:constrainted_transport} shows a schematic of a single grid cell with the location of the ion charge density, ion current, and electric and magnetic fields. The charge and current density are sampled from the particles and deposited at the centre of each grid cell as $\rho_{\rm i}(i, j, k)$ and $\J_{\rm i}(i, j, k)$, respectively. We note that $(i, j, k)$ represents the coordinate of the cell centre denoted as $\vec{r}_g$, which is described in \Sec{sec:interpolation}. 

In the following, we discretize the electromagnetic fields on the grid, where $i \rightarrow i + 1$ represents $x \rightarrow x + \dx$, where $\dx$ is the grid cell length along the $x$ direction, and analogously for $j$ and $k$ in the $y$ and $z$ direction, respectively. The electric fields, $\E$, required to update the magnetic fields, $\B$, are calculated on the edge centres of the grid cell using the Ohm's law, \Eq{eqn:ohms_law_hp_hyperres}. The $x$-component of the electric field is calculated on the edge centre parallel to the $x$-axis, $\ex(i, j-1/2, k-1/2)$. The $y$ and $z$-components of the electric field are defined in analogy to the $x$-component. These are shown by the blue squares on the edge centre of the grid cell in \Fig{fig:constrainted_transport}. 

The magnetic fields are calculated on the face-center of the grid cell and represent the face-averaged value of the magnetic fields. The $x$-component of the magnetic field is calculated on the face of the grid cell parallel to the $y$-$z$ plane, $\bx(i-1/2, j, k)$. The magnetic field's $y$ and $z$-components are defined similarly to the $x$-component. These points are depicted by the grey circles in \Fig{fig:constrainted_transport}.

Next, we discretize Faraday's law and update the magnetic fields in time. In one time step, i.e., $t \rightarrow t + \dt$, the evolution of the magnetic field components is given by
\begin{equation}
\begin{aligned}
    \bx^{t+\dt}(i-1/2, j, k) = \bx^{t}(i-1/2, j, k) \\ 
    + \frac{\dt}{\dz} \big[ \widetilde{E}^{t+\dt/2}_y(i-1/2, j, k+1/2) - \widetilde{E}^{t+\dt/2}_y(i-1/2, j, k-1/2)\big] \\ 
    - \frac{\dt}{\dy} \big[\widetilde{E}^{t+\dt/2}_z(i-1/2, j+1/2, k) - \widetilde{E}^{t+\dt/2}_z(i-1/2, j-1/2, k) \big],
\end{aligned}
\label{eqn:bx_CT}
\end{equation}
\begin{equation}
\begin{aligned}
    \by^{t+\dt}(i, j-1/2, k) = \by^{t}(i, j-1/2, k) \\
    + \frac{\dt}{\dx} \big[ \widetilde{E}^{t+\dt/2}_z(i+1/2, j-1/2, k) - \widetilde{E}^{t+\dt/2}_z(i-1/2, j-1/2, k) \big] \\ 
    - \frac{\dt}{\dz} \big[ \widetilde{E}^{t+\dt/2}_x(i, j-1/2, k+1/2) - \widetilde{E}^{t+\dt/2}_x(i, j-1/2, k-1/2) \big],
\end{aligned}
\label{eqn:by_CT}
\end{equation}
\begin{equation}
\begin{aligned}
    \bz^{t+\dt}(i, j, k-1/2) = \bz^{t}(i, j, k-1/2) \\
    + \frac{\dt}{\dy} \big[\widetilde{E}^{t+\dt/2}_x(i, j+1/2, k-1/2) - \widetilde{E}^{t+\dt/2}_x(i, j-1/2, k-1/2)\big] \\ 
    - \frac{\dt}{\dx} \big[ \widetilde{E}^{t+\dt/2}_y(i+1/2, j, k-1/2) - \widetilde{E}^{t+\dt/2}_y(i-1/2, j, k-1/2)\big].
\end{aligned}
\label{eqn:bz_CT}
\end{equation}

We also require the cell-centred values of the magnetic fields to interpolate these fields to the particle positions. The cell-centered values of the magnetic field at $t + \dt$ are obtained by averaging the face-centred values.

The divergence of the magnetic field from the CT method calculated at the cell centre is thus
\begin{equation}
\begin{aligned}
    \nabla \cdot \B^{t+\dt}(i, j, k) = \frac{1}{\dx} \big[ \bx^{t+\dt}(i+1/2, j, k) - \bx^{t+\dt}(i-1/2, j, k) \big] \\
    + \frac{1}{\dy} \big[ \by^{t+\dt}(i, j+1/2, k) - \by^{t+\dt}(i, j-1/2, k) \big] \\
    + \frac{1}{\dz} \big[ \bz^{t+\dt}(i, j, k+1/2) - \bz^{t+\dt}(i, j, k-1/2) \big].
\end{aligned}
\label{eqn:divB_CT}
\end{equation}
Due to the staggered positions of the electric and magnetic fields in the construction of the stencil, we find $\nabla \cdot \B^{t+\dt}(i, j, k) = 0$ when Eqs.~\ref{eqn:bx_CT}--\ref{eqn:bz_CT} are plugged into \Eq{eqn:divB_CT}. The construction of the CT method ensures that the analytical expression, $\nabla \cdot (\nabla \times \vec{C}) = 0$, is satisfied for any vector $\vec{C}$, and therefore it can be used to ensure the solenoidality of the magnetic field, to machine precision.

For comparison with the CT method, we also describe the cell-centred finite difference method to update magnetic fields in \App{app:ccfd}, which we show also ensures that the magnetic fields are divergence-free. Similar to the CT method, the cell-centred finite difference method ensures the solenoidality of magnetic fields (although on a larger stencil, which has exactly twice the linear extent of the CT method) using the construction of the numerical stencil. Our code includes both the CT and the cell-centred finite difference method to update magnetic fields, but we use the CT method as the default.

\subsection{Predictor-predictor-corrector integration scheme}
\label{sec:integration_scheme}
Hybrid PIC codes use different integration schemes to advance the particle positions, velocities, and electromagnetic fields in time. Examples of these are the predictor-corrector method and the cyclic leapfrog (CL) with the current-advance method (CAM-CL) \citep{Holmstrom2009, Holmstrom2011, Winske+2022}. Here we use the predictor-predictor-corrector algorithm \citep{Kunz+2014b}, which requires two predictions for the electric and magnetic fields at an intermediate time-step, $t + \dt/2$. The predictor-predictor-corrector method is second-order accurate and can stably propagate Alfv\'{e}n and whistler waves, as demonstrated in \Sec{sec:wave_test}. A complete description of the predictor-predictor-corrector integration scheme is provided in \cite{Kunz+2014a}. 

As described in \Sec{sec:Boris_push}, to update the particle velocity, $\vel^{t} \rightarrow \vel^{t+\dt}$, we require the electromagnetic fields $\widetilde{\E}^{t+\dt/2}$ and $\widetilde{\B}^{t+\dt/2}$ at the intermediate time-step, $t + \dt/2$. Furthermore, the constrained transport method also requires the electric field, $\widetilde{\E}^{t+\dt/2}$, to advance the magnetic fields, $\B^{t} \rightarrow \B^{t+\dt}$, as discussed in \Sec{sec:CT}. Therefore, we need estimates of $\widetilde{\E}^{t+\dt/2}$ and $\widetilde{\B}^{t+\dt/2}$. 

The electric field is calculated from the Ohm's law, \Eq{eqn:ohms_law_hp_hyperres}, and is a function of the charge density, ion current and magnetic field, $\E(\rho_{\rm i}, \J_{\rm i}, \B)$. In the first predictor step of the predictor-predictor-corrector algorithm, $\B^{t + \dt}_{\rm pred, 1}$ is calculated using $\E^{t}$. The first prediction for the electric field, $\E^{t + \dt}_{\rm pred, 1}$, is calculated using $\rho_{\rm i}^{t}$, $\J_{\rm i}^{t}$ and $\B^{t + \dt}_{\rm pred, 1}$. Once $\E^{t+\dt}_{\rm pred, 1}$ and $\B^{t+\dt}_{\rm pred, 1}$ are estimated, the electromagnetic fields at $t + \dt/2$ are calculated as 
\begin{align} \label{eqn:et+1/2}
        & \widetilde{\E}^{t+\dt/2} = \frac{1}{2} (\E^{t} +\E^{t+\dt}_{\rm pred, 1}),\\ \label{eqn:bt+1/2}
        & \widetilde{\B}^{t+\dt/2} = \frac{1}{2} (\B^{t} +\B^{t+\dt}_{\rm pred, 1}).
\end{align}
The second prediction for the magnetic field, $\B^{t+\dt}_{\rm pred, 2}$, is calculated using $\widetilde{\E}^{t+\dt/2}$. We advance the particles from $t \rightarrow t + \dt$ using $\widetilde{\E}^{t+\dt/2}$ and $\widetilde{\B}^{t+\dt/2}$ to obtain $\rho_{\rm i}^{t+\dt}$ and $\J_{\rm i}^{t+\dt}$, which we use along with $\B^{t+\dt}_{\rm pred, 2}$ to predict $\E^{t+\dt}_{\rm pred, 2}$. Advancing the particles is computationally the most expensive step in the algorithm, but using an additional particle-advancement step helps us better estimate the electromagnetic fields directly from the motion of the particles. The quantities $\widetilde{\E}^{t+\dt/2}$ and $\widetilde{\B}^{t+\dt/2}$ are then re-calculated using $\E^{t+\dt}_{\rm pred, 2}$ and $\B^{t+\dt}_{\rm pred, 2}$, similar to \Eq{eqn:et+1/2}-\ref{eqn:bt+1/2},
\begin{align}
        & \widetilde{\E}^{t+\dt/2} = \frac{1}{2} (\E^{t} +\E^{t+\dt}_{\rm pred, 2}),\\ 
        & \widetilde{\B}^{t+\dt/2} = \frac{1}{2} (\B^{t} +\B^{t+\dt}_{\rm pred, 2}).
\end{align}
Finally, ${\B}^{t+\dt}$ is calculated using $\widetilde{\E}^{t+\dt/2}$ and the particle positions and velocities are updated using $\widetilde{\E}^{t+\dt/2}$ and $\widetilde{\B}^{t+\dt/2}$.

\subsection{Corrections for interpolated electromagnetic fields}
\label{sec:E&B_corrections}
In the absence of thermo-electric and resistive effects, the electric and magnetic fields are perpendicular to each other during their calculation on the computational grid. However, when these fields are interpolated to the particles, this may no longer be true due to interpolation errors. To address this issue, we perform the following correction to the interpolated electric field at the particle positions \citep{Lehe2009, Kunz+2014a, AchikanathChirakkara+2023},
\begin{equation}
    \E_{\rm nores}^{*} = \E_{\rm nores} + \left[ (\E_{\rm nores} \cdot \B)_{\rm G \rightarrow P} - \E_{\rm nores} \cdot \B  \right] \frac{\B}{|\B|^{2}},
\end{equation}
where all the terms in the above equation are fields interpolated to the particle positions.
$(\E_{\rm nores} \cdot \B)_{\rm G \rightarrow P}$ is the dot product of $\E_{\rm nores}$ and $\B$, calculated on the grid and interpolated to the particle positions.
$\E_{\rm nores}^{*}$ is the corrected electric field, which the particles experience. 

\subsection{Hyper-resistivity}
\label{sec:hyperres}
Hyper-resistivity is introduced in Hybrid PIC codes to remove the propagation of grid-scale fluctuations by damping magnetic field noise on grid-cell scales (hence the use of a 4th-order spatial derivative). To do this, we add a hyper-diffusivity ($\eta_{\rm hyper}$) term on the right-hand side of the Ohm's law, \Eq{eqn:ohms_law_hp},
\begin{equation}
    \E = \frac{\left( \J - \J_{\rm i} \right) \times \B}{\rho_{\rm i}} - \Big( \frac{k_{\rm B} T_{\rm e}}{q_{\rm e}} \Big) \frac{\nabla \rho_{\rm i}}{\rho_{\rm i}} + \mu_{0} \eta \J - \mu_{0} \eta_{\rm hyper} \nabla^{2} \J.
\label{eqn:ohms_law_hp_hyperres}
\end{equation}
The choice for the resistivity and hyper-resistivity depends on the physical problem and the grid resolution \citep[for more details, see Sec.~2.5.2 in][]{AchikanathChirakkara+2023}.

\subsection{Turbulent driving}
\label{sec:driving}
To drive turbulence, we add a turbulent acceleration field $\vec{f}$ to \Eq{eqn:part_evolution1},
\begin{equation}
    \frac{\dd \vel_{p} }{\dd t} = \frac{q_{\rm i}}{m_{\rm i}}\left(\E_{\rm nores} + \vel_{p} \times \B \right) + \vec{f}.
\end{equation}
The turbulent acceleration field $\vec{f}$ is constructed in Fourier space and evolved by an Ornstein-Uhlenbeck process, using \texttt{TurbGen} \citep{Federrath+2010,FederrathEtAl2022ascl}.
We drive turbulence on large scales in our numerical simulations. This turbulent energy injection is restricted to large scales (usually half the computational domain). This energy then self-consistently cascades to smaller and smaller scales, where it is ultimately dissipated as heat \citep{Frisch1995,Federrath+2021}. Thus, the injection of turbulent energy via driving leads to an increase in the temperature of the ions.

The turbulent velocity fluctuations excited via driving can be quantified in terms of the sonic Mach number,
\begin{equation}
    \Mach = \frac{\Vturb}{\Vtherm},
\end{equation}
where $\Vturb$ is the turbulent speed and $\Vtherm$ is the thermal speed. As a result of energy dissipation on small scales, the temperature increases in the absence of cooling, and therefore, the thermal speed increases and the Mach number of the plasma decreases. This means that statistically-steady turbulence cannot be maintained in the absence of cooling, which is problematic, because the properties of the turbulence are sensitive to the Mach number \citep{CFetal11, Seta&Federrath2021b, AchikanathEtAl2021, Seta&Federrath2022}. This problem is even more significant for turbulence in the supersonic regime ($\Mach > 1$), where $\Vtherm$ increases faster compared to the subsonic regime ($\Mach < 1$), as more turbulent energy is dissipated in the supersonic case, due to dissipation in shocks.

\subsection{Plasma cooling}
\label{sec:cooling}
\begin{figure*}
    \includegraphics[width=1.0\linewidth,trim={0.05cm 0.0cm 0.0cm 0.0cm},clip]{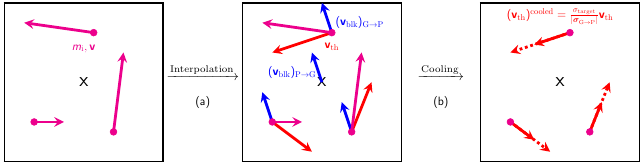}
    \caption{Schematic diagram of the cooling method. A single grid cell with 3~particles is shown for simplicity. (a) Using interpolation operations, the particle velocity is decomposed into the bulk, $(\vel_{\rm blk})_{\rm G \rightarrow P}$, and the particle thermal velocity, $\vel_{\rm th}$, as described in \Eq{eq:vdecomp}. (b) Then $\vel_{\rm th} = \vel - (\vel_{\rm blk})_{\rm G \rightarrow P}$, is rescaled to the target value, $\sigma_{\rm target}$, and $(\vel_{\rm blk})_{\rm G \rightarrow P}$ reattached (not shown in the schematic), as described in \Eq{eqn:cooling}. This method ensures isothermal conditions locally and globally throughout the plasma.}
    \label{fig:cooling_chart}
\end{figure*}

As explained in the previous section, maintaining isothermal plasma conditions is critical for certain physics applications and controlled numerical experiments. Therefore, to maintain an isothermal plasma, we introduce a cooling method, with a schematic diagram shown in \Fig{fig:cooling_chart}. The goal of this cooling method is to achieve a constant (in space and time) target ion thermal speed,
\begin{equation}
    \sigma_{\rm target} = \Bigg( \frac{N_{d} k_{\rm B} T_{\rm target}}{m_{\rm i}} \Bigg)^{1/2},
\end{equation}
where $T_{\rm target}$ is the target ion temperature and $N_{d} = 1, 2$ or 3, depending on the dimensionality of the problem. We perform this cooling on a user-controlled timescale, $(\Delta t)_{\rm cool}$ (in units of $\tcool$), which is tested and calibrated in Sec.~\ref{sec:cooling_tests}. When $T_{\rm target} = T_{\rm e}$, cooling ensures that electrons and ions have the same temperature.

We begin by noting that the total particle velocity can be decomposed as
\begin{equation} \label{eq:vdecomp}
    \vel = (\vel_{\rm blk})_{\rm G \rightarrow P} + \vel_{\rm th},
\end{equation}
i.e., the sum of the local bulk flow velocity, $(\vel_{\rm blk})_{\rm G \rightarrow P}$, and the particle thermal velocity, $\vel_{\rm th}$. To cool the plasma, we first perform ${\rm P \rightarrow G}$ interpolations using \Eq{eq:interp} with $Q (\vec{r}_{p}) = m_{\rm i} \vel_p$ and $Q (\vec{r}_{p}) = m_{\rm i} \vel_p^{2}$, where both quantities are vectors with each component representing the 3~spatial coordinate directions, i.e., $\vel_p=(v_{p,x},\,v_{p,y},\,v_{p,z})$ and $\vel_p^{2}=(v_{p,x}^2,\,v_{p,y}^2,\,v_{p,z}^2)$, respectively. Following this, the two vector quantities are interpolated back from the computational grid to the particle locations as $(m_{\rm i}\vel)_{\rm G \rightarrow P}$ and $(m_{\rm i}\vel^{2})_{\rm G \rightarrow P}$, respectively. The ion mass density on the computational grid, $\rho_{m}$, is also interpolated from the grid to the particle positions (${\rm G \rightarrow P}$) as $(\rho_{m})_{\rm G \rightarrow P}$ using \Eq{eq:interp} with $Q (\vec{r}_{g}) = \rho_{m}$. These back-and-forth interpolation operations are performed to obtain the bulk velocities and thermal speed at the particle positions. At each particle position, we then have
\begin{align} \label{eqn:blk_def}
        & (\vel_{\rm blk})_{\rm G \rightarrow P} = \frac{(m_{\rm i}\vel)_{\rm G \rightarrow P}}{(\rho_{m})_{\rm G \rightarrow P}}, \\
        & (\vel^{2})_{\rm G \rightarrow P} = \frac{(m_{\rm i}\vel^{2})_{\rm G \rightarrow P}}{(\rho_{m})_{\rm G \rightarrow P}}.
\end{align}
Using the above, we can calculate 
\begin{equation} \label{eqn:temp_def}
    {\bm \sigma}_{\rm G \rightarrow P} = \left((\vel^2)_{\rm G \rightarrow P} - (\vel_{\rm blk})_{\rm G \rightarrow P}^{2}\right)^{1/2},
\end{equation}
where ${\bm \sigma}_{\rm G \rightarrow P}$ is the thermal speed in the $x$, $y$ and $z$ directions at the particle locations. 

Having gathered all this information, we are now ready to perform the cooling step. From Eq.~(\ref{eq:vdecomp}), each particle's thermal velocity, $\vel_{\rm th} = \vel - (\vel_{\rm blk})_{\rm G \rightarrow P}$, which is re-scaled (corrected) by the proportional mismatch between the measured ${\bm \sigma}_{\rm G \rightarrow P}$ from Eq.~(\ref{eqn:temp_def}) and the target thermal speed, then the bulk speed is reattached, such that the total particle velocity after the cooling step is set to
\begin{equation}
\label{eqn:cooling}
\begin{aligned}
\vel \to \vel_{\rm cooled} &= \frac{\sigma_{\rm target}}{{|\bm \sigma}_{\rm G \rightarrow P}|} \vel_{\rm th} + (\vel_{\rm blk})_{\rm G \rightarrow P} \\
& = \frac{\sigma_{\rm target}}{{|\bm \sigma}_{\rm G \rightarrow P}|} (\vel - (\vel_{\rm blk})_{\rm G \rightarrow P}) + (\vel_{\rm blk})_{\rm G \rightarrow P}.
\end{aligned}
\end{equation}
This re-scaling adjusts the thermal velocity component of each particle, while keeping the direction of the thermal velocity component at each particle position unchanged (i.e., using the magnitude, $|{\bm \sigma}_{\rm G \rightarrow P}|$ of the three~${\bm \sigma}_{\rm G \rightarrow P}$ components in Eq.(\ref{eqn:cooling}), to maintain a constant target plasma temperature.

The cooling is performed every $(\Delta t)_{\rm cool}$, in units of $\tcool$, which is defined as
\begin{equation} \label{eqn:tcool_def}
    \tcool = \frac{\tth}{\Mach},
\end{equation} 
using the thermal crossing time, $\tth = \dl/\Vtherm$, where $\dl = {\rm min} (\dx, \dy, \dz)$ is the minimum side length of a computational grid cell and $\Vtherm$ is the thermal speed. Cooling is done after the predictor-predictor-corrector integration scheme (see \Sec{sec:integration_scheme}) is complete. This method ensures isothermal plasma conditions locally (on the scale of individual grid cells), and by extension also globally.

\subsection{Time-step constraints}
\label{sec:timesteps}
To resolve the different physical timescales involved in weakly collisional plasmas, the following time-step constraints must be fulfilled: 
\begin{enumerate}
    \item Larmor gyration: to resolve the Larmor motion of ions in the presence of a magnetic field, one needs to use a time step that is significantly smaller than the timescale required for a charged particle to perform one gyration about the magnetic field,
    \begin{equation}
        \tL = \frac{2 \pi m_{\rm i}}{q_{\rm i} |\B|}, \label{eq:tL}
    \end{equation}
    where $m_{\rm i}$ and $q_{\rm i}$ are the mass and charge of the ions, respectively. The required time step to resolve Larmor gyration is problem-dependent, and will be tested in Sec.~\ref{sec:particle_test}.
    \item Particle speed: the particle speeds set a time-step constraint determined by the time taken by the fastest moving particle in the computational domain to travel across the grid cell of minimum side length $\dl = {\rm min} (\dx, \dy, \dz)$. This is given by
    \begin{equation}
        t_{\rm part} = \frac{\dl}{{|\vel|}_{\rm max}}, \label{eq:t_vpart}
    \end{equation}
    where $|\vel|_{\rm max}$ is the speed of the fastest particle within the domain.
    \item Alfv\'{e}n waves: the timescale for the propagation of Alfv\'{e}n waves across a computational grid with cells of minimum side length $\dl$ is
    \begin{equation}
        t_{\rm  \text{Alfv\'{e}n}} = \frac{\dl}{v_\mathrm{A}},
    \end{equation}
    with the Alfv\'en speed $v_\mathrm{A} = |\B|/\sqrt{\mu_{0} m_{\rm i} n_{\rm i}}$,     where $n_{\rm i}$ and $\mu_{0}$ are the number density and the vacuum magnetic permeability, respectively.
    \item Whistler waves: the timescale for the propagation of whistler waves across a grid cell is
    \begin{equation}
        t_{\rm whistler} = \frac{\dl}{v_{\rm whistler}},
    \end{equation}
    where $v_{\rm whistler} = \pi|\B|/(q_{\rm i}\mu_{0}n_{\rm i}\dl)$ is the speed of whistler waves with wavenumber, $k = 2\pi/2\dl$. Note that $t_{\rm whistler}\propto\dl^2$.
    \item Ohmic resistivity: the magnetic diffusivity of the plasma, $\eta$, sets the resistive timescale
    \begin{equation}
        t_{\eta} = \frac{(\dl)^2}{\eta}.
    \end{equation}
    \item Hyper-resistivity: the higher-order numerical 
    diffusivity sets an additional dissipative time step, the hyper-resistive timescale
    \begin{equation}
        t_{\eta_{\rm hyper}} = \frac{(\dl)^4}{\eta_{\rm hyper}},
    \end{equation}
    where $\eta_{\rm hyper}$ is the hyper-diffusivity.
    \item Thermal crossing: the thermal crossing timescale is the time taken by a sound wave to cross a grid cell, defined as
    \begin{equation}
        \tth = \frac{\dl}{\Vtherm},
    \end{equation}
    where $\Vtherm$ is the thermal speed of the plasma. 
\end{enumerate}

The overall time step is calculated as the minimum of the all above time-step constraints, 

\begin{equation}
\begin{aligned}
    \dt = {\rm min} \{ s_{\rm L} t_{\rm Larmor}, s_{\rm part} t_{\rm part}, s_{\rm A} t_{\rm  \text{Alfv\'{e}n}}, s_{\rm w}t_{\rm whistler}, s_{\rm \eta}t_{\eta},\\
    s_{\rm \eta_{\rm hyper}}t_{\eta_{\rm hyper}}, s_{\rm th}\tth \},
\end{aligned}
\end{equation}
where $s_{\rm L}$, $s_{\rm part}$, $s_{\rm A}$, $s_{\rm w}$, $s_{\rm \eta}$, $s_{\rm \eta_{\rm hyper}}$, and $s_{\rm th}$, are the safety factors for the Larmor, particle speed, \text{Alfv\'{e}n}, whistler, resistive, hyper-resistive, and thermal time steps, respectively. We introduce these safety factors to ensure that each time step is sufficiently resolved. The default value for all the safety factors is 0.1, except for the particle-speed time step, which is $s_{\rm part}=0.5$. Depending on the physical problem, these safety factors can be adjusted.

\section{Particle, wave, and wave-particle interaction tests}
\label{sec:tests}
To test the accuracy of our code, we perform the following standard tests and compare our numerical results with the analytical solutions. First, we test the Larmor gyration of charged particles in \Sec{sec:particle_test}. In \Sec{sec:wave_test} we test the propagation of Alfv\'{e}n and whistler waves. Finally, we test whether the code can capture physical instabilities arising from particle-wave interactions in \Sec{sec:landau_damping}.

\subsection{Charged particle in a uniform magnetic field}
\label{sec:particle_test}
We test our implementation of the Boris integrator by studying the motion of a charged particle in a uniform magnetic field. We set up an external (non-evolving) magnetic field, $\B = B_{0}\,\hat{z}$, where $B_{0} = 10^{-8}$ Tesla, and give the charged particles an initial velocity perpendicular to the direction of the magnetic field, $\vel = v_{0}\,\hat{x}$, where $v_{0} = 1.0\,\m\,\s^{-1}$. Due to the effect of the Lorentz force, the charged particles will gyrate in a circle in the $x$-$y$ plane with the Larmor radius, $\rL = m_{\rm i} v_{0} / q_{\rm i} B_{0}$, where $m_{\rm i}$ and $q_{\rm i}$ are the mass and charge of the particle, respectively. The analytical solution for the evolution of the $x$-coordinate of the particle's position is $x(t) = \rL\,\sin{(\Omega_{\rm i} t)}$, where $\Omega_{\rm i} = q_{\rm i} B_{0}/m_{\rm i}$ is the gyration frequency of the charged particle. The total kinetic energy of the particle does not change as the magnetic field does not do any work on the particle. 

\begin{figure}
    \centering
    \includegraphics[width=1.0\linewidth]{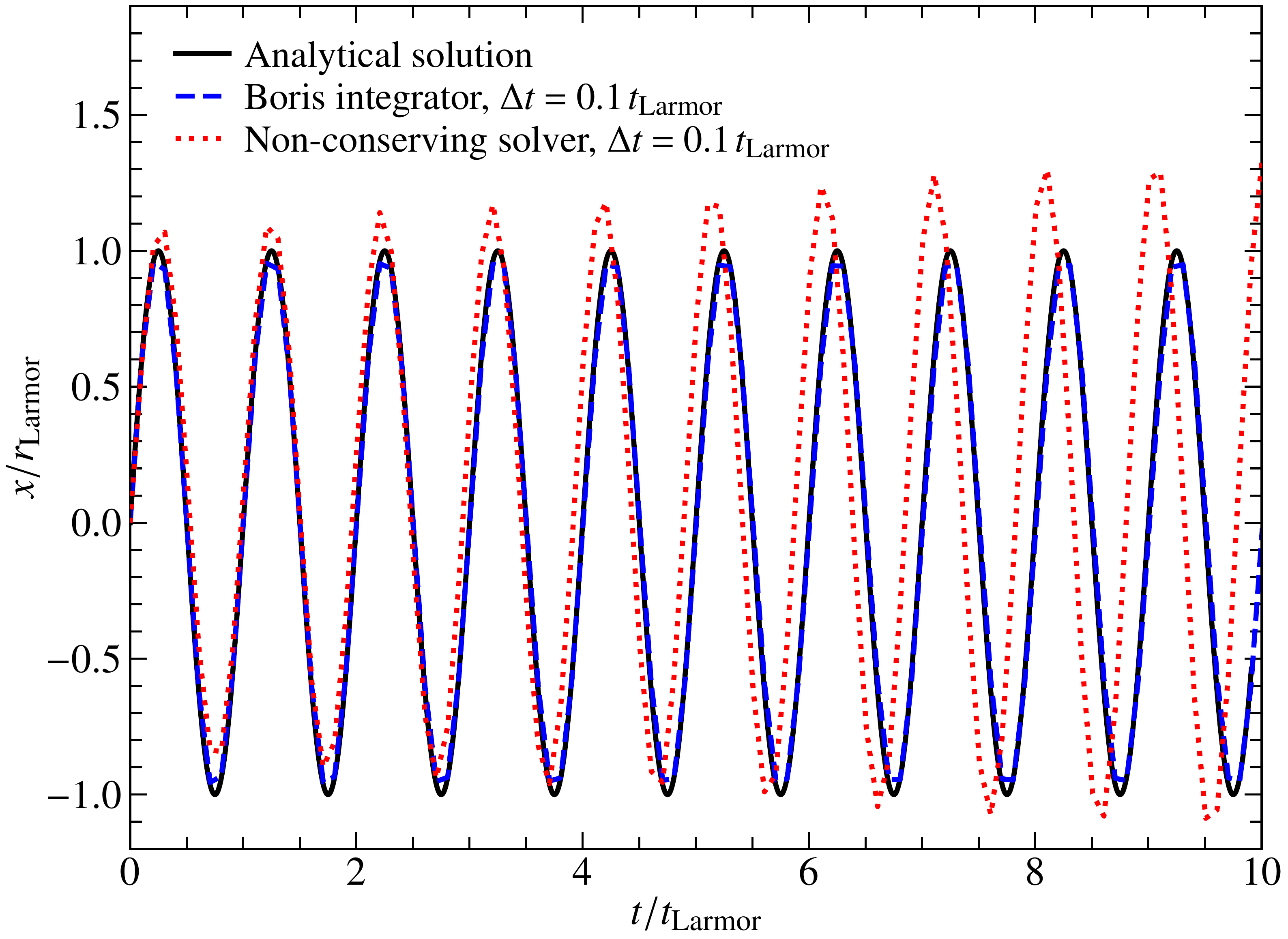}
    \caption{The $x$-coordinate of a charged particle undergoing gyrations in the presence of a uniform magnetic field, normalised to the Larmor radius ($\rL$), as a function of time in units of the Larmor time ($\tL$). We plot this for simulations with $\dt = 0.1\,\tL$ for the Boris integrator and a non-conserving particle integrator. The solid black line depicts the analytical solution. We find that the gyration radius of our numerical solutions does not change with the Boris integration method, while with the non-conserving particle integrator, the gyration radius increases over time.}
    \label{fig:particle_test_x}
\end{figure}

We pick the time step, $\dt = 0.1\,\tL$, where $\tL = 2 \pi / \Omega_{\rm i}$ is the time taken by the charged particle to complete one gyration. \Fig{fig:particle_test_x} shows the $x$-coordinate of the particle, normalised to $\rL$, as a function of time, normalised to the Larmor time ($\tL$, defined in Eq.~\ref{eq:tL}), along with the analytical solution. We also perform this test with a non-conserving cyclic leapfrog particle integrator \citep{Holmstrom2009}. 

We find that the amplitude (radius) of the Larmor gyration is stable in the Boris integration method, as the energy is conserved up to the machine precision level in this case. However, with a non-conserving particle integrator (dotted red line), the radius of gyration increases over time, due to the particle gaining energy as a result of integration inaccuracies. Thus, we conclude that using the Boris integrator provides the basic requirement for stably modelling Larmor gyration, and thus we use this integration scheme as the default for all PIC simulations.

\begin{figure}
    \centering
    \includegraphics[scale=0.355]{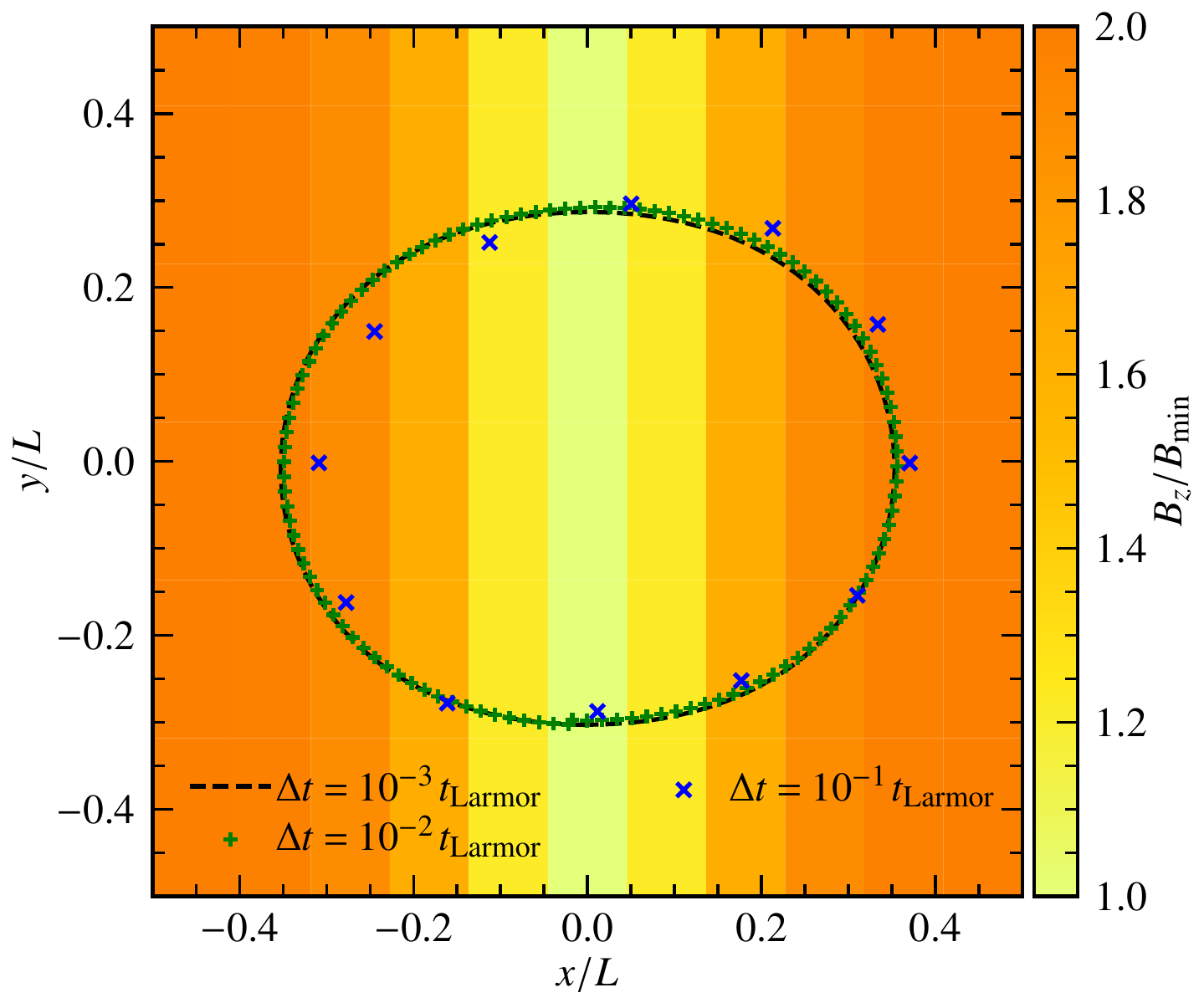}
    \caption{Orbit of a charged particle in a spatially varying magnetic field, $B_{z}(x)$, described by \Eq{eqn:nonuniform_B}, using the Boris integrator for varying time steps, $\dt/\tL = 10^{-3}, 10^{-2}$, and $10^{-1}$. Using a relatively large $\dt/\tL = 10^{-1}$, we find that the trajectory of the particle deviates significantly from the cases that use $\dt/\tL = 10^{-2}$ and $10^{-3}$. While the required time step depends on the physical problem, this test shows that it is important to pick a time step that can suitably resolve the trajectory of the particles.}
    \label{fig:particle_test_nonuniformB}
\end{figure}

While the Boris integrator provides stable Larmor (circular) orbits for any choice of time step $\dt$, for a uniform magnetic field it may not do so for a spatially varying magnetic field, and instead, the choice of time step may play a role. This is tested next, where we follow the motion of a charged particle in a non-uniform magnetic field,
\begin{equation} \label{eqn:nonuniform_B}
    B_{z} (x) = B_{\rm min} \left\{2 - \exp\left[-\left(\frac{x}{2\dx}\right)^2\right]\right\},
\end{equation}
where $x$ is the $x$-coordinate of the grid cell centre, $\dx$ is the cell length in the x-direction and $B_{\rm min}$ is the magnetic field strength at $x=0$. This profile is shown by the colour map in \Fig{fig:particle_test_nonuniformB}. The charged particle is given an initial velocity in the $x$-direction, similar to the test in Fig.~\ref{fig:particle_test_x}, and we plot the results up to $1\,\tL$, where $\tL$ is calculated based on the maximum magnetic field strength, $B_{\rm max} = 2 B_{\rm min}$. We vary the time step, $\dt/\tL = 10^{-3}, 10^{-2}$ and $10^{-1}$, and show the results in \Fig{fig:particle_test_nonuniformB}. The near-perfect orbit is obtained for a very small time step--here $\dt/\tL=10^{-3}$ (dashed line). For a relatively large time step, $\dt/\tL = 10^{-1}$ (blue crosses), we see that the orbit is significantly distorted, with a clear orbital precision. Using $\dt/\tL = 10^{-2}$ (green pluses), the trajectory is fairly close to the optimal orbit. This test demonstrates that it is important to choose a suitable time step to achieve the desired accuracy in the trajectories of the charged particles, especially when the magnetic field is non-uniform, which is almost always the case for PIC applications.

\subsection{Propagation of waves in collisionless plasma}
\label{sec:wave_test}
By using perturbation analysis on the Hybrid-kinetic equations, one can obtain the dispersion relation for the propagation of circularly polarized waves. Right-handed waves follow the dispersion relation $\omega_{\rm R}^{2} = k^{2}(1+\omega_{\rm R})$ and left-handed waves follow the dispersion relation $\omega_{\rm L}^{2} = k^{2}(1-\omega_{\rm L})$ \citep{Kulsrud2005}. We note that in these equations, the frequency, $\omega$, is normalised to the ion gyration frequency, $\Omega_{\rm i}$, and the wavenumber, $k$, is normalised to the ion inertial length, $d_{\rm i}$. In the low-wavenumber limit, $k \ll 1$, both the right- and left-handed waves follow the Alfv\'en wave dispersion relation, $\omega^{2} = k^{2}$.
In the high-wavenumber limit, $k \gg 1$, the dispersion relation for the right-handed waves is that of whistler waves, $\omega_{\rm R} = k^{2}$, and for the left-handed waves, it is that of the ion-cyclotron waves, $\omega_{\rm L} \rightarrow 1$ asymptotically as $k \rightarrow \infty$.

To test this with our numerical code, we initialise a mean magnetic field, $B_{0}$, along the $x$-axis and oscillatory magnetic fields along the $y$ and $z$-axes, $B_{y} = \delta B \sin(kx)$ and $B_{z} = \delta B \cos(kx)$, where $k = 2\pi/L_{x}$, $\delta B = 10^{-3} B_{0}$ and $L_{x}$ is the length of the computational domain in the $x$-direction. Our test simulations have cold ions and electrons, $T_{\rm i} = T_{\rm e} = 0$. The Ohmic dissipation and hyper-resistivity have been switched off in these tests. We choose $\nppc = 100$ particles per cell and periodic boundary conditions. We also resolve the Alfv\'en and whistler wave time steps and use a safety factor of 0.1 for both these time steps (see \Sec{sec:timesteps} for details). 

\subsubsection{Wavelength dependence}
\label{sec:wave_tests_results}
\begin{figure}
    \centering
    \includegraphics[width=1.0\linewidth]{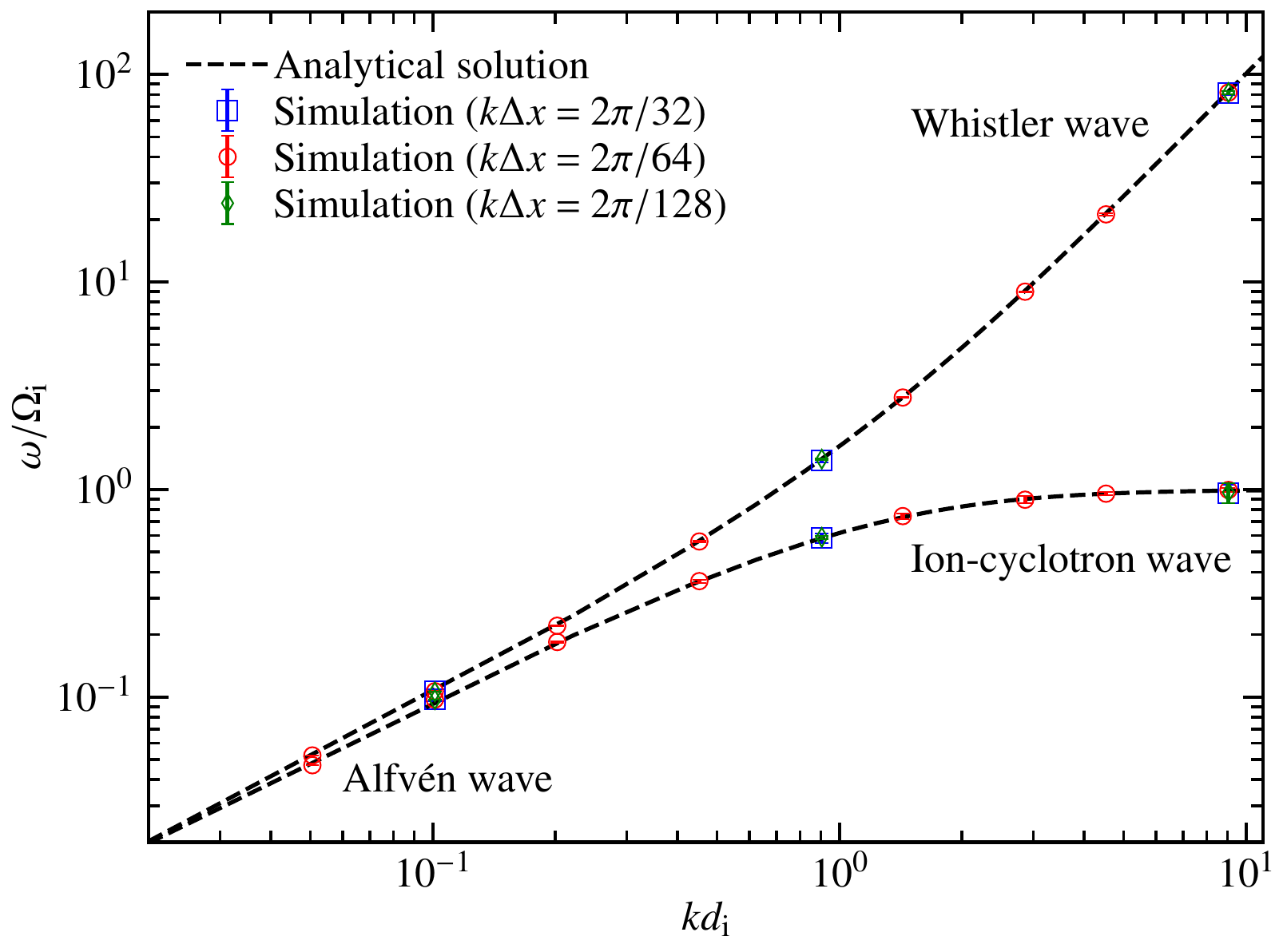}
    \caption{Dispersion relationship of Alfv\'{e}n, whistler and ion-cyclotron waves, i.e., frequency of the propagating wave, $\omega$, normalised to the ion gyration frequency, $\Omega_{\rm i}$, for different values of the wavenumber $k$ (in units of the ion inertial length, $d_{\rm i}$), for a wide range of wavenumbers, $k d_{\rm i} = [0.05, 9]$ with $k\dx = 2\pi/64$ (i.e. 64 grid cells per wavelength), where $\dx$ is the grid-cell length in the direction of wave propagation. The red points show the numerical results with our new code, $\ahkash$. The black dashed curves show the analytical solutions for the left-handed and right-handed waves. In the low-wavenumber regime ($k d_{\rm i} \ll 1$), the Alfv\'{e}n wave dispersion relation ($\omega/\Omega_{\rm i} = k d_{\rm i}$) is followed and at higher wavenumbers ($kd_{\rm i} \gg 1$), the whistler waves ($\omega/\Omega_{\rm i} = (k d_{\rm i})^{2}$) and the ion-cyclotron waves ($\omega = \Omega_{\rm i}$) appear.  The measured errors in the frequencies are very small ($\lesssim 4 \%$).  For the tests with $k d_{\rm i} = 0.1, 0.9$ and 9, we vary the grid resolution, $k \dx = 2\pi/32$ and $2\pi/128$ and find that the numerical solutions are converged with $k \dx = 2 \pi/32$ in different $k d_{\rm i}$ regimes. We find that our results are in excellent agreement with the analytical solutions and show stable and accurate propagation of waves.}
    \label{fig:wave_propagation}
\end{figure}

The initial wave propagates across the computational domain and to accurately extract the frequencies of the waves, we perform a Fourier analysis on the time evolution data. We then use a lognormal fitting function to fit the frequency peaks in the data. The frequency we report is the mean of the lognormal distribution and the standard deviation of the distribution is a measure of its uncertainty. In \Fig{fig:wave_propagation}, we show the frequency of the waves normalised to the ion gyration frequency, $\omega/\Omega_{\rm i}$, for different values of the wavenumber normalised to the ion inertial length, $k d_{\rm i} = [0.05,9]$. We also plot the analytical solutions for the dispersion relation and find excellent agreement with the analytical solutions across all $k d_{\rm i}$ regimes. This shows that our code can stably and accurately capture the propagation of waves.

We pick $k\dx = 2\pi/64$ in our tests, where $\dx$ is the grid-cell length in the direction of wave propagation and $k = 2\pi/\ell$, where $\ell$ is the wavelength. This implies there are 64~cells per wavelength, $\ell = 64 \dx$. The tests with $k d_{\rm i} = 0.1, 0.9$ and 9 are repeated with varying grid resolution, $\dx=\ell/32$ or $k\dx=2\pi/32$ and $\dx=\ell/128$ or $k\dx=2\pi/128$. In \Fig{fig:wave_propagation}, we show that the numerical results with $k\dx = 2\pi/32$ are converged in different $k d_{\rm i}$ regimes.

\subsubsection{Grid resolution criteria}
\begin{figure}
    \centering
    \includegraphics[width=1.0\linewidth]{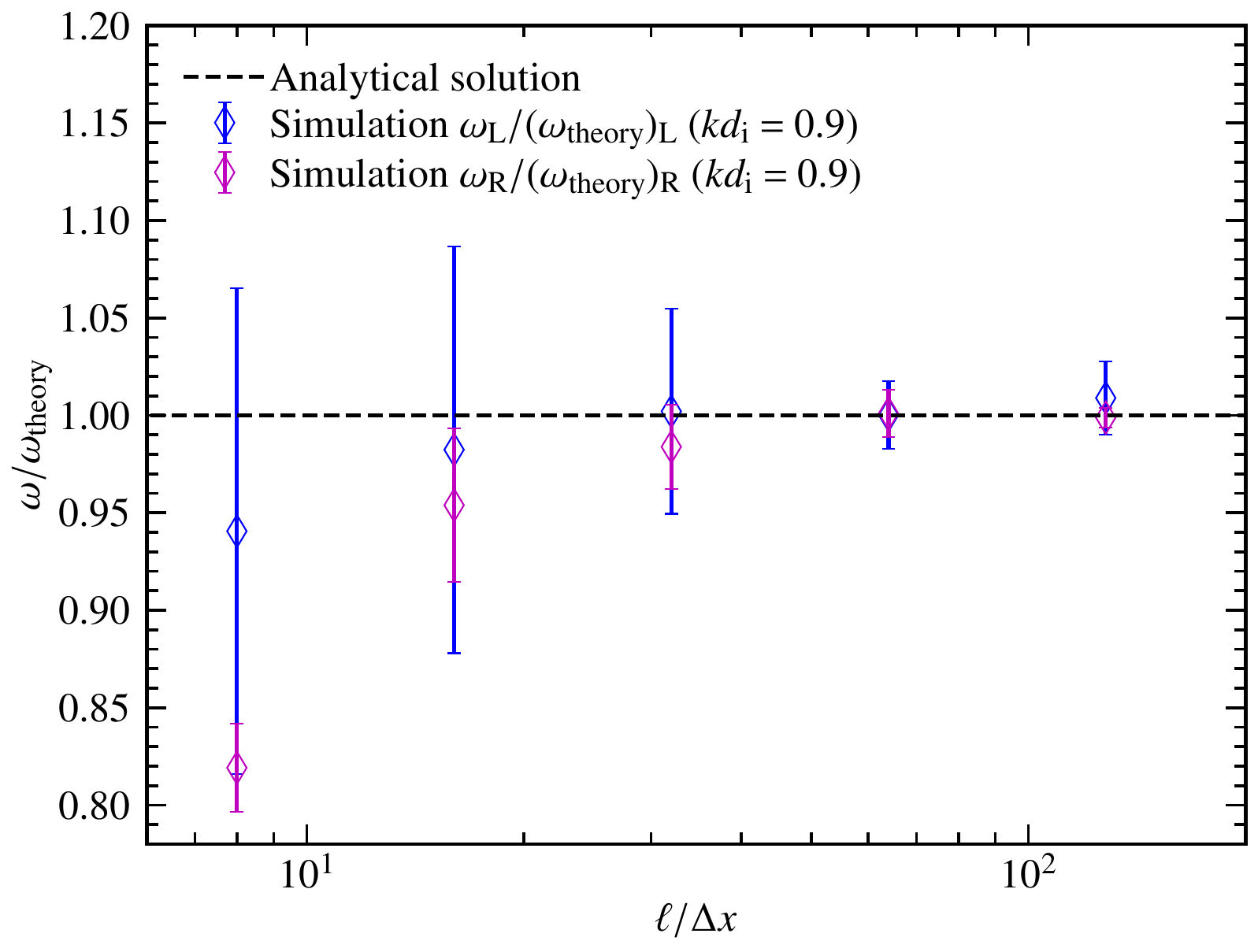}
    \caption{The frequency of the propagating wave, $\omega$, normalised to the theoretical frequency, $\omega_{\rm theory}$, for the wave test with $kd_{\rm i} = 0.9$, as a function of grid resolution, $\ell/\dx$, where $\ell = 2\pi/k$ is the wavelength and $\dx$ is the grid-cell length. For tests with $\dx = \ell/8$ and $\dx = \ell/16$, the analytical solutions are not obtained for the right-handed waves. As the grid resolution is increased $\dx \gtrsim \ell/32$, the analytical solutions are recovered, and the numerical solutions are converged.} 
    \label{fig:wave_propagation_res_test}
\end{figure}

In this section, we study in more detail the effect of the grid resolution on the wave test. We vary the grid-cell length in the direction of wave propagation, $\dx = \ell/8$\,--\,$\ell/128$, where $\ell = 2\pi/k$ is the wavelength, and measure the frequency of wave propagation, $\omega$, normalised to the ion gyration frequency, $\Omega_{\rm i}$, for the test with $kd_{\rm i} = 0.9$. The results are shown in \Fig{fig:wave_propagation_res_test}. We find that with $\ell = (32$\,--\,$128)\,\dx$, the analytical solutions are recovered (to within one sigma) and the numerical solutions are converged with $\ell = 32\dx$. Using a lower grid resolution, the analytical solutions are not obtained, and we also find spurious frequencies start to dominate in the Fourier analysis.

\subsection{Wave-particle interaction}
\label{sec:landau_damping}
Here we test the decay of ion acoustic waves due to Landau damping using the $\delta f$ method (see \Sec{sec:deltaf}). Landau damping of ion acoustic waves follows the dispersion relation $d Z (\zeta) / d \zeta = 2\,T_{\rm i}/T_{\rm e}$, where $Z (\zeta)$ is the plasma dispersion function and $\zeta$ is the phase velocity, which has a real (oscillatory) and an imaginary (decaying) part, and $T_{\rm i}$ and $T_{\rm e}$ are the ion and electron temperature, respectively. The decay rate of the wave depends on the ratio of the ion to electron temperature. We will compare our numerical solutions with the solutions of the analytical dispersion relation, $d Z/d \zeta = 2\,T_{\rm i}/T_{\rm e}$, which we solve numerically using the Newton-Raphson method for a given $T_{\rm i}/T_{\rm e}$.

The length of our computational domain is $L_{x} = 16\,d_{\rm i}$ along the $x$-axis, where $d_{\rm i}$ is the ion inertial length. We initialise the velocity of the ions (particles) using a Maxwell–Boltzmann distribution with temperature $T_{\rm i}$ and fixed electron temperature $T_{\rm e}$. We initialise the density with the profile $1 + a \cos(kx)$, where we choose a small perturbation amplitude, $a = 0.01$ and the wavenumber of the perturbation, $k = 2\pi/L_{x}$. We used $\nppc = 10^{6}$ particles per cell and periodic boundary conditions.

\subsubsection{Results}
\begin{figure}
    \centering
    \includegraphics[width=1.0\linewidth]{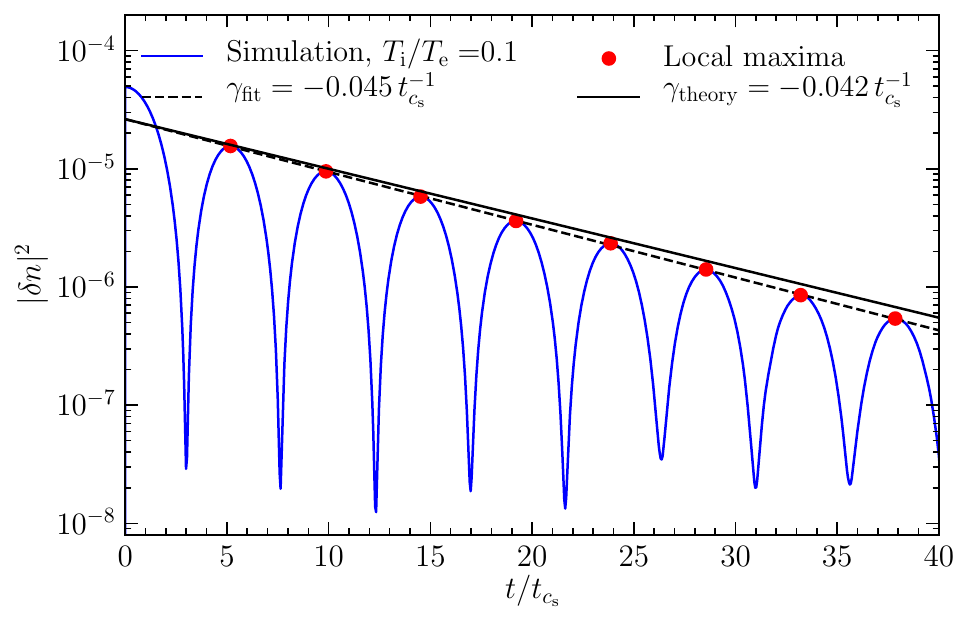}
    \caption{Landau damping of ion acoustic waves shown by the decay of the squared amplitude of the density fluctuation as a function of time, normalised to the sound-crossing time, $t_{c_\mathrm{s}}$, for a ratio of the ion temperature to the electron temperature of $T_{\rm i}/T_{\rm e} = 0.1$. The local maxima used for fitting are shown by the red dots. We fit for the Landau damping decay rate, shown by the dashed line and find a decay rate of $\gamma_{\rm fit} = (4.5 \pm 0.2) \times 10^{-2}\,t_{c_\mathrm{s}}^{-1}$, which shows good agreement with the theoretical value, $\gamma_{\rm theory}$.}
    \label{fig:dn}
\end{figure}

\Fig{fig:dn} shows the squared amplitude of the density fluctuations as a function of time, normalised to the sound-crossing timescale, $t_{c_\mathrm{s}} = 1/(k c_\mathrm{s})$, where $c_\mathrm{s}$ is the sound speed based on $T_{\rm e}$. We set the ratio of the ion to electron temperature, $T_{\rm i}/T_{\rm e} = 0.1$. We find that the density fluctuations decay exponentially with time, by locating the local maxima of the density fluctuations and fitting an exponential decay function. We find a decay rate of $\gamma_{\rm fit} = (4.5 \pm 0.2) \times 10^{-2}\,t_{c_\mathrm{s}}^{-1}$. We measure the decay rates between successive data points and the decay rate we report is the mean of all the intervals and the error reported is the standard deviation of the decay rate across all the intervals. We compare this with the solutions of the dispersion relation and find good agreement with the theoretical solution,  $\gamma_{\rm theory} = 0.042\,t_{c_\mathrm{s}}^{-1}$.

\subsubsection{Particle resolution criterion}
\begin{figure}
    \centering
    \includegraphics[width=1.0\linewidth]{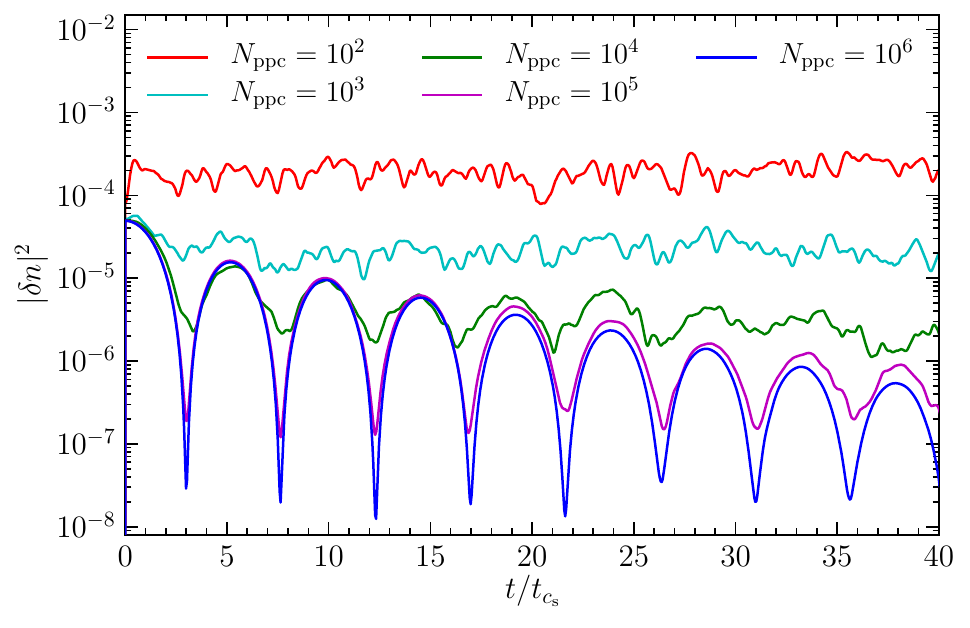}
    \caption{Same as \Fig{fig:dn}, but for $\nppc = 10^2, 10^3, 10^4, 10^5$ and $10^{6}$ particles per cell. For $\nppc \gtrsim 10^4$, the Landau damping of ion acoustic waves are resolved, while particle noise dominates below $\nppc\sim10^4$.}
    \label{fig:dn_nppc}
\end{figure}

Here we repeat the Landau damping test with $\nppc = 10^2, 10^3, 10^4$ and $10^5$ particles per cell. The results are shown in \Fig{fig:dn_nppc}. We find that with low particle count $\nppc \lesssim 10^4$, particle noise dominates, while for higher $\nppc$ the Landau damping of ion acoustic waves are resolved.

\section{Particle-Grid interpolation tests} \label{sec:interpolation_tests}

After confirming that the code agrees with analytical solutions for standard particle, wave, and particle-wave problems, we perform collisionless turbulent dynamo tests to understand the effect of using different interpolation kernels and smoothing in our numerical simulations. The turbulent dynamo mechanism converts turbulent kinetic energy into magnetic energy, exponentially amplifying magnetic fields \citep{Kazantsev1968, Moffatt1978, Ruzmaikin1988}. The properties of the turbulent dynamo are sensitive to the Mach number, nature of turbulence driving and the magnetic and kinetic Reynolds number of the plasma \citep{Schekochihin+2004, Haugen+2004a, Haugen+2004b, CFetal11, Schoberetal2012, Seta+2020, Seta&Federrath2020, AchikanathEtAl2021, KrielEtAl2022, Gent+2023} and recently numerical studies have extended this to the collisionless regime using Hybrid-kinetic codes \citep{Rinconetal2016, St-Onge&Kunz2018, AchikanathChirakkara+2023}.

\subsection{Turbulence driving setup and initial conditions} \label{sec:turb_driving}
In all the numerical tests presented in this and the following sections, we drive turbulence with a random acceleration field, $\vec{f}$, following an Ornstein-Uhlenbeck process \citep{EswaranPope1988, Federrath+2010,FederrathEtAl2022ascl}. The acceleration field is constructed in Fourier space to contain only large-scale modes, $kL/2\pi = [1,3]$, where $L$ is the size of the cubic computational domain. The turbulence driving amplitude is modelled as a parabolic function that peaks at $k\Lturb/2\pi = 2$ and is zero at $kL/2\pi = 1$ and 3, where $\Lturb$ is the peak driving scale of the turbulence. The turbulent velocity dispersion, $\Vturb$, is controlled by adjusting the amplitude of the driving. We inject purely solenoidal acceleration modes $(\nabla \cdot \vec{f} = 0)$ \citep{Federrath+2010} into the plasma for these tests. For all the tests in this section, we initialize the velocity of the particles using a Maxwellian distribution with temperature $T_{\rm i}$, which is equal to $T_{\rm e}$.

All the numerical tests in this section are performed on a periodic three-dimensional computational domain with $\ngrid^{3} = 128^{3}$ grid cells. We drive the turbulence to produce a steady-state (fully-developed) turbulent Mach number of  $\mathcal{M} = 0.2$. Using the steady-state kinetic energy, we initialise the magnetic field such that the ratio of magnetic to turbulent kinetic energy is $(E_{\rm mag}/E_{\rm kin})_{\rm init} = 10^{-8}$, and the initial ratio of the Larmor radius to the box size is $\initmagnetisation = 10^{2}$ (see Sec.~2.5 of \citet{AchikanathChirakkara+2023} for further details). The convergence of the turbulent dynamo simulations with respect to the grid and particle resolution is studied in App.~D of \citet{AchikanathChirakkara+2023}. The magnetic Reynolds number of the plasma, defined as
\begin{equation}
    \Rm = \frac{\Vturb \Lturb}{\eta},
\end{equation}
is set to $\Rm = 500$. The hyper-resistive magnetic Reynolds number is defined as
\begin{equation}
    \Rmhyper = \frac{\Vturb (\Lturb)^{3}}{\eta_{\rm hyper}},
\end{equation}
and is set to $\Rmhyper = 10^{7}$. For all the tests in this section, we use cooling (tested and discussed separately in detail in \Sec{sec:cooling_tests} below) with a cooling frequency of $(\Delta t)_{\rm cool}=0.1 \, \tcool$ (see Eq.~\ref{eqn:tcool_def} and Sec.~\ref{sec:cooling}). 

\subsection{Comparison of global quantities in interpolation tests}
We test three interpolation kernels: the nearest-grid-point (NGP), the cloud-in-cell (CIC), and the triangular-shaped-cloud (TSC) interpolation functions, detailed in \Sec{sec:interpolation} and \App{app:weight_function}. We also vary the number of smoothing passes, $\nsmooth=0, 1, 2$, and $4$, as discussed in \Sec{sec:smoothing}. Finally, we investigate the influence of different numbers of particles per cell, $\nppc=50, 100, 200$, and $400$.

\begin{figure}
\begin{center}
\def\arraystretch{0}
\setlength{\tabcolsep}{0pt}
\begin{tabular}{r}
\vspace{-0.15cm}
\includegraphics[width=0.992\linewidth]{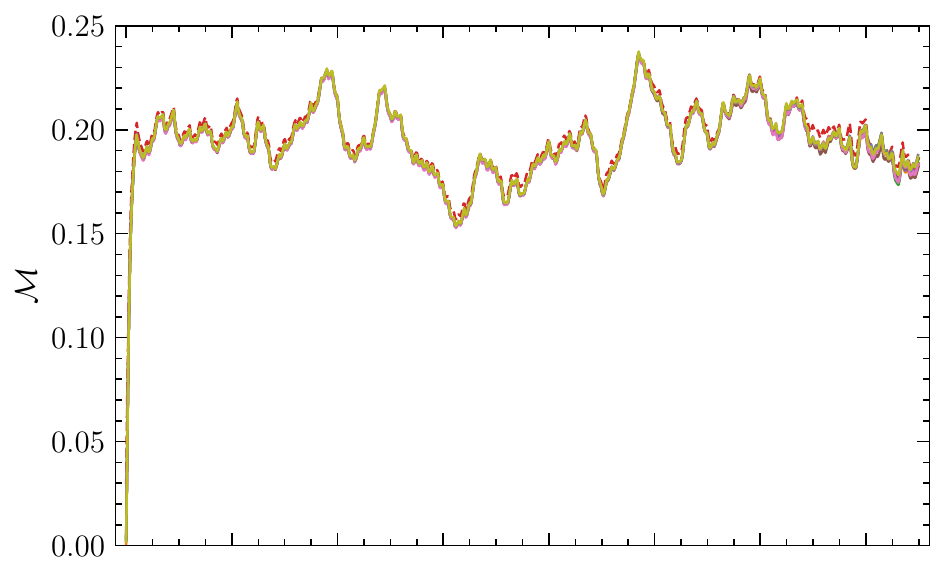} \\
\includegraphics[width=1.0\linewidth]{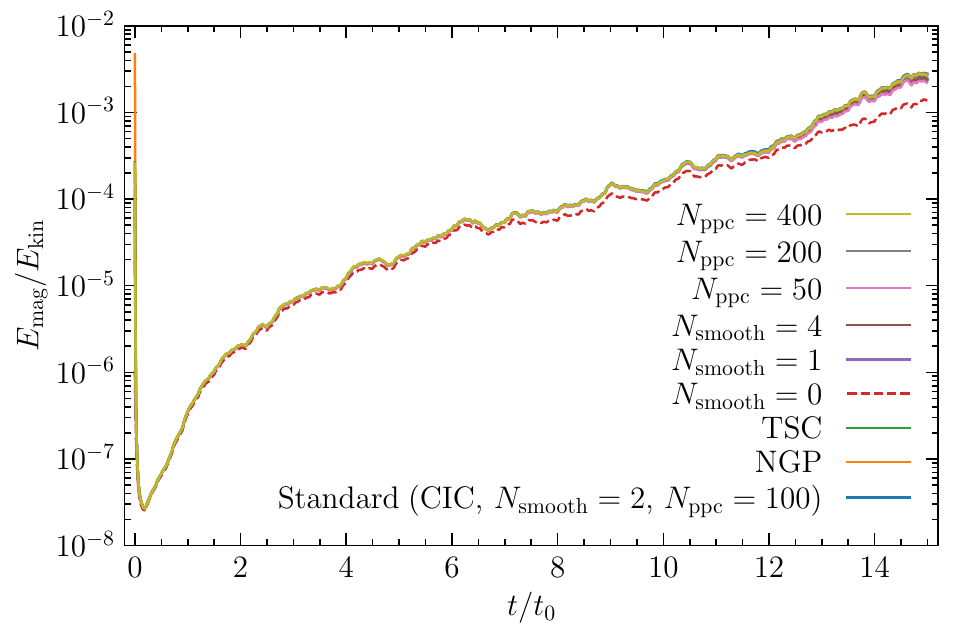}
\end{tabular}
\end{center}
\caption{Particle$\to$grid interpolation tests using collisionless turbulent dynamo simulations. The Mach number, $\mathcal{M}$, is shown in the top panel, and the magnetic-to-kinetic energy ratio, $\ratio$, is shown in the bottom panel, as a function of time normalised to the turbulent eddy turnover time, $\ted$.  For the standard set of numerical parameters, we use the cloud-in-cell (CIC) interpolation kernel with $\nppc=100$ particles per cell and two smoothing passes, $\nsmooth = 2$. We find that using nearest-grid-point (NGP) or triangular-shaped-cloud (TSC) interpolation functions does not noticeably change the box-averaged solutions for the Mach number or the magnetic energy (provided $\nsmooth=2$). We also find that changing the number of particles per cell, $\nppc=50, 200$, or 400, does not change the box-averaged solutions. Changing the number of smoothing operations when $\nsmooth \geq 1$ also has no significant impact on the box-averaged solutions. However, without any smoothing, $\nsmooth=0$ (red dashed line), the growth in magnetic energy is slightly underestimated.}
\label{fig:interp_tevol}
\end{figure}

Fig.~\ref{fig:interp_tevol} shows the time evolution (in units of the turbulent turnover time, $\ted = \Lturb/\Vturb$) of the sonic Mach number ($\Mach$; top panel) and the magnetic-to-kinetic energy ratio ($\ratio$; bottom panel), for all the interpolation tests considered here. We see that all test simulations yield very similar results, with the sonic Mach number reaching a steady state of $\Mach\sim0.2$. We also see that after an initial transient phase of about 2--3~$\ted$, the magnetic energy grows exponentially in all tests. These are typical characteristics of small-scale turbulent dynamo growth during the so-called `kinematic phase' \citep{B&S2005,Schekochihin+2004,Federrath2016jpp,Seta&Federrath2020,KrielEtAl2022,AchikanathChirakkara+2023}.

Comparing the different tests in Fig.~\ref{fig:interp_tevol}, we find that all runs are virtually identical in terms of their box-averaged global evolution, except for the simulation that does not use any smoothing ($\nsmooth=0$) after particle$\to$grid interpolations (red dashed line).

\subsection{Spatial structure of interpolated fields}
\begin{figure*}
\begin{center}
\def\arraystretch{0}
\setlength{\tabcolsep}{0pt}
\begin{tabular}{llll}
\includegraphics[height=0.3\linewidth]{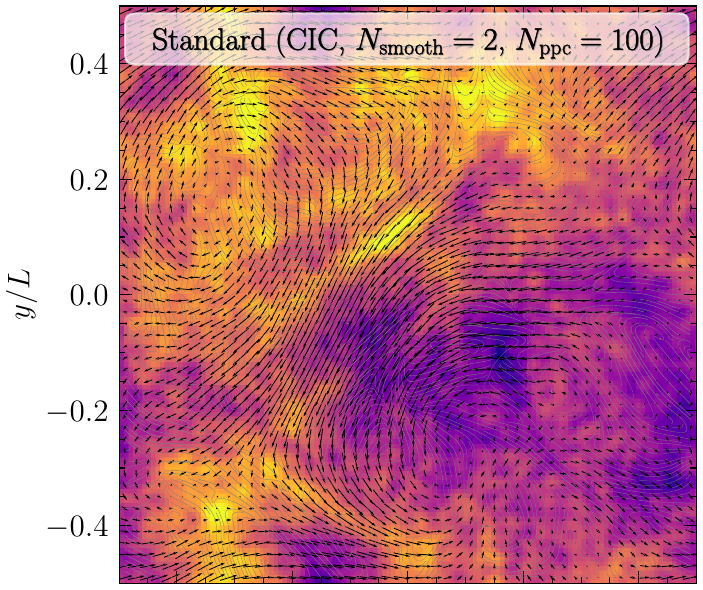} &
\includegraphics[height=0.3\linewidth]{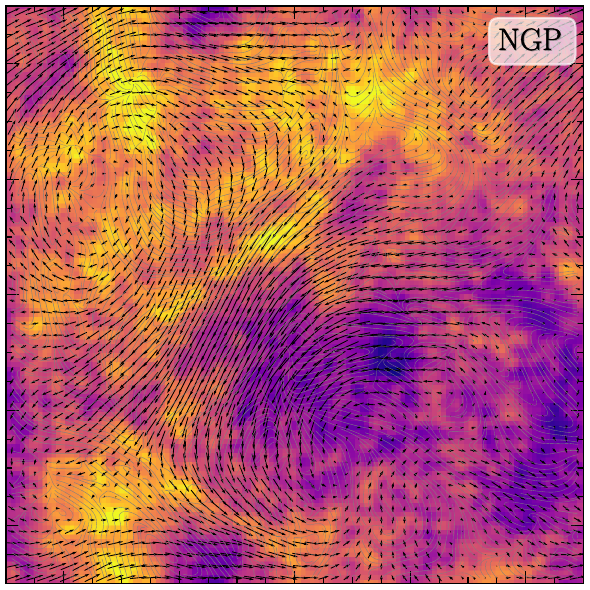} &
\includegraphics[height=0.3\linewidth]{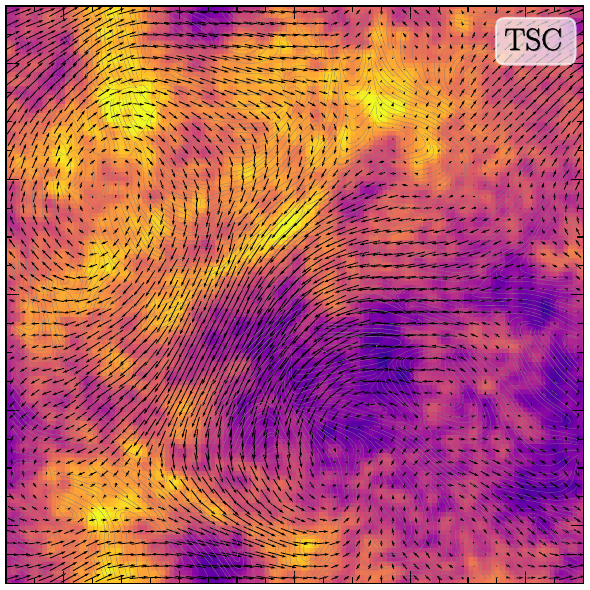} &
\multirow{3}{*}{\parbox{1.0\linewidth}{\vspace{-5.45cm}\includegraphics[height=0.9135\linewidth]{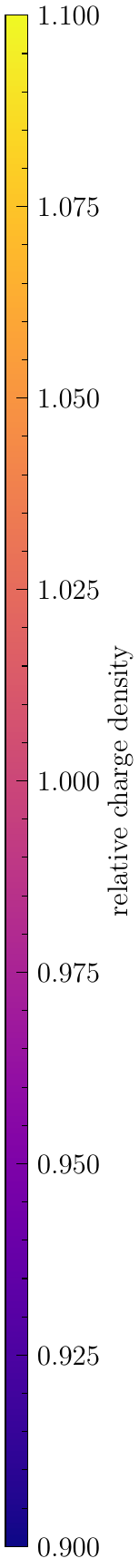}}}
\\
\includegraphics[height=0.3\linewidth]{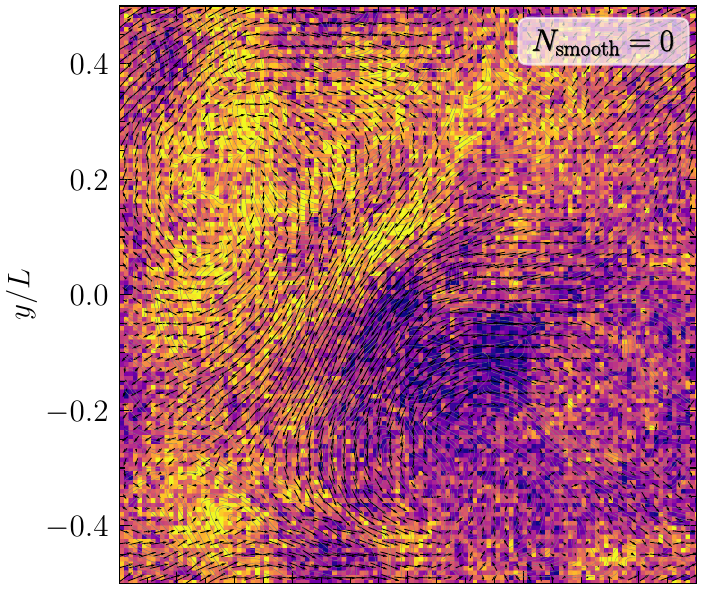} &
\includegraphics[height=0.3\linewidth]{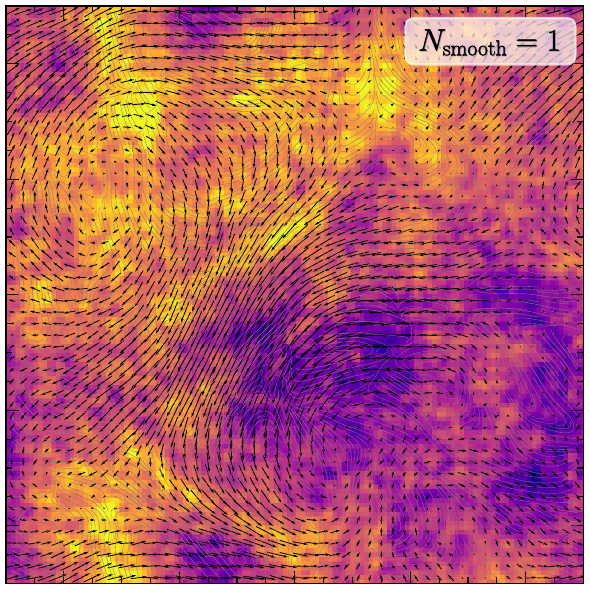} &
\includegraphics[height=0.3\linewidth]{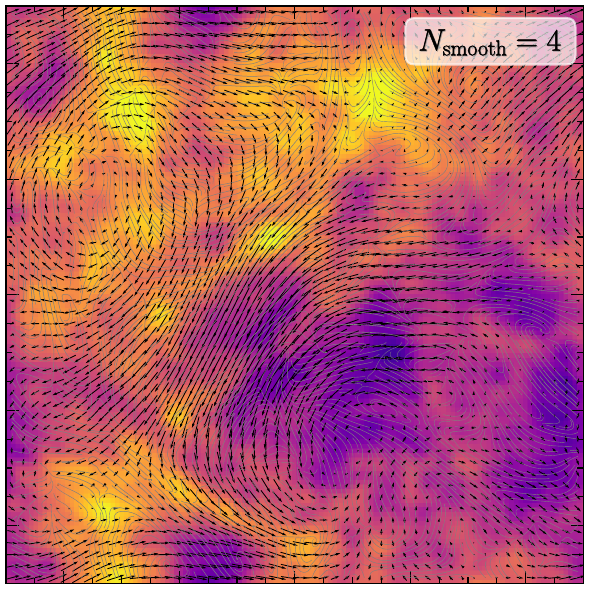}
& \\
\includegraphics[height=0.338\linewidth]{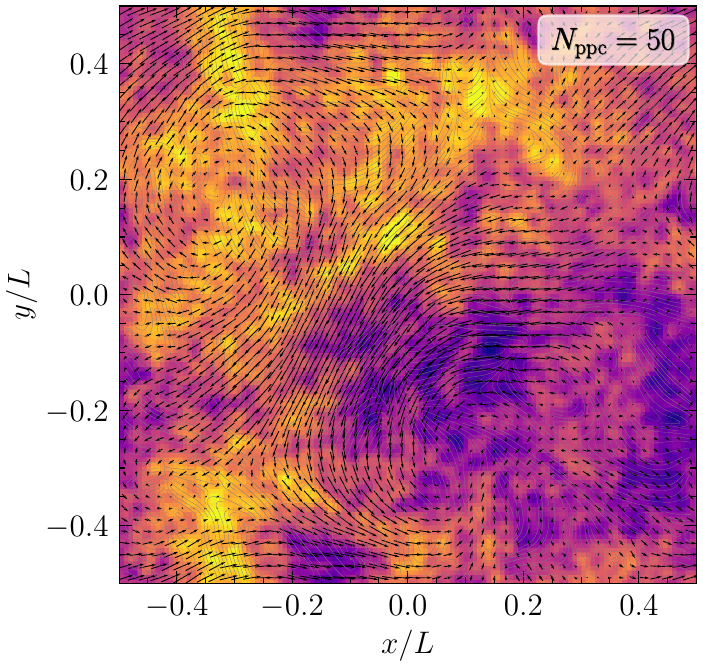} &
\includegraphics[height=0.338\linewidth]{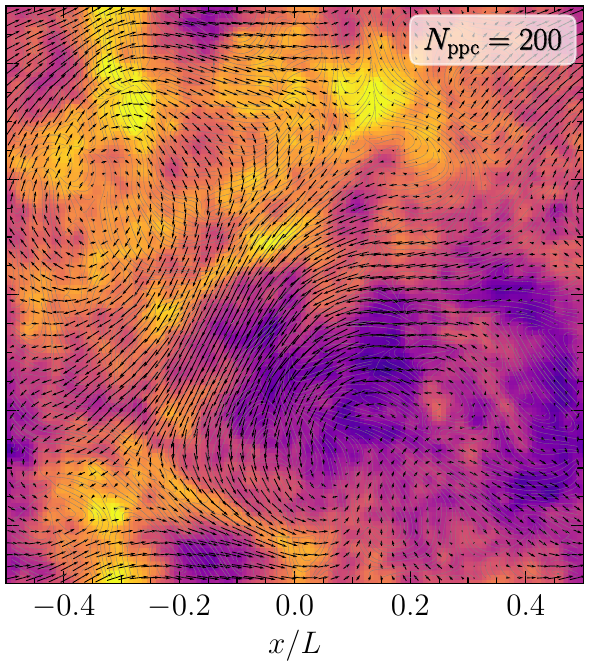} &
\includegraphics[height=0.338\linewidth]{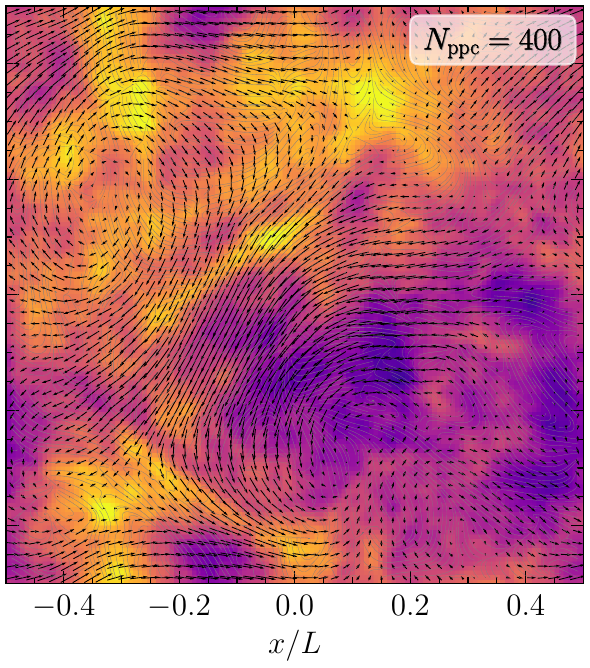}
&
\end{tabular}
\end{center}
\caption{Slice plots of the relative charge density (colour map) at $10\,\ted$ for the interpolation test simulations shown in Fig.~\ref{fig:interp_tevol}. The superimposed vectors and streamlines show the velocity and magnetic field, respectively. For the standard test (top left panel), we use the cloud-in-cell (CIC) interpolation kernel with 2~smoothing passes ($\nsmooth = 2$), and $\nppc = 100$ particles per cell. First row: comparison of different interpolation kernels, including the nearest-grid-point (NGP) and triangular-shaped-cloud (TSC) interpolation kernels, with $\nsmooth = 2$ and $\nppc = 100$. Second row: tests without smoothing ($\nsmooth = 0$), with $\nsmooth = 1$, and $\nsmooth = 4$, using the CIC interpolation kernel and $\nppc = 100$. Third row: tests with different numbers of particles per cell, $\nppc = 50, 200$, and $400$. These tests show that it is important to smooth the particle-sampled fields with at least $\nsmooth=1$ smoothing pass (somewhat better results still, are obtained here with $\nsmooth=2$), and to work with a sufficiently large number of particles per cell ($\nppc \sim 100$) to limit the effects of particle interpolation noise.}
\label{fig:interp_images}
\end{figure*}

\Fig{fig:interp_images} shows the relative charge density at $t=10\,\ted$ for all the interpolation test simulations above. The top left panel shows the test with the standard set of numerical parameters (CIC, $\nsmooth=2$, $\nppc=100$), which serves as the main reference simulation. Comparing the tests in the first row (CIC vs.~NGP vs.~TSC), we find no significant difference in the spatial distribution of the charge density. While the NGP method on its own is generally inferior compared to the CIC and TSC interpolation methods, the fact that we perform $\nsmooth=2$ smoothing passes after these interpolations nullifies any specific differences between NGP, CIC, and TSC.

The 2nd row of \Fig{fig:interp_images} shows a comparison of $\nsmooth=0$ (no post-interpolation smoothing), $\nsmooth=1$, and $4$. We see that not smoothing after interpolation (middle row, left-hand panel) leaves a very noisy density distribution. This substantially improves already with a single smoothing pass (central panel). Using 4 smoothing passes ($\nsmooth=4$; right panel) leads to a relatively smooth distribution. However, neither $\nsmooth=1$ nor 4 are substantially different from the standard $\nsmooth=2$ results. While the spatial details of the interpolated fields depend on the number of smoothing passes, we saw in Fig.~\ref{fig:interp_tevol} that the choice of smoothing passes (except for $\nsmooth=0$) has no significant impact on the global evolution of the system. 

Finally, the last row of Fig.~\ref{fig:interp_images} shows the effect of varying the number of particles per cell, $\nppc=50$, 200, 400. While using a higher $\nppc$ improves the spatial smoothness of the charge density field, we found no significant impact on the global evolution of the system in Fig.~\ref{fig:interp_tevol}. It is worth noting that the test with $\nsmooth=2$ and $\nppc=400$ (bottom right panel) is visually very similar to the case with $\nsmooth=4$ and $\nppc=100$ (middle row, right-hand panel). This suggests that a balanced combination of $\nppc$ and $\nsmooth$ can lead to comparably good results for a desirably low $\nppc$, as it saves computational resources. However, in general, we suggest that the number of smoothing passes should be kept sufficiently low, in particular in supersonic flows, which would result in excessive smearing of discontinuities (shocks). An optimal setting of $\nsmooth$ and $\nppc$ is therefore problem-dependent and requires experimentation and testing in any PIC simulation.

\section{Cooling tests}
\label{sec:cooling_tests}
In this section, we demonstrate the working of our cooling method for ions described in \Sec{sec:cooling}.

\subsection{Cooling vs.~no cooling}
Here we investigate the requirement of plasma cooling to model subsonic and supersonic turbulent flows. We run the following numerical simulations on a triply periodic uniform computational domain with $\ngrid^{3} = 128^{3}$ grid cells and $\nppc = 100$ particles per cell for subsonic, $\Mach_{\rm target} = 0.2$, and supersonic, $\Mach_{\rm target} = 2$ target Mach numbers, without and with cooling. The turbulence driving used here is described in \Sec{sec:turb_driving}. We note that magnetic fields have not been included for all the tests described in this section as they play no role in the cooling functionality.

\subsubsection{Time evolution of global thermodynamic and turbulent quantities}
\begin{figure}
    \centering
    \includegraphics[scale=0.235]{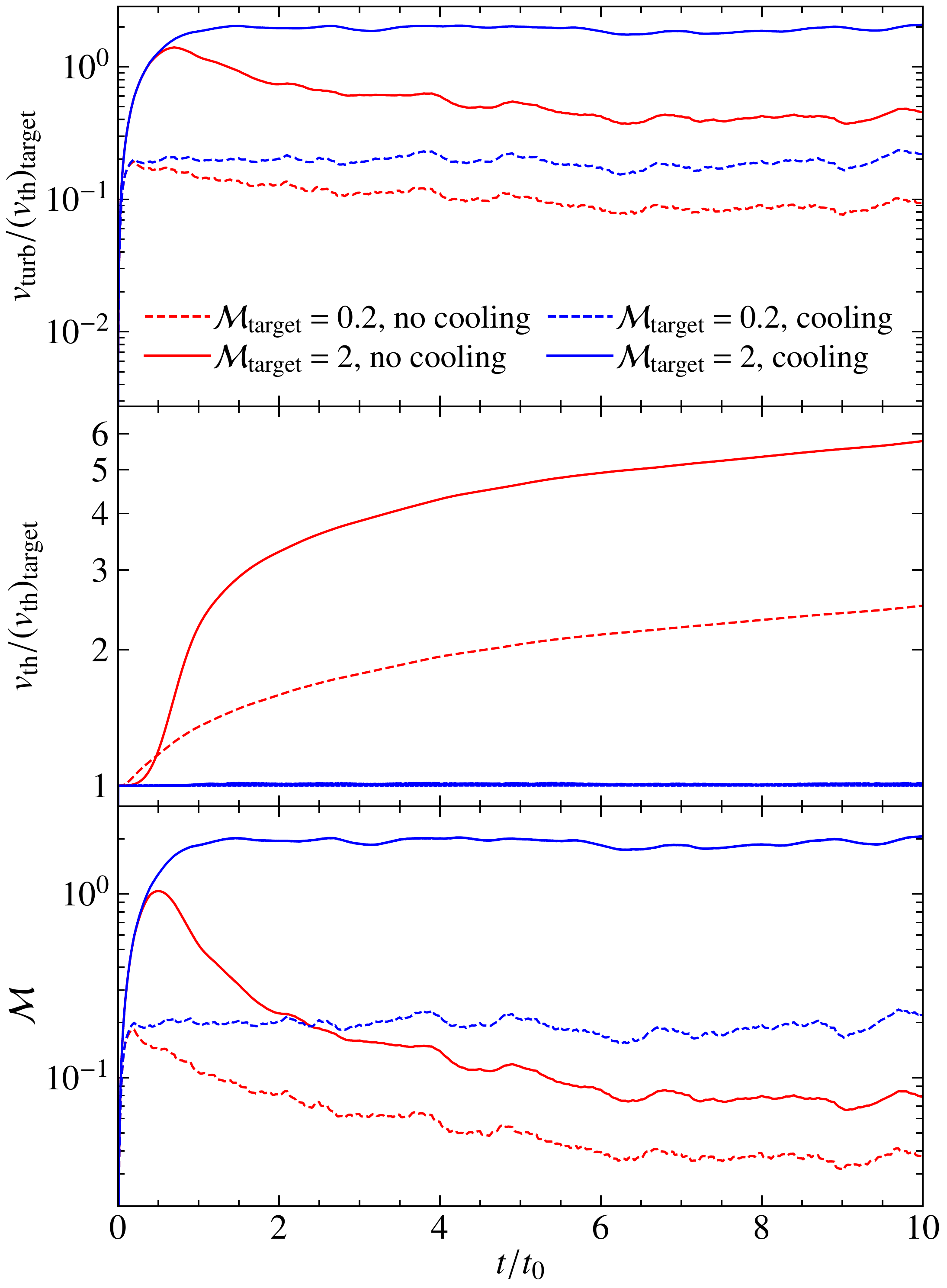}
    \caption{Turbulent speed normalised to the target thermal speed, $\Vturb/\Vthtarget$ (top panel), thermal speed normalised to the target thermal speed, $\Vtherm/\Vthtarget$ (middle panel), and the sonic Mach number, $\Mach$ (bottom panel), as a function of time, normalised to the eddy turn-over time, $\ted$, for turbulence driving simulations on $\ngrid^{3} = 128^{3}$ grid cells with $\nppc = 100$ particles per cells with cooling (blue) and without cooling (red), in tests with a target Mach number of 0.2 (dashed lines) and 2 (solid lines), respectively. We find that without cooling it is impossible to maintain the target Mach number. This is primarily the result of heating due to turbulent dissipation and shocks (for the supersonic test), demonstrating the necessity for cooling to model steady-state turbulent flows.} 
    \label{fig:cooling}
\end{figure}

The time evolution of the cooling experiments is shown in \Fig{fig:cooling}. The top panel shows the turbulent speed, normalised to the fixed target thermal speed, $\Vturb/\Vthtarget$, in our numerical tests as a function of time. We find that without cooling (red curves), it is difficult to maintain the target turbulent speeds, $\Vturb/\Vthtarget = 0.2$ in the subsonic regime and $\Vturb/\Vthtarget = 2$ in the supersonic regime. After an initial maximum, $\Vturb$ decreases in the tests without cooling (red curves). When cooling is included, we find that the target $\Vturb$ is reached and maintained throughout the simulations. The middle panel of \Fig{fig:cooling} shows the thermal speed, normalised to the target thermal speed, $\Vtherm/\Vthtarget$. For the tests without cooling, we find that $\Vtherm$ increases monotonically as turbulence energy dissipates, while with cooling, $\Vtherm$ is maintained at its target level. We also find that the thermal speed of the plasma increases faster in the supersonic regime when cooling is not included. For both the tests with cooling in the subsonic and supersonic regime, $(\Delta t)_{\rm cool} = 0.1 \, \tcool$, where $\tcool$ is the cooling timescale (more details on the role of $(\Delta t)_{\rm cool}$ below). The ratio of the data in the top and bottom panels is shown in the bottom panel of \Fig{fig:cooling}, i.e., the sonic Mach number, $\Mach$, which reflects the issues discussed separately for $\Vturb$ and $\Vtherm$ without cooling. Over time, the plasma becomes increasingly subsonic, which makes studying maintained supersonic flows practically impossible without cooling. However, even for the subsonic regime, turbulence cannot be maintained at a steady Mach number without cooling. This demonstrates the necessity for cooling in applications such as steady-state turbulence.

\subsubsection{Statistical and spatial distribution of the temperature}
\begin{figure}
    \centering
    \includegraphics[trim={0.26cm 0.1cm 0.25cm 0.1cm},clip,scale=0.24]{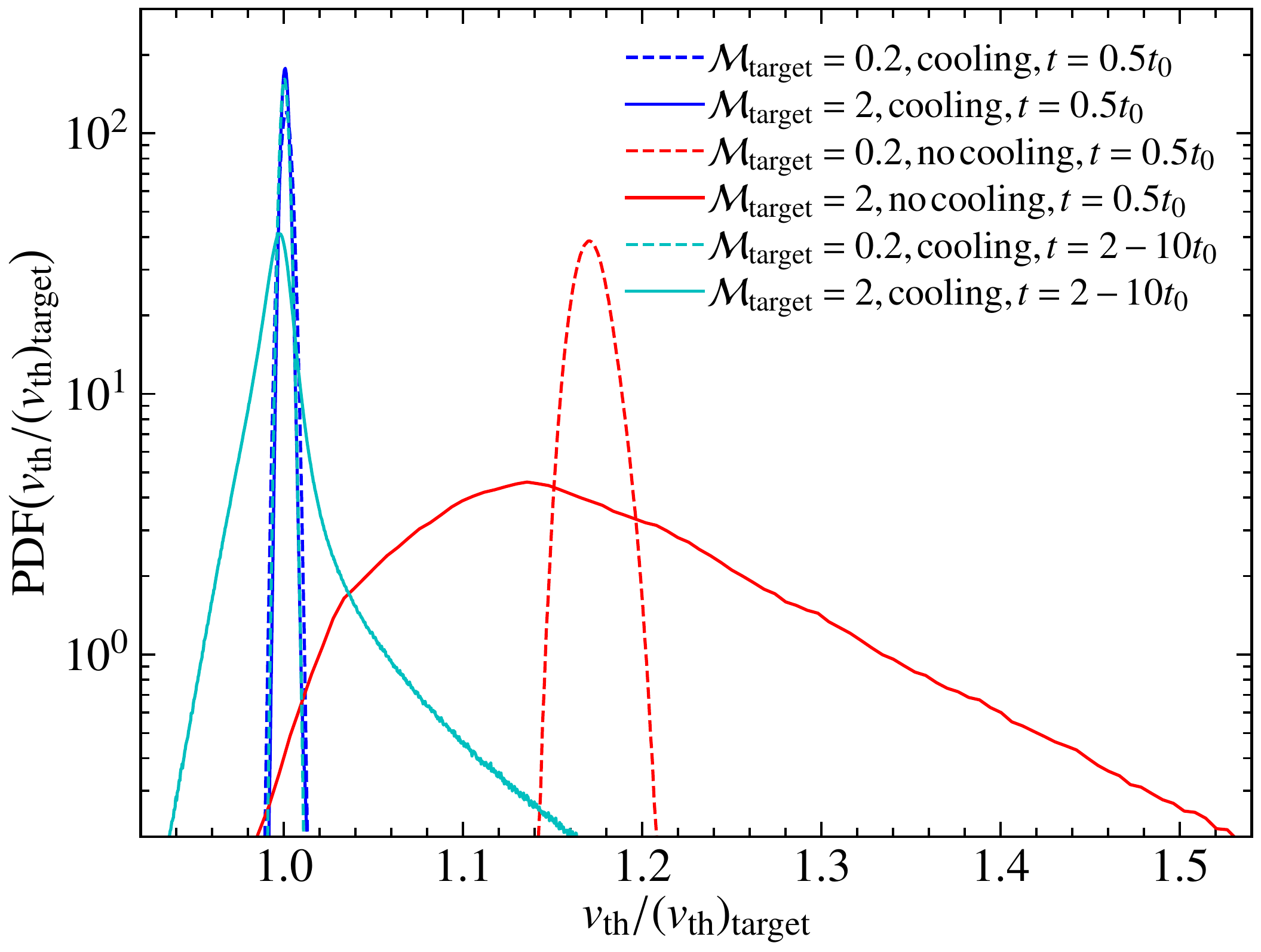}
    \caption{Probability density functions (PDFs) of the thermal speed, $\Vtherm$, normalised to the target thermal speed, $\Vthtarget$, for simulations with target Mach number, $\Mach_{\rm target} = 0.2$ (dashed lines) and $\Mach_{\rm target} = 2$ (solid lines), without cooling (red) and with cooling at $t/\ted=0.5$ (blue) and time-averaged in $t/\ted=[2,10]$ (cyan). The tests without cooling show very broad distributions, i.e., not only is the target average temperature overestimated, but the plasma is also spatially at very different temperatures. For the tests with cooling, the target thermal speed is reached and maintained (see time-averaged curves shown in cyan) to within 0.6\% accuracy in both the subsonic and supersonic regimes. The spatial fluctuations remain reasonably small with a standard deviation of 0.3\% and 4.3\% in the subsonic and supersonic regimes, respectively. The higher residual fluctuations in the supersonic regime are due to the presence of shocks.}
    \label{fig:vth_pdfs}
\end{figure}

Here we investigate the distribution and spatial structure of the plasma temperature in more detail. \Fig{fig:vth_pdfs} shows the probability density function (PDF) of the thermal speed, $\Vtherm$, normalised to the target thermal speed, $\Vthtarget$. For the subsonic and supersonic test without cooling at $t = 0.5 \, \ted$, we find $\Vtherm/\Vthtarget = 1.172 \pm 0.010$ and $1.189 \pm 0.117$, respectively. While the mean values for both these tests are similar with about 17--19\% higher than the target, the spread of the thermal speed is much higher for the supersonic simulation. This can be attributed to more turbulent kinetic energy available to heat the plasma and the presence of shocks dissipating large amounts of energy in the supersonic regime. By contrast, when cooling is included, we find $\Vtherm/\Vthtarget = 1.001 \pm 0.003$ and  $1.000 \pm 0.002$ for the same subsonic and supersonic test comparison, respectively. For the simulations with cooling, we also show the PDFs averaged between $2 - 10 \, \ted$,  for which we find $\Vtherm/\Vthtarget = 1.000 \pm 0.003$ and $1.006 \pm 0.043$, respectively, i.e., the target thermal speeds are reasonably well maintained.

\begin{figure*}
    \centering
    \begin{subfigure}{0.498\textwidth}
        \centering
        \includegraphics[trim={0.1cm 0.05cm 0.25cm 0.05cm},clip,scale=0.53]{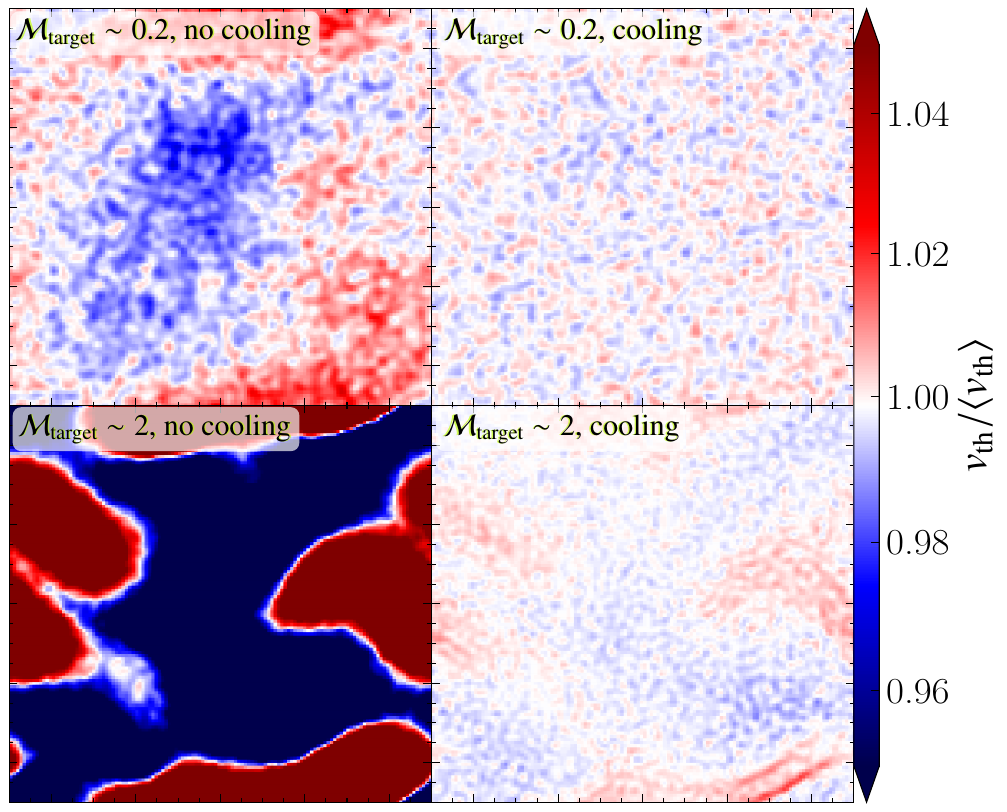}
        \caption{Thermal speed relative to the average thermal speed.}
        \label{fig:vth_slices}
    \end{subfigure}
    \begin{subfigure}{0.498\textwidth}
        \centering
        \includegraphics[trim={0.1cm 0.05cm 0.25cm 0.05cm},clip,scale=0.53]{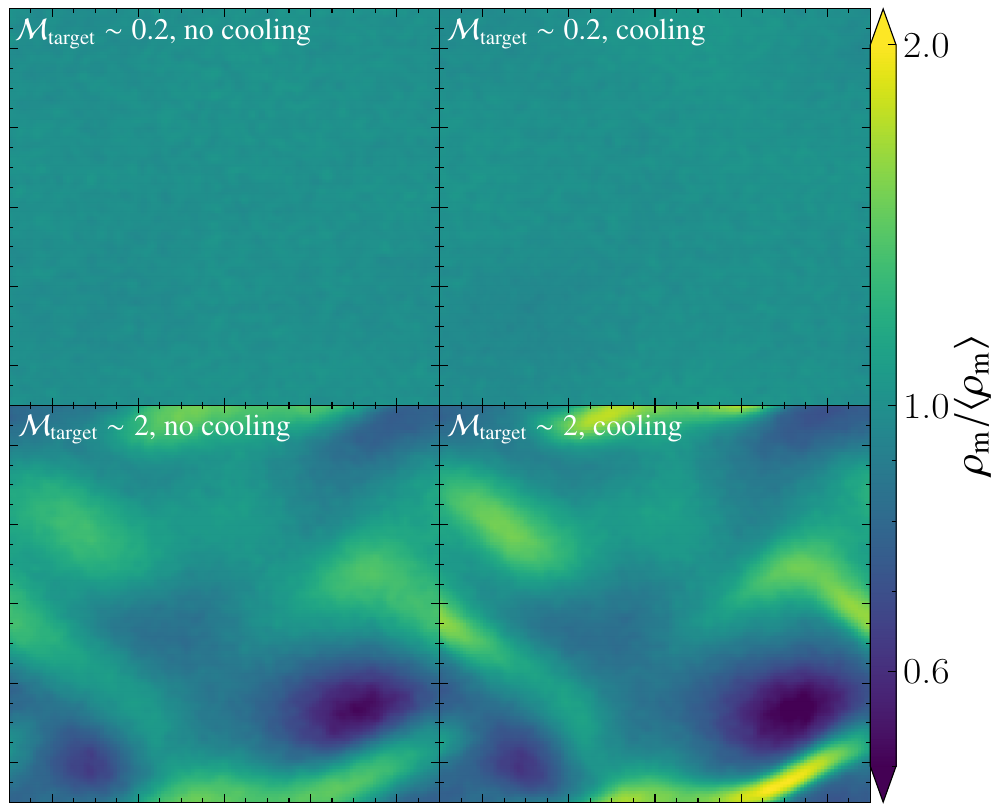}
        \caption{Plasma density relative to the average density.}
        \label{fig:dens_slices}
    \end{subfigure}
    \caption{(a) Slice plots of the thermal speed, $\Vtherm$, normalised to the mean thermal speed, $\langle \Vtherm \rangle$, at $t/\ted=0.5$, for the same test simulations as in \Fig{fig:cooling} (as indicated in the panel labels). We see that without cooling, the thermal speed is inhomogeneous, particularly in the supersonic regime. Cooling homogenises the thermal speed, showing that cooling is required locally to achieve reasonable isothermal conditions throughout the plasma. (b) Same as (a), but for the ion mass density, $\rho_{\rm m}$, normalised to the mean ion mass density, $\langle \rho_{\rm m} \rangle$. The fluctuations in the mass density are much smaller in the subsonic tests when compared to the supersonic tests, as expected. Shocked structures with large density gradients are apparent in the supersonic regime, which correlates with regions of higher thermal speeds, showing that shocks dissipate kinetic energy efficiently.}
    \label{fig:cooling_slices}
\end{figure*}

The respective spatial distribution of the thermal speed is shown in \Fig{fig:cooling_slices}. We see that without cooling, the distribution of the thermal speed is inhomogeneous. In the case of supersonic turbulence, this distribution is significantly more inhomogeneous as more energy is injected into the plasma and dissipation occurs primarily in shocks. Comparing the over- and under-densities in the bottom right-hand panels we see that they are associated with the under- and over-heated regions in the bottom left-hand panels, respectively, i.e., adiabatic heating and cooling occur in relatively over- and under-dense regions, respectively. Once cooling is introduced, the resulting thermal speed distribution is fairly homogeneous. This shows that it is important to cool the plasma locally (not just globally), to obtain a spatially homogeneous isothermal plasma.

\subsection{Choosing the cooling frequency}
In order to keep the plasma isothermal, we need to apply the cooling operator at regular time intervals, $(\Delta t)_{\rm cool}\propto\tcool$ (see Sec.~\ref{sec:cooling}). Here we investigate the required cooling frequency, $(\Delta t)_{\rm cool}^{-1}$, to maintain approximate isothermal plasma conditions in the subsonic and supersonic regime of turbulence.

\subsubsection{Subsonic tests}
\begin{figure}
    \centering
    \includegraphics[scale=0.235]{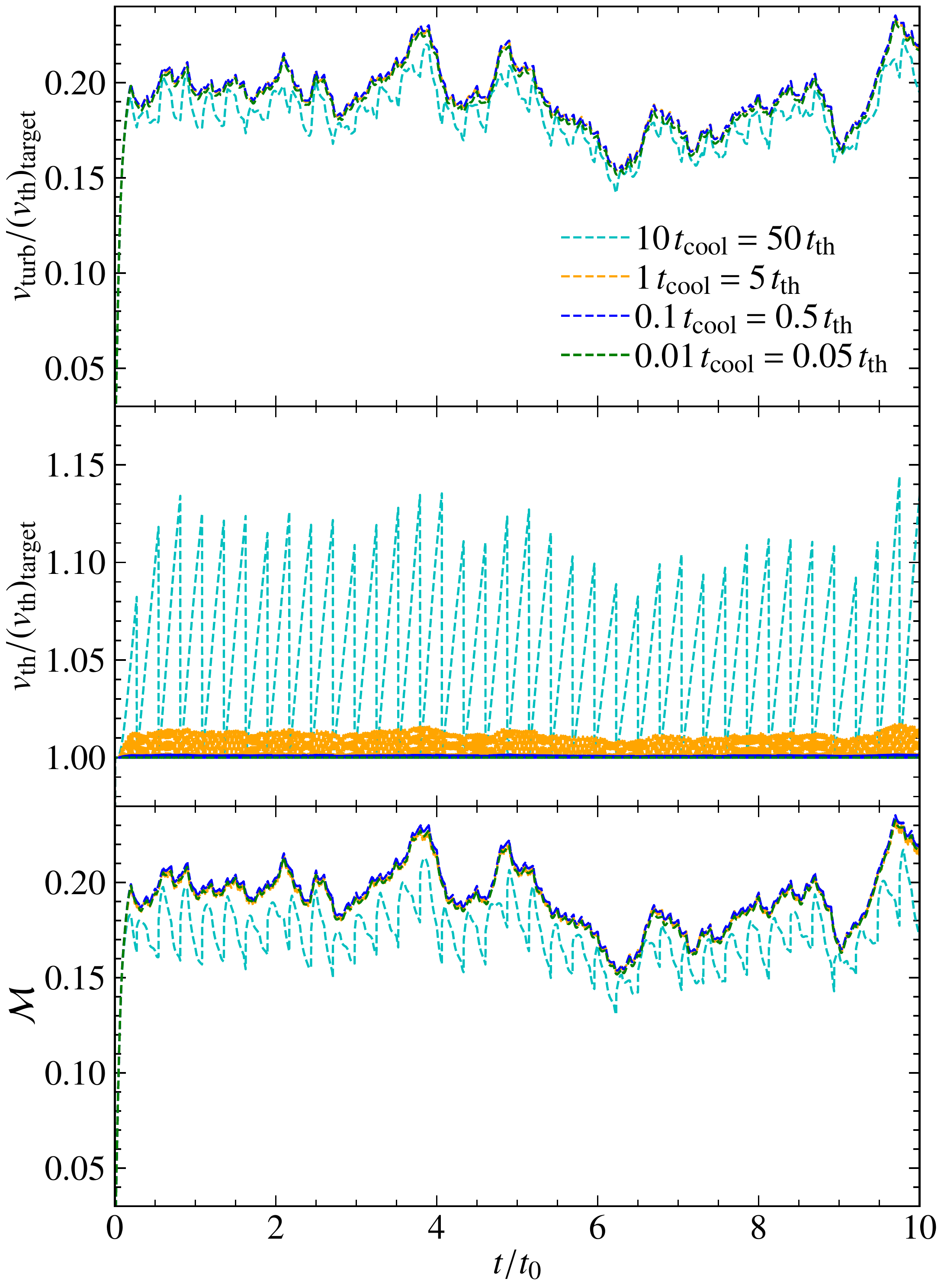}
    \caption{Same as \Fig{fig:cooling}, but for subsonic turbulence with target $\Mach = 0.2$ and different cooling times, $(\Delta t)_{\rm cool} = 0.01$, $0.1$, $1$ and $10\,\tcool$, where $\tcool$ is the cooling timescale. We see that the target thermal speed is overestimated with $(\Delta t)_{\rm cool} = 1$ and $10\,\tcool$. By contrast, the thermal speed is converged and maintained close to the target value for $(\Delta t)_{\rm cool} \lesssim0.1\,\tcool$.}
    \label{fig:cooling_velcorr_subsonic}
\end{figure}

In \Fig{fig:cooling_velcorr_subsonic} we vary the cooling time for the subsonic turbulence simulations presented above, $(\Delta t)_{\rm cool} = 0.01$, $0.1$, $1$ and $10\,\tcool$. We perform these simulations with a target $\Mach = \Vturb/\Vthtarget = 0.2$ and list the results in \Tab{table:cooling_sims}. The figure legend also lists the cooling timescale in units of the thermal crossing time, $\tcool = \tth/\Mach = 5\,\tth$ (see \Eq{eqn:tcool_def}). The middle panel of \Fig{fig:cooling_velcorr_subsonic} shows that for larger values of the cooling time, the thermal speeds start increasing, while for lower values of $(\Delta t)_{\rm cool}$, $\Vtherm$ is maintained close to the target thermal speed. We find that for the turbulent speed and the Mach number, shown by the top and the bottom panel of \Fig{fig:cooling_velcorr_subsonic}, respectively, the numerical solutions converge and the target $\Vturb$ and $\Mach$ are attained with $(\Delta t)_{\rm cool} \sim 0.1\,\tcool = 0.5\,\tth$.

\begin{table}
\centering
\caption{List of simulations shown in \Fig{fig:cooling_velcorr_subsonic} and \Fig{fig:cooling_velcorr_supersonic} with the corresponding model name, turbulent speed normalised to the target thermal speed, $\Vturb/\Vthtarget$, thermal speed normalised to the target thermal speed, $\Vtherm/\Vthtarget$, and the sonic Mach number, $\Mach$. These quantities are measured for $t/\ted = 2-10$, where $\ted$ is the turbulent turnover time. This shows that the thermal speed
and Mach number are converged (to within one sigma) as the cooling frequency is increased, in both the subsonic and the supersonic regimes of turbulence.}
\begin{tabular}{lccc}
\hline
Model & $\Vturb/(\Vtherm)_{\rm target}$ & $\Vtherm/(\Vtherm)_{\rm target}$ & $\Mach=\Vturb/\Vtherm$\\
\hline
\texttt{M0.2tcool10} & $0.18 \pm 0.02$ & $1.06 \pm 0.03$ & $0.17 \pm 0.02$\\
\texttt{M0.2tcool1} & $0.19 \pm 0.02$ & $1.01 \pm 0.00$ & $0.19 \pm 0.02$\\
\texttt{M0.2tcool0.1} & $0.19 \pm 0.02$ & $1.00 \pm 0.00$ & $0.19 \pm 0.02$\\
\texttt{M0.2tcool0.01} & $0.19 \pm 0.02$ & $1.00 \pm 0.00$ & $0.19 \pm 0.02$\\
\texttt{M2tcool10} & $1.75 \pm 0.10$ & $1.28 \pm 0.18$ & $1.39 \pm 0.20$\\
\texttt{M2tcool1} & $2.08 \pm 0.10$ & $1.03 \pm 0.02$ & $2.01 \pm 0.10$\\
\texttt{M2tcool0.1} & $2.00 \pm 0.11$ & $1.01 \pm 0.00$ & $1.99 \pm 0.11$\\
\texttt{M2tcool0.01} & $1.90 \pm 0.10$ & $1.00 \pm 0.00$ & $1.90 \pm 0.10$\\
\texttt{M2tcool0.001} & $2.02 \pm 0.10$ & $1.00 \pm 0.00$ & $2.02 \pm 0.10$\\
 \hline
\end{tabular}
\label{table:cooling_sims}
\end{table}

\subsubsection{Supersonic tests}

\begin{figure}
    \centering
    \includegraphics[scale=0.235]{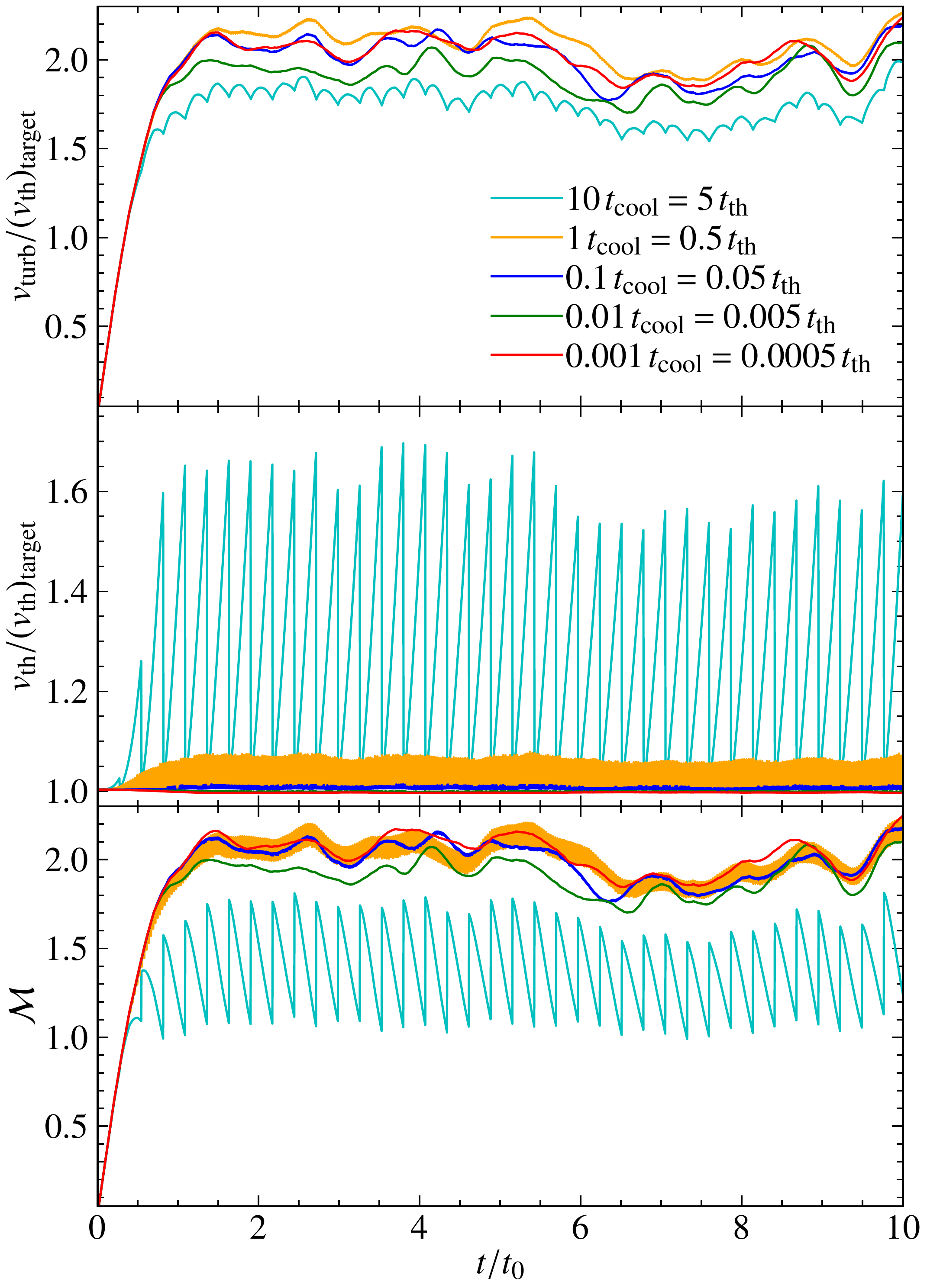}
    \caption{Same as \Fig{fig:cooling_velcorr_subsonic}, but for supersonic turbulence with a target $\Mach = 2$. As for the subsonic case, a cooling time of $(\Delta t)_{\rm cool} \sim 0.1 \,\tcool$, is required to maintain $\Vtherm$ at its target level and therefore maintain the target Mach number. We also perform a test with $(\Delta t)_{\rm cool} = 10^{-3} \,\tcool$, to demonstrate the convergence of the thermal speed and the Mach number as the cooling frequency is increased in the supersonic regime and tabulate the results in \Tab{table:cooling_sims}.}
    \label{fig:cooling_velcorr_supersonic}
\end{figure}

We repeat the same experiment with a supersonic target Mach number, $\Mach = \Vturb/\Vthtarget = 2$, and show the results in \Fig{fig:cooling_velcorr_supersonic}. We list the results in \Tab{table:cooling_sims}.
We find that for $(\Delta t)_{\rm cool} = 1$ and $10 \,\tcool$, the plasma heats up very quickly and the target thermal speed is not maintained. Performing the cooling with a higher frequency, by using $(\Delta t)_{\rm cool} = 0.1 \,\tcool = 0.05 \, \tth$, the target thermal speed is acquired and maintained and the target Mach number is achieved. 

\subsubsection{Summary}
The thermal crossing time, $\tth$, is fixed between the subsonic and supersonic tests, as both have been configured to reach the same target thermal speed. Since the Mach number is 10 times higher in the supersonic runs, we need a 10 times higher cooling frequency (in absolute terms) in the supersonic tests, compared to the subsonic tests. However, when expressed in terms of the cooling time, which takes the target Mach number into account (Eq.~\ref{eqn:tcool_def}), the required cooling frequency is the same in the supersonic and subsonic regimes. That is, we find that for $(\Delta t)_{\rm cool} \sim 0.1 \,\tcool$, i.e., using a safety factor of $\sim0.1$, the target thermal speed and Mach number are reasonably achieved and maintained throughout.

\section{Code optimisation and performance}
\label{sec:code_performance}
In this section, we discuss the performance and scalability of our code. We discuss the `hybrid precision' method introduced in \citet{Federrath+2021}. We extend and test the method for Hybrid PIC in \Sec{sec:hybrid_precision}. Parallel scaling tests are provided in \Sec{sec:scaling}. 

\subsection{Hybrid precision}
\label{sec:hybrid_precision}
Hybrid PIC simulations can be extremely computationally demanding, as they require both high grid-cell counts and high particle counts. As we have seen in Sec.~\ref{sec:interpolation_tests}, $\nppc\sim100$~particles per cell is a standard requirement for sufficient interpolation accuracy. With a grid-cell count of just $100^3$~cells, this means that even at such moderate grid resolutions, we very quickly end up with an enormous number of resolution elements\footnote{For example, if we wanted to do a Hybrid PIC turbulence simulation with $1000^3$ grid cells and $\nppc=100$, we would have $1000^3\times100=10^{11}$ resolution elements to compute on. This is not only significant in terms of memory requirements, but the compute time per time step is also significantly higher for particle operations compared to MHD grid-cell operations.}. Therefore, it is important to investigate and develop numerical methods that will help reduce the computational resources required to perform these simulations, without compromising the accuracy of the numerical solutions.

One such approach for optimisation is to store and compute solutions on single precision (4~byte) floating-point numbers instead of the standard double precision (8~byte per floating-point number). This reduces the memory footprint of an application by a factor of 2. The time for computations and parallel message passing interface (MPI) communications on single precision numbers is also approximately halved compared to double precision, as long as communication times are dominated by bandwidth and not latency\footnote{This depends on the size of the problem and the number of cores used. However, for normal production applications, communication is often bandwidth limited, as communication is only required once per time step, and the data-package size scales with the number of particles, noting that practically all applications use $\nppc\gg1$, and thus, the total number of particles determines the relevant problem size rather than the number of grid cells. Packing MPI messages and communicating many quantities per particle means that applications using this Hybrid PIC code are usually bandwidth-dominated and therefore benefit from using hybrid precision compared to double precision.}. However, using single precision without safeguards may lead to substantial inaccuracies in the numerical solution\footnote{We note that any computation is necessarily only accurate to machine precision -- the question is whether the quantity of interest is sufficiently accurate in the context of a specific application.}. Thus, we will use the `hybrid precision' approach introduced in \citet{Federrath+2021}. The basic approach is to store and communicate all quantities that occupy large arrays (such as any quantity that is defined per grid cell or per particle) in single precision arrays. However, computations that require double precision for sufficient accuracy are carried out in double precision by explicit promotion to double precision during those operations. The operations that require double precision include summing up the mass, charge and momentum of all the particles. Additionally, the time step is promoted to double precision throughout the code. All global integral quantities for grid variables, such as electric and magnetic energy, as well as particle variables like the component-wise mean particle velocity and the particle Larmor radius, are also calculated in double precision. As shown in \citet{Federrath+2021}, this approach cuts down memory storage and compute time requirements by nearly a factor of $\sim2$ overall, while retaining sufficient (near double precision) accuracy for all relevant quantities.

\begin{figure}
    \centering
    \includegraphics[scale=0.22]{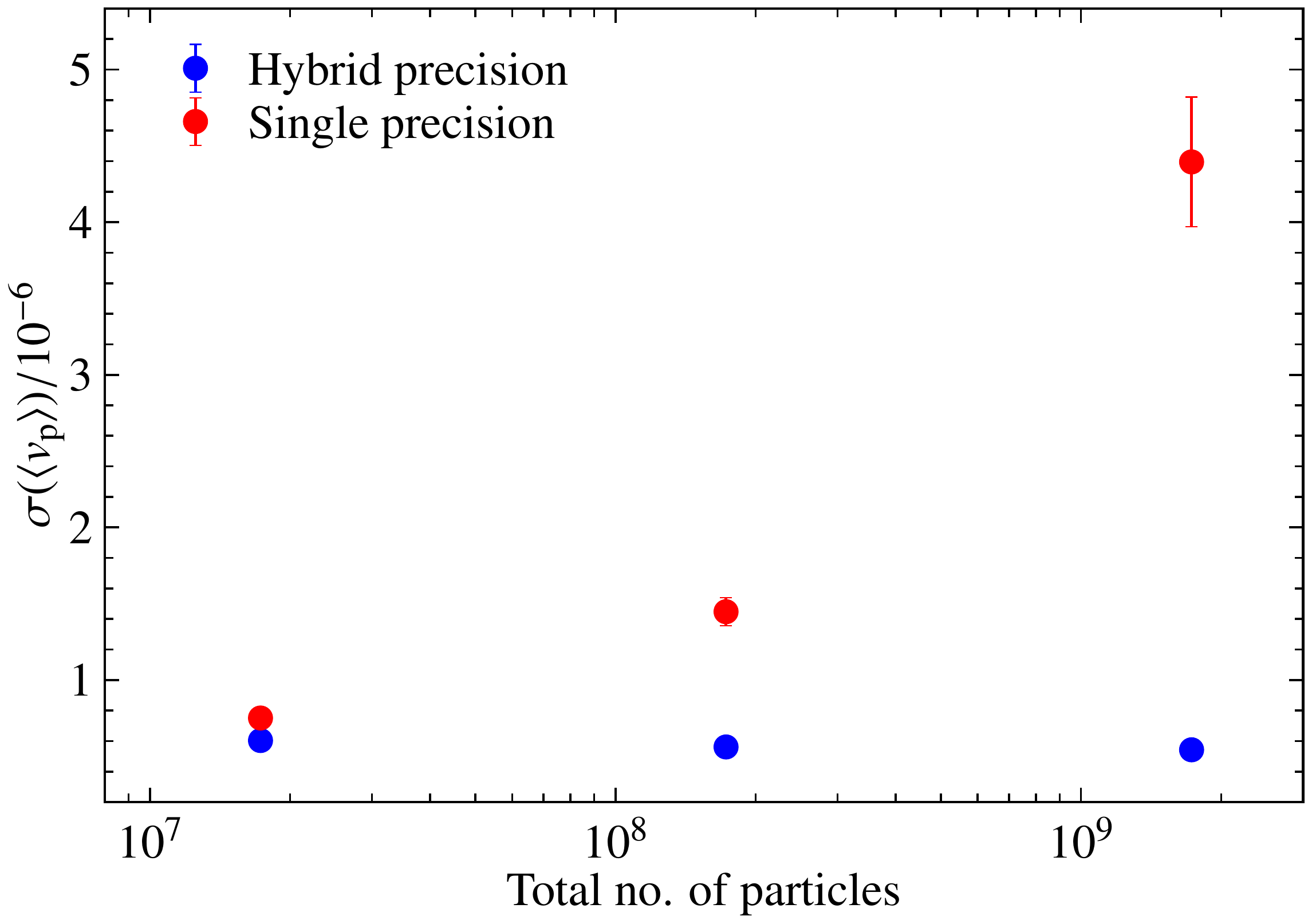}
    \caption{The time- and particle-averaged variations in the average particle velocity (averaged over all particles and across the $x$, $y$ and $z$ components), normalised to the expected precision level with single precision $\approx 10^{-6}$, as a function of the total number of particles, $N$, for numerical tests with single precision (red) and hybrid precision (blue). The error bars are calculated as the standard deviation of the $x$, $y$ and $z$ components of $\sigma(\vpavg)$. Using single precision, the variation in the average particle speed increases, becoming worse with increasing number of particles. By contrast, with hybrid precision, the error is bounded, as desired.}
    \label{fig:sigma_v}
\end{figure}

Here we implement and test this hybrid precision approach for Hybrid PIC, and compare the results of this with pure single precision and pure double precision solutions. To do this, we use the same basic turbulence setup as in the previous two sections, as these represent complex and relevant use cases of the code. \Fig{fig:sigma_v} shows the fluctuations in the average particle velocity over time, as a function of the total number of particles, $N$, with single and hybrid precision. The fluctuations in the average particle velocity, $\sigma(\vpavg)$, are calculated as the standard deviation of the average particle speed, $\vpavg$, over time ($\approx 0 - 1 \, \ted$). The average particle speed and its fluctuations are expected to be zero to within machine precision, as there are no bulk flows in the plasma for the turbulent driving setup. Using single precision, we find that $\sigma(\vpavg)/10^{-6}$ increases strongly as the number of particles is increased, where $10^{-6}$ is the expected precision level for single precision. This makes using single precision undesirable, as with a higher number of particles the numerical solutions become progressively worse and unreliable. On the other hand, using hybrid precision, we find $\sigma(\vpavg)/10^{-6} \lesssim 1$, and this value does not change as the number of particles is increased. Thus, the absolute error with hybrid precision remains bound to the level of single precision, while with pure single precision, the errors increase very quickly.

We note that while \Fig{fig:sigma_v} shows a specific metric for comparison, other metrics show a similar behaviour. Thus, hybrid precision provides a stable and accurate (to within $\lesssim10^{-6}$) solution, while single precision does not. We also note that pure double precision outperforms hybrid precision in this metric, as expected (the fluctuations in the average particle speed are zero to within double precision, as expected). However, the main goal here is to establish that in the hybrid-precision approach (while naturally only providing an absolute accuracy between that of single and double precision), the accuracy of the solution is stable with hybrid precision, while it quickly deteriorates with pure single precision.

\begin{figure}
    \centering
    \includegraphics[scale=0.23]{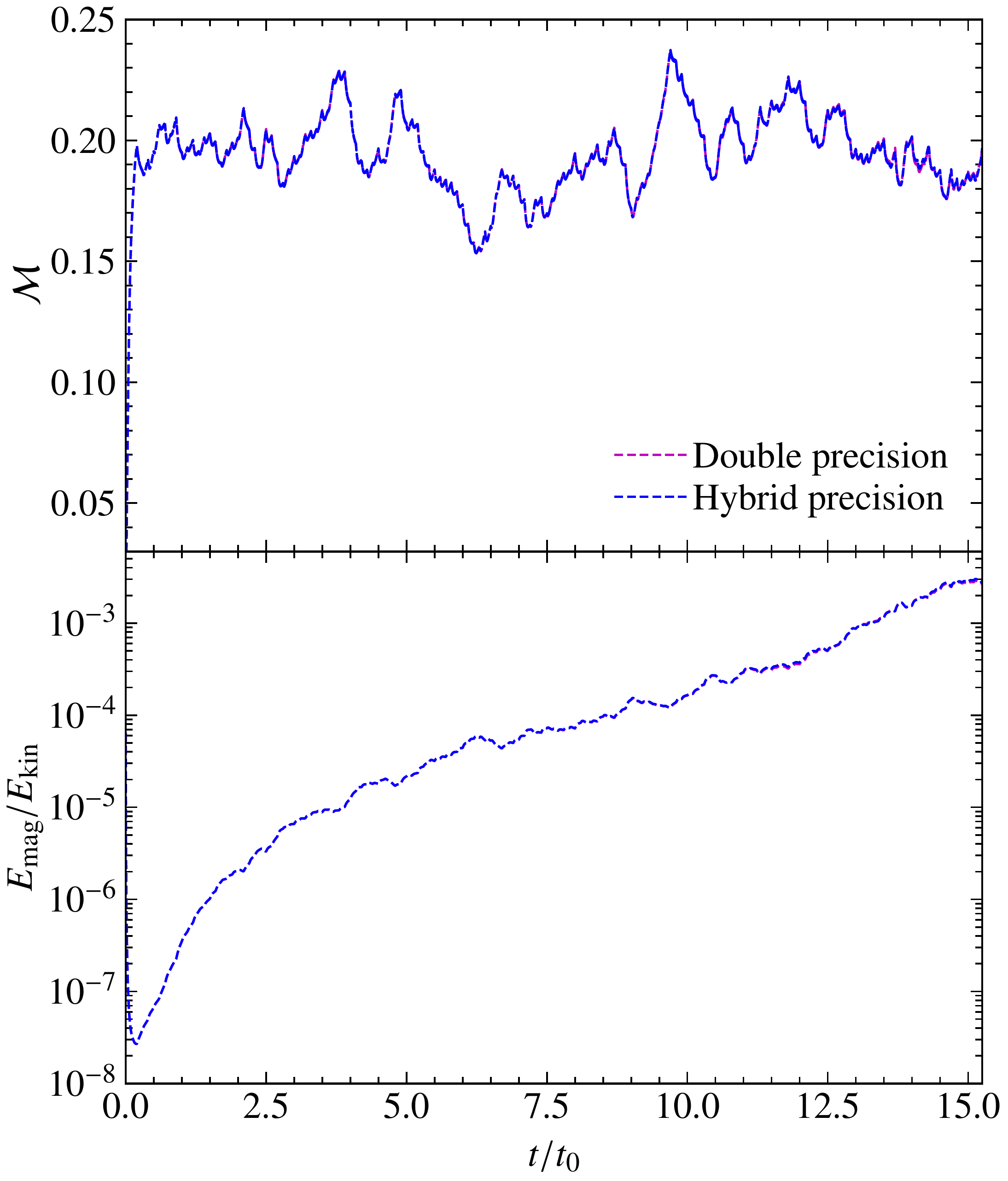}
    \caption{Same as \Fig{fig:interp_tevol}, but for simulations comparing double and hybrid precision. The numerical solutions for the Mach number, $\mathcal{M}$, and the ratio of magnetic energy to kinetic energy, $\ratio$, show excellent agreement.} 
    \label{fig:hptest_turb}
\end{figure}

To further demonstrate that the relevant physical solution is practically as accurate with hybrid precision as it is with pure double precision, we show the Mach number and the magnetic-to-kinetic energy ratio in tests with double and hybrid precision in \Fig{fig:hptest_turb}. We find excellent agreement between the double and hybrid precision runs. We also checked other quantities, including probability density functions and power spectra, which all show excellent agreement as well (see \App{app:precision_spectra_pdf}). The following subsection compares the computational cost of hybrid vs.~double precision, and quantifies the parallel scaling of the code.

\subsection{Parallel code scaling}
\label{sec:scaling}
\begin{figure}
    \centering
    \includegraphics[scale=0.215]{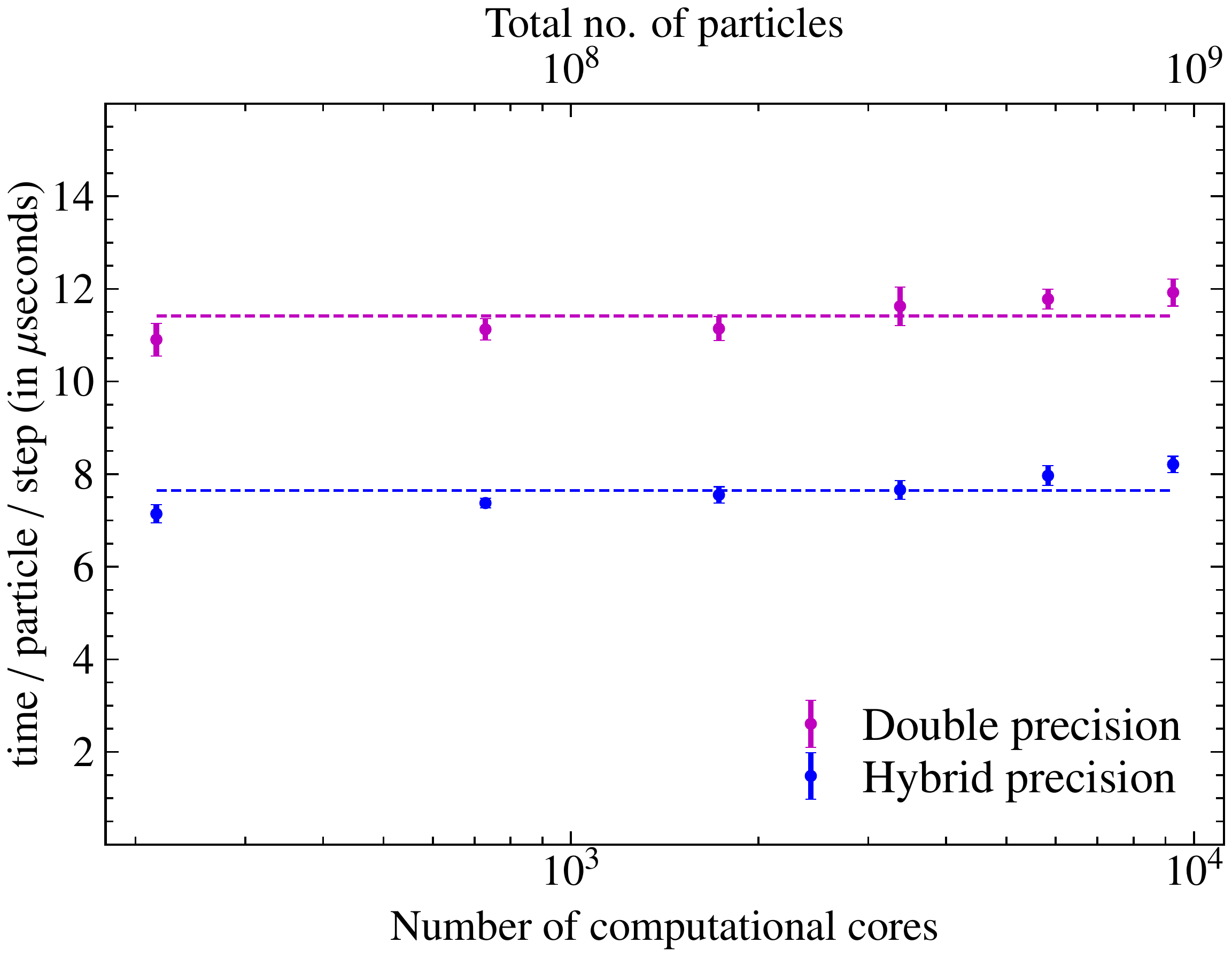}
    \caption{Time taken by the code to perform one-time step (averaged over a total of 100~time steps), normalised by the total number of particles (top x-axis) as a function of the number of compute cores for the turbulence simulations described in \Sec{sec:hybrid_precision}, with double precision (magenta) and hybrid precision (blue). The respective dashed lines represent fits with a constant (ideal weak scaling). Both double and hybrid precision cases show excellent parallel scaling up to $\sim10^4$ compute cores. Moreover, hybrid precision simulations are a factor of $\sim1.5$ faster than pure double precision calculations. The error bars are obtained from the standard deviation of sub-intervals of 10~time steps over the total time-step interval.}
    \label{fig:scaling_cooling}
\end{figure}

Here we test how the performance of our code scales with an increasing number of computational cores for numerical tests with hybrid and double precision, discussed in \Sec{sec:hybrid_precision}. We show the time per particle per time step as a function of the number of compute cores in \Fig{fig:scaling_cooling}. We note that this is a weak scaling test, i.e., the number of particles grows proportionally with the number of compute cores (see top x-axis for the absolute number of particles). Therefore, ideal scaling is indicated by a constant time per particle per time step. We calculate the time by taking an average of over 100~time steps, where the error bars have been estimated as the standard deviation over intervals of 10~time steps in the total time step range. The dashed lines show fits with a constant, i.e., ideal scaling.

We find excellent (near ideal) parallel scaling of the code with both double and hybrid precision up to $\sim10^4$ compute cores. \Fig{fig:scaling_cooling} also shows that the hybrid precision simulations are a factor of $\approx 1.5$ faster than the double precision calculations. The theoretical speed-up could be as high as factor~2, as discussed in Sec.~\ref{sec:hybrid_precision}\footnote{Measuring the time for guard-cell updates alone, we indeed observe a factor of~$\sim2$ speed-up with hybrid- vs.~double-precision, confirming that MPI communication itself is primarily bandwidth-limited rather than latency-limited, as discussed in Sec.~\ref{sec:hybrid_precision}.}, but significant care must be taken in promoting relevant single-precision operations to double precision, in order to maintain sufficient overall accuracy of the code (c.f., Fig.~\ref{fig:hptest_turb}) Therefore, a reduction of the required compute time by $\sim33\%$ with hybrid precision compared to pure double precision is highly beneficial for computationally-expensive Hybrid PIC simulations.

\section{Summary and Conclusions}
\label{sec:summary}
We introduced $\ahkash$ -- a new Hybrid particle-in-cell (PIC) code within the $\texttt{FLASH}$ astrophysics code framework \citep{Fryxelletal2000,Dubey+2008a,Dubey+2008b} to study weakly collisional plasmas. The Hybrid PIC code includes state-of-the-art numerical methods for time integration of particle trajectories, i.e., the Boris integrator \citep[][\Sec{sec:Boris_push}]{Boris1970,Zenitani&Umeda2018}, and the predictor-predictor-corrector integration algorithm \citep[][\Sec{sec:integration_scheme}]{Kunz+2014a}. We use the constrained transport (CT) method to ensure that magnetic fields remain divergence-free \citep[][\Sec{sec:CT}]{Yee1966}. The code supports various grid$\to$particle and particle$\to$grid interpolation schemes, as well as post-interpolation smoothing to reduce finite particle noise (see \Sec{sec:interpolation&filtering}). It further supports the `$\delta f$ method' to study instabilities in weakly collisional plasmas (see \Sec{sec:deltaf}). The code implements a series of time step constraints to ensure that physical plasma time scales are appropriately resolved (see \Sec{sec:timesteps}).

We perform several tests to demonstrate the ability and accuracy of the new code to model standard physical problems, such as the motion of a charged particle in the presence of magnetic fields (\Sec{sec:particle_test}), the propagation of Alfv\'{e}n and whistler waves (\Sec{sec:wave_test}), and Landau damping of ion acoustic waves (\Sec{sec:landau_damping}).

In Sec.~\ref{sec:interpolation_tests} we compare the quality of various particle$\to$grid interpolation schemes. We find that global quantities are largely independent of the details of the interpolation scheme, provided that at least 1~smoothing pass is performed after the particle deposition step. However, structural details can vary significantly with the interpolation scheme and with the number of particles per cell ($\nppc$). We suggest that a reasonable compromise between computational feasibility and accuracy is achieved with the cloud-in-cell interpolation and 2~post-deposition smoothing passes for $\nppc\sim100$~particles per cell, noting that detailed requirements are case-specific and problem-dependent.

The new code further supports turbulence driving \citep{Federrath+2010,Federrath+2021}, modelled with the Ornstein-Uhlenbeck process (\Sec{sec:driving}). To study steady-state turbulence in weakly collisional plasma, we have to maintain isothermal conditions across the computational domain and to do so we introduced a novel cooling method in the code (\Sec{sec:cooling}). This allows the modelling of fully developed, steady-state turbulence, including turbulent dynamos in weakly collisional plasmas, whose properties are sensitive to the Mach number of the plasma. In particular, we show that the cooling method is essential for modelling steady-state supersonic turbulence in Hybrid PIC. We demonstrate and quantify the ability of the cooling method to keep the plasma both locally and globally isothermal (\Sec{sec:cooling_tests}).

Since Hybrid PIC simulations can be computationally demanding, we extend the hybrid precision method introduced in \citet{Federrath+2021} for grid-based hydrodynamics to our Hybrid PIC code (Sec.~\ref{sec:hybrid_precision}). This method uses a hybrid approach that promotes critical computations to double precision arithmetic (8~bytes per floating-point number), but otherwise stores and communicates all quantities in single precision (4~bytes per floating-point number). This provides a factor of $\sim1.5$ speed-up over a pure double precision calculation (c.f.~Fig~\ref{fig:scaling_cooling}), but retains sufficient accuracy and precision throughout the calculations (c.f.~Figs.\ref{fig:sigma_v} and~\ref{fig:hptest_turb}). Finally, Fig.~\ref{fig:scaling_cooling} shows that the new code exhibits excellent parallel scalability up to 10,000~compute cores.

We plan to use $\ahkash$ to study astrophysical problems, especially the physics of the turbulent dynamo in supersonic plasmas and its application to the intracluster medium of galaxy clusters.

\section*{Acknowledgements}
We thank Matthew W.~Kunz for sharing his expertise on developing Hybrid PIC codes and for many useful discussions that have been very beneficial to this work. R.~A.~C.~acknowledges that this work was supported by an NCI HPC-AI Talent Program 2023 Scholarship (project gp08), with computational resources provided by NCI Australia, an NCRIS-enabled capability supported by the Australian Government. C.~F.~acknowledges funding provided by the Australian Research Council (Discovery Project DP230102280), and the Australia-Germany Joint Research Cooperation Scheme (UA-DAAD). We further acknowledge high-performance computing resources provided by the Leibniz Rechenzentrum and the Gauss Centre for Supercomputing (grants~pr32lo, pr48pi and GCS Large-scale project~10391), the Australian National Computational Infrastructure (grant~ek9) and the Pawsey Supercomputing Centre (project~pawsey0810) in the framework of the National Computational Merit Allocation Scheme and the ANU Merit Allocation Scheme. The simulation software, \texttt{FLASH}, was in part developed by the Flash Centre for Computational Science at the University of Chicago and the Department of Physics and Astronomy at the University of Rochester.

\section*{Data Availability}
The Hybrid particle-in-cell code $\ahkash$ has been implemented in our private fork of the \texttt{FLASH} code. The code and results from our simulations will be shared on reasonable request to the corresponding author.

\bibliographystyle{mnras}
\bibliography{Reffiles_radhika.bib}

\begin{thebibliography}{}
\makeatletter
\relax
\def\mn@urlcharsother{\let\do\@makeother \do\$\do\&\do\#\do\^\do\_\do\%\do\~}
\def\mn@doi{\begingroup\mn@urlcharsother \@ifnextchar [ {\mn@doi@} {\mn@doi@[]}}
\def\mn@doi@[#1]#2{\def\@tempa{#1}\ifx\@tempa\@empty \href {http://dx.doi.org/#2} {doi:#2}\else \href {http://dx.doi.org/#2} {#1}\fi \endgroup}
\def\mn@eprint#1#2{\mn@eprint@#1:#2::\@nil}
\def\mn@eprint@arXiv#1{\href {http://arxiv.org/abs/#1} {{\tt arXiv:#1}}}
\def\mn@eprint@dblp#1{\href {http://dblp.uni-trier.de/rec/bibtex/#1.xml} {dblp:#1}}
\def\mn@eprint@#1:#2:#3:#4\@nil{\def\@tempa {#1}\def\@tempb {#2}\def\@tempc {#3}\ifx \@tempc \@empty \let \@tempc \@tempb \let \@tempb \@tempa \fi \ifx \@tempb \@empty \def\@tempb {arXiv}\fi \@ifundefined {mn@eprint@\@tempb}{\@tempb:\@tempc}{\expandafter \expandafter \csname mn@eprint@\@tempb\endcsname \expandafter{\@tempc}}}

\bibitem[\protect\citeauthoryear{{Achikanath Chirakkara}, {Federrath}, {Trivedi}  \& {Banerjee}}{{Achikanath Chirakkara} et~al.}{2021}]{AchikanathEtAl2021}
{Achikanath Chirakkara} R.,  {Federrath} C.,  {Trivedi} P.,   {Banerjee} R.,  2021, \mn@doi [\prl] {10.1103/PhysRevLett.126.091103}, \href {https://ui.adsabs.harvard.edu/abs/2021PhRvL.126i1103A} {126, 091103}

\bibitem[\protect\citeauthoryear{{Achikanath Chirakkara}, {Seta}, {Federrath}  \& {Kunz}}{{Achikanath Chirakkara} et~al.}{2024}]{AchikanathChirakkara+2023}
{Achikanath Chirakkara} R.,  {Seta} A.,  {Federrath} C.,   {Kunz} M.~W.,  2024, \mn@doi [\mnras] {10.1093/mnras/stad3967}, \href {https://ui.adsabs.harvard.edu/abs/2024MNRAS.528..937A} {528, 937}

\bibitem[\protect\citeauthoryear{{Arzamasskiy}, {Kunz}, {Chandran}  \& {Quataert}}{{Arzamasskiy} et~al.}{2019}]{Arzamasskiy+2019}
{Arzamasskiy} L.,  {Kunz} M.~W.,  {Chandran} B. D.~G.,   {Quataert} E.,  2019, \mn@doi [\apj] {10.3847/1538-4357/ab20cc}, \href {https://ui.adsabs.harvard.edu/abs/2019ApJ...879...53A} {879, 53}

\bibitem[\protect\citeauthoryear{{Bagdonat}}{{Bagdonat}}{2004}]{Bagdonat2004}
{Bagdonat} T.~B.,  2004, PhD thesis, Technical University of Braunschweig, Germany

\bibitem[\protect\citeauthoryear{{Bagdonat} \& {Motschmann}}{{Bagdonat} \& {Motschmann}}{2002}]{Bagdonat&Motschmann2002}
{Bagdonat} T.,  {Motschmann} U.,  2002, \mn@doi [Journal of Computational Physics] {10.1006/jcph.2002.7203}, \href {https://ui.adsabs.harvard.edu/abs/2002JCoPh.183..470B} {183, 470}

\bibitem[\protect\citeauthoryear{{Birdsall} \& {Langdon}}{{Birdsall} \& {Langdon}}{1991}]{Birdsall&Langdon1991}
{Birdsall} C.~K.,  {Langdon} A.~B.,  1991, {Plasma Physics via Computer Simulation}

\bibitem[\protect\citeauthoryear{Boris}{Boris}{1970}]{Boris1970}
Boris J.~P.,  1970, Proceeding of Fourth Conference on Numerical Simulations of Plasmas

\bibitem[\protect\citeauthoryear{{Brandenburg} \& {Dobler}}{{Brandenburg} \& {Dobler}}{2010}]{Pencil}
{Brandenburg} A.,  {Dobler} W.,  2010, {Pencil: Finite-difference Code for Compressible Hydrodynamic Flows} (\mn@eprint {ascl} {1010.060})

\bibitem[\protect\citeauthoryear{{Brandenburg} \& {Subramanian}}{{Brandenburg} \& {Subramanian}}{2005}]{B&S2005}
{Brandenburg} A.,  {Subramanian} K.,  2005, \mn@doi [\physrep] {10.1016/j.physrep.2005.06.005}, \href {https://ui.adsabs.harvard.edu/abs/2005PhR...417....1B} {417, 1}

\bibitem[\protect\citeauthoryear{{Bruno} \& {Carbone}}{{Bruno} \& {Carbone}}{2013}]{Bruno&Carbone2013}
{Bruno} R.,  {Carbone} V.,  2013, \mn@doi [Living Reviews in Solar Physics] {10.12942/lrsp-2013-2}, \href {https://ui.adsabs.harvard.edu/abs/2013LRSP...10....2B} {10, 2}

\bibitem[\protect\citeauthoryear{{Cerutti}, {Werner}, {Uzdensky}  \& {Begelman}}{{Cerutti} et~al.}{2013}]{Zeltron2013}
{Cerutti} B.,  {Werner} G.~R.,  {Uzdensky} D.~A.,   {Begelman} M.~C.,  2013, \mn@doi [\apj] {10.1088/0004-637X/770/2/147}, \href {https://ui.adsabs.harvard.edu/abs/2013ApJ...770..147C} {770, 147}

\bibitem[\protect\citeauthoryear{{Denton} \& {Kotschenreuther}}{{Denton} \& {Kotschenreuther}}{1995}]{Denton&Kotschenreuther1995}
{Denton} R.~E.,  {Kotschenreuther} M.,  1995, \mn@doi [Journal of Computational Physics] {10.1006/jcph.1995.1136}, \href {https://ui.adsabs.harvard.edu/abs/1995JCoPh.119..283D} {119, 283}

\bibitem[\protect\citeauthoryear{Derouillat et~al.,}{Derouillat et~al.}{2018}]{Smilei2018}
Derouillat J.,  et~al., 2018, \mn@doi [Computer Physics Communications] {https://doi.org/10.1016/j.cpc.2017.09.024}, 222, 351

\bibitem[\protect\citeauthoryear{{Dubey}, {Reid}  \& {Fisher}}{{Dubey} et~al.}{2008a}]{Dubey+2008a}
{Dubey} A.,  {Reid} L.~B.,   {Fisher} R.,  2008a, \mn@doi [Physica Scripta Volume T] {10.1088/0031-8949/2008/T132/014046}, \href {https://ui.adsabs.harvard.edu/abs/2008PhST..132a4046D} {132, 014046}

\bibitem[\protect\citeauthoryear{{Dubey} et~al.,}{{Dubey} et~al.}{2008b}]{Dubey+2008b}
{Dubey} A.,  et~al., 2008b, in {Pogorelov} N.~V.,  {Audit} E.,   {Zank} G.~P.,  eds,  Astronomical Society of the Pacific Conference Series Vol. 385, Numerical Modeling of Space Plasma Flows. p.~145

\bibitem[\protect\citeauthoryear{{Eswaran} \& {Pope}}{{Eswaran} \& {Pope}}{1988}]{EswaranPope1988}
{Eswaran} V.,  {Pope} S.~B.,  1988, Computers and Fluids, \href {http://adsabs.harvard.edu/abs/1988CF.....16..257E} {16, 257}

\bibitem[\protect\citeauthoryear{{Event Horizon Telescope Collaboration} et~al.,}{{Event Horizon Telescope Collaboration} et~al.}{2019}]{EHTCollab2019}
{Event Horizon Telescope Collaboration} et~al., 2019, \mn@doi [\apjl] {10.3847/2041-8213/ab0ec7}, \href {https://ui.adsabs.harvard.edu/abs/2019ApJ...875L...1E} {875, L1}

\bibitem[\protect\citeauthoryear{{Federrath}}{{Federrath}}{2013}]{Federrath2013}
{Federrath} C.,  2013, \mn@doi [\mnras] {10.1093/mnras/stt1644}, \href {https://ui.adsabs.harvard.edu/abs/2013MNRAS.436.1245F} {436, 1245}

\bibitem[\protect\citeauthoryear{{Federrath}}{{Federrath}}{2016}]{Federrath2016jpp}
{Federrath} C.,  2016, \mn@doi [Journal of Plasma Physics] {10.1017/S0022377816001069}, \href {http://adsabs.harvard.edu/abs/2016JPlPh..82f5301F} {82, 535820601}

\bibitem[\protect\citeauthoryear{{Federrath} \& {Banerjee}}{{Federrath} \& {Banerjee}}{2015}]{Federrath&Banerjee2015}
{Federrath} C.,  {Banerjee} S.,  2015, \mn@doi [\mnras] {10.1093/mnras/stv180}, \href {https://ui.adsabs.harvard.edu/abs/2015MNRAS.448.3297F} {448, 3297}

\bibitem[\protect\citeauthoryear{{Federrath}, {Roman-Duval}, {Klessen}, {Schmidt}  \& {Mac Low}}{{Federrath} et~al.}{2010}]{Federrath+2010}
{Federrath} C.,  {Roman-Duval} J.,  {Klessen} R.~S.,  {Schmidt} W.,   {Mac Low} M.~M.,  2010, \mn@doi [\aap] {10.1051/0004-6361/200912437}, \href {https://ui.adsabs.harvard.edu/abs/2010A&A...512A..81F} {512, A81}

\bibitem[\protect\citeauthoryear{{Federrath}, {Chabrier}, {Schober}, {Banerjee}, {Klessen}  \& {Schleicher}}{{Federrath} et~al.}{2011}]{CFetal11}
{Federrath} C.,  {Chabrier} G.,  {Schober} J.,  {Banerjee} R.,  {Klessen} R.~S.,   {Schleicher} D.~R.~G.,  2011, \mn@doi [\prl] {10.1103/PhysRevLett.107.114504}, \href {https://ui.adsabs.harvard.edu/abs/2011PhRvL.107k4504F} {107, 114504}

\bibitem[\protect\citeauthoryear{{Federrath}, {Klessen}, {Iapichino}  \& {Beattie}}{{Federrath} et~al.}{2021}]{Federrath+2021}
{Federrath} C.,  {Klessen} R.~S.,  {Iapichino} L.,   {Beattie} J.~R.,  2021, \mn@doi [Nature Astronomy] {10.1038/s41550-020-01282-z}, \href {https://ui.adsabs.harvard.edu/abs/2021NatAs...5..365F} {5, 365}

\bibitem[\protect\citeauthoryear{{Federrath}, {Roman-Duval}, {Klessen}, {Schmidt}  \& {Mac Low}}{{Federrath} et~al.}{2022}]{FederrathEtAl2022ascl}
{Federrath} C.,  {Roman-Duval} J.,  {Klessen} R.~S.,  {Schmidt} W.,   {Mac Low} M.~M.,  2022, {TG: Turbulence Generator}, Astrophysics Source Code Library, record ascl:2204.001 (\mn@eprint {ascl} {2204.001})

\bibitem[\protect\citeauthoryear{{Ferri{\`e}re}}{{Ferri{\`e}re}}{2020}]{Ferriere2020}
{Ferri{\`e}re} K.,  2020, \mn@doi [Plasma Physics and Controlled Fusion] {10.1088/1361-6587/ab49eb}, \href {https://ui.adsabs.harvard.edu/abs/2020PPCF...62a4014F} {62, 014014}

\bibitem[\protect\citeauthoryear{{Frisch}}{{Frisch}}{1995}]{Frisch1995}
{Frisch} U.,  1995, {Turbulence}

\bibitem[\protect\citeauthoryear{{Fryxell} et~al.,}{{Fryxell} et~al.}{2000}]{Fryxelletal2000}
{Fryxell} B.,  et~al., 2000, \mn@doi [\apjs] {10.1086/317361}, \href {https://ui.adsabs.harvard.edu/abs/2000ApJS..131..273F} {131, 273}

\bibitem[\protect\citeauthoryear{{Ganse} et~al.,}{{Ganse} et~al.}{2023}]{Vlasiator2023}
{Ganse} U.,  et~al., 2023, \mn@doi [Physics of Plasmas] {10.1063/5.0134387}, \href {https://ui.adsabs.harvard.edu/abs/2023PhPl...30d2902G} {30, 042902}

\bibitem[\protect\citeauthoryear{{Gargat{\'e}}, {Bingham}, {Fonseca}  \& {Silva}}{{Gargat{\'e}} et~al.}{2007}]{Gargat+2007}
{Gargat{\'e}} L.,  {Bingham} R.,  {Fonseca} R.~A.,   {Silva} L.~O.,  2007, \mn@doi [Computer Physics Communications] {10.1016/j.cpc.2006.11.013}, \href {https://ui.adsabs.harvard.edu/abs/2007CoPhC.176..419G} {176, 419}

\bibitem[\protect\citeauthoryear{{Gatuzz}, {Sanders}, {Dennerl}, {Pinto}, {Fabian}, {Tamura}, {Walker}  \& {ZuHone}}{{Gatuzz} et~al.}{2022a}]{Gatuzz+2022a}
{Gatuzz} E.,  {Sanders} J.~S.,  {Dennerl} K.,  {Pinto} C.,  {Fabian} A.~C.,  {Tamura} T.,  {Walker} S.~A.,   {ZuHone} J.,  2022a, \mn@doi [\mnras] {10.1093/mnras/stab2661}, \href {https://ui.adsabs.harvard.edu/abs/2022MNRAS.511.4511G} {511, 4511}

\bibitem[\protect\citeauthoryear{{Gatuzz} et~al.,}{{Gatuzz} et~al.}{2022b}]{Gatuzz+2022b}
{Gatuzz} E.,  et~al., 2022b, \mn@doi [\mnras] {10.1093/mnras/stac846}, \href {https://ui.adsabs.harvard.edu/abs/2022MNRAS.513.1932G} {513, 1932}

\bibitem[\protect\citeauthoryear{{Gatuzz} et~al.,}{{Gatuzz} et~al.}{2023}]{Gatuzz+2023}
{Gatuzz} E.,  et~al., 2023, \mn@doi [\mnras] {10.1093/mnras/stad1132}, \href {https://ui.adsabs.harvard.edu/abs/2023MNRAS.522.2325G} {522, 2325}

\bibitem[\protect\citeauthoryear{{Gent}, {Mac Low}, {Korpi-Lagg}  \& {Singh}}{{Gent} et~al.}{2023}]{Gent+2023}
{Gent} F.~A.,  {Mac Low} M.-M.,  {Korpi-Lagg} M.~J.,   {Singh} N.~K.,  2023, \mn@doi [\apj] {10.3847/1538-4357/acac20}, \href {https://ui.adsabs.harvard.edu/abs/2023ApJ...943..176G} {943, 176}

\bibitem[\protect\citeauthoryear{{Haugen}, {Brandenburg}  \& {Mee}}{{Haugen} et~al.}{2004a}]{Haugen+2004a}
{Haugen} N. E.~L.,  {Brandenburg} A.,   {Mee} A.~J.,  2004a, \mn@doi [\mnras] {10.1111/j.1365-2966.2004.08127.x}, \href {https://ui.adsabs.harvard.edu/abs/2004MNRAS.353..947H} {353, 947}

\bibitem[\protect\citeauthoryear{{Haugen}, {Brandenburg}  \& {Mee}}{{Haugen} et~al.}{2004b}]{Haugen+2004b}
{Haugen} N. E.~L.,  {Brandenburg} A.,   {Mee} A.~J.,  2004b, \mn@doi [\mnras] {10.1111/j.1365-2966.2004.08127.x}, \href {https://ui.adsabs.harvard.edu/abs/2004MNRAS.353..947H} {353, 947}

\bibitem[\protect\citeauthoryear{{Hitomi Collaboration} et~al.,}{{Hitomi Collaboration} et~al.}{2016}]{HitomiCollaboration2016}
{Hitomi Collaboration} et~al., 2016, \mn@doi [\nat] {10.1038/nature18627}, \href {https://ui.adsabs.harvard.edu/abs/2016Natur.535..117H} {535, 117}

\bibitem[\protect\citeauthoryear{{Holmstrom}}{{Holmstrom}}{2009}]{Holmstrom2009}
{Holmstrom} M.,  2009, arXiv e-prints, \href {https://ui.adsabs.harvard.edu/abs/2009arXiv0911.4435H} {p. arXiv:0911.4435}

\bibitem[\protect\citeauthoryear{{Holmstr{\"o}m}}{{Holmstr{\"o}m}}{2011}]{Holmstrom2011}
{Holmstr{\"o}m} M.,  2011, in {Pogorelov} N.~V.,  {Audit} E.,   {Zank} G.~P.,  eds,  Astronomical Society of the Pacific Conference Series Vol. 444, 5th International Conference of Numerical Modeling of Space Plasma Flows (ASTRONUM 2010). p.~211 (\mn@eprint {arXiv} {1010.3291})

\bibitem[\protect\citeauthoryear{{Horowitz}, {Shumaker}  \& {Anderson}}{{Horowitz} et~al.}{1989}]{Horowitz+1989}
{Horowitz} E.~J.,  {Shumaker} D.~E.,   {Anderson} D.~V.,  1989, \mn@doi [Journal of Computational Physics] {10.1016/0021-9991(89)90234-9}, \href {https://ui.adsabs.harvard.edu/abs/1989JCoPh..84..279H} {84, 279}

\bibitem[\protect\citeauthoryear{{Howes}}{{Howes}}{2010}]{Howes2010}
{Howes} G.~G.,  2010, \mn@doi [\mnras] {10.1111/j.1745-3933.2010.00958.x}, \href {https://ui.adsabs.harvard.edu/abs/2010MNRAS.409L.104H} {409, L104}

\bibitem[\protect\citeauthoryear{{Howes}, {Klein}  \& {TenBarge}}{{Howes} et~al.}{2014}]{Howes+2014}
{Howes} G.~G.,  {Klein} K.~G.,   {TenBarge} J.~M.,  2014, \mn@doi [\apj] {10.1088/0004-637X/789/2/106}, \href {https://ui.adsabs.harvard.edu/abs/2014ApJ...789..106H} {789, 106}

\bibitem[\protect\citeauthoryear{{Kazantsev}}{{Kazantsev}}{1968}]{Kazantsev1968}
{Kazantsev} A.~P.,  1968, Soviet Journal of Experimental and Theoretical Physics, \href {https://ui.adsabs.harvard.edu/abs/1968JETP...26.1031K} {26, 1031}

\bibitem[\protect\citeauthoryear{{Klein} \& {Howes}}{{Klein} \& {Howes}}{2015}]{Klein&Howes2015}
{Klein} K.~G.,  {Howes} G.~G.,  2015, \mn@doi [Physics of Plasmas] {10.1063/1.4914933}, \href {https://ui.adsabs.harvard.edu/abs/2015PhPl...22c2903K} {22, 032903}

\bibitem[\protect\citeauthoryear{{Kriel}, {Beattie}, {Seta}  \& {Federrath}}{{Kriel} et~al.}{2022}]{KrielEtAl2022}
{Kriel} N.,  {Beattie} J.~R.,  {Seta} A.,   {Federrath} C.,  2022, \mn@doi [\mnras] {10.1093/mnras/stac969}, \href {https://ui.adsabs.harvard.edu/abs/2022MNRAS.tmp..948K} {}

\bibitem[\protect\citeauthoryear{{Kulsrud}}{{Kulsrud}}{2005}]{Kulsrud2005}
{Kulsrud} R.~M.,  2005, {Plasma physics for astrophysics}

\bibitem[\protect\citeauthoryear{{Kunz}, {Schekochihin}, {Cowley}, {Binney}  \& {Sanders}}{{Kunz} et~al.}{2011}]{Kunz+2011}
{Kunz} M.~W.,  {Schekochihin} A.~A.,  {Cowley} S.~C.,  {Binney} J.~J.,   {Sanders} J.~S.,  2011, \mn@doi [\mnras] {10.1111/j.1365-2966.2010.17621.x}, \href {https://ui.adsabs.harvard.edu/abs/2011MNRAS.410.2446K} {410, 2446}

\bibitem[\protect\citeauthoryear{{Kunz}, {Schekochihin}  \& {Stone}}{{Kunz} et~al.}{2014a}]{Kunz+2014b}
{Kunz} M.~W.,  {Schekochihin} A.~A.,   {Stone} J.~M.,  2014a, \mn@doi [\prl] {10.1103/PhysRevLett.112.205003}, \href {https://ui.adsabs.harvard.edu/abs/2014PhRvL.112t5003K} {112, 205003}

\bibitem[\protect\citeauthoryear{{Kunz}, {Stone}  \& {Bai}}{{Kunz} et~al.}{2014b}]{Kunz+2014a}
{Kunz} M.~W.,  {Stone} J.~M.,   {Bai} X.-N.,  2014b, \mn@doi [Journal of Computational Physics] {10.1016/j.jcp.2013.11.035}, \href {https://ui.adsabs.harvard.edu/abs/2014JCoPh.259..154K} {259, 154}

\bibitem[\protect\citeauthoryear{{Kunz}, {Jones}  \& {Zhuravleva}}{{Kunz} et~al.}{2022}]{Kunz+2022}
{Kunz} M.~W.,  {Jones} T.~W.,   {Zhuravleva} I.,  2022, arXiv e-prints, \href {https://ui.adsabs.harvard.edu/abs/2022arXiv220502489K} {p. arXiv:2205.02489}

\bibitem[\protect\citeauthoryear{{Le}, {Stanier}, {Yin}, {Wetherton}, {Keenan}  \& {Albright}}{{Le} et~al.}{2023}]{Le+2023}
{Le} A.,  {Stanier} A.,  {Yin} L.,  {Wetherton} B.,  {Keenan} B.,   {Albright} B.,  2023, \mn@doi [Physics of Plasmas] {10.1063/5.0146529}, \href {https://ui.adsabs.harvard.edu/abs/2023PhPl...30f3902L} {30, 063902}

\bibitem[\protect\citeauthoryear{{Lehe}, {Parrish}  \& {Quataert}}{{Lehe} et~al.}{2009}]{Lehe2009}
{Lehe} R.,  {Parrish} I.~J.,   {Quataert} E.,  2009, \mn@doi [\apj] {10.1088/0004-637X/707/1/404}, \href {https://ui.adsabs.harvard.edu/abs/2009ApJ...707..404L} {707, 404}

\bibitem[\protect\citeauthoryear{{Lipatov}}{{Lipatov}}{2002}]{Lipatov2002}
{Lipatov} A.~S.,  2002, {The hybrid multiscale simulation technology: an introduction with application to astrophysical and laboratory plasmas}

\bibitem[\protect\citeauthoryear{{Mignone}, {Bodo}, {Massaglia}, {Matsakos}, {Tesileanu}, {Zanni}  \& {Ferrari}}{{Mignone} et~al.}{2007}]{Pluto2007}
{Mignone} A.,  {Bodo} G.,  {Massaglia} S.,  {Matsakos} T.,  {Tesileanu} O.,  {Zanni} C.,   {Ferrari} A.,  2007, \mn@doi [\apjs] {10.1086/513316}, \href {https://ui.adsabs.harvard.edu/abs/2007ApJS..170..228M} {170, 228}

\bibitem[\protect\citeauthoryear{{Moffatt}}{{Moffatt}}{1978}]{Moffatt1978}
{Moffatt} H.~K.,  1978, {Magnetic field generation in electrically conducting fluids}

\bibitem[\protect\citeauthoryear{{M{\"u}ller}, {Simon}, {Motschmann}, {Sch{\"u}le}, {Glassmeier}  \& {Pringle}}{{M{\"u}ller} et~al.}{2011}]{Muller+2011}
{M{\"u}ller} J.,  {Simon} S.,  {Motschmann} U.,  {Sch{\"u}le} J.,  {Glassmeier} K.-H.,   {Pringle} G.~J.,  2011, \mn@doi [Computer Physics Communications] {10.1016/j.cpc.2010.12.033}, \href {https://ui.adsabs.harvard.edu/abs/2011CoPhC.182..946M} {182, 946}

\bibitem[\protect\citeauthoryear{{N{\"a}ttil{\"a}}}{{N{\"a}ttil{\"a}}}{2022}]{Runko2022}
{N{\"a}ttil{\"a}} J.,  2022, \mn@doi [\aap] {10.1051/0004-6361/201937402}, \href {https://ui.adsabs.harvard.edu/abs/2022A&A...664A..68N} {664, A68}

\bibitem[\protect\citeauthoryear{{Parker} \& {Lee}}{{Parker} \& {Lee}}{1993}]{Parker&Lee1993}
{Parker} S.~E.,  {Lee} W.~W.,  1993, \mn@doi [Physics of Fluids B] {10.1063/1.860870}, \href {https://ui.adsabs.harvard.edu/abs/1993PhFlB...5...77P} {5, 77}

\bibitem[\protect\citeauthoryear{{Pencil Code Collaboration} et~al.,}{{Pencil Code Collaboration} et~al.}{2021}]{Pencil2021}
{Pencil Code Collaboration} et~al., 2021, \mn@doi [The Journal of Open Source Software] {10.21105/joss.02807}, \href {https://ui.adsabs.harvard.edu/abs/2021JOSS....6.2807P} {6, 2807}

\bibitem[\protect\citeauthoryear{{Qin}, {Zhang}, {Xiao}, {Liu}, {Sun}  \& {Tang}}{{Qin} et~al.}{2013}]{Qin+2013}
{Qin} H.,  {Zhang} S.,  {Xiao} J.,  {Liu} J.,  {Sun} Y.,   {Tang} W.~M.,  2013, \mn@doi [Physics of Plasmas] {10.1063/1.4818428}, \href {https://ui.adsabs.harvard.edu/abs/2013PhPl...20h4503Q} {20, 084503}

\bibitem[\protect\citeauthoryear{{Rincon}, {Califano}, {Schekochihin}  \& {Valentini}}{{Rincon} et~al.}{2016}]{Rinconetal2016}
{Rincon} F.,  {Califano} F.,  {Schekochihin} A.~A.,   {Valentini} F.,  2016, \mn@doi [Proceedings of the National Academy of Science] {10.1073/pnas.1525194113}, \href {https://ui.adsabs.harvard.edu/abs/2016PNAS..113.3950R} {113, 3950}

\bibitem[\protect\citeauthoryear{{Rosin}, {Schekochihin}, {Rincon}  \& {Cowley}}{{Rosin} et~al.}{2011}]{Rosin+2011}
{Rosin} M.~S.,  {Schekochihin} A.~A.,  {Rincon} F.,   {Cowley} S.~C.,  2011, \mn@doi [\mnras] {10.1111/j.1365-2966.2010.17931.x}, \href {https://ui.adsabs.harvard.edu/abs/2011MNRAS.413....7R} {413, 7}

\bibitem[\protect\citeauthoryear{Ruzmaikin, Shukurov  \& Sokoloff}{Ruzmaikin et~al.}{1988}]{Ruzmaikin1988}
Ruzmaikin A.,  Shukurov A.,   Sokoloff D.,  1988, Magnetic Fields of Galaxies.
Academic Press, Dordrecht

\bibitem[\protect\citeauthoryear{{Schekochihin} \& {Cowley}}{{Schekochihin} \& {Cowley}}{2006}]{Schekochihin&Cowley2006}
{Schekochihin} A.~A.,  {Cowley} S.~C.,  2006, \mn@doi [Physics of Plasmas] {10.1063/1.2179053}, \href {https://ui.adsabs.harvard.edu/abs/2006PhPl...13e6501S} {13, 056501}

\bibitem[\protect\citeauthoryear{{Schekochihin}, {Cowley}, {Taylor}, {Maron}  \& {McWilliams}}{{Schekochihin} et~al.}{2004}]{Schekochihin+2004}
{Schekochihin} A.~A.,  {Cowley} S.~C.,  {Taylor} S.~F.,  {Maron} J.~L.,   {McWilliams} J.~C.,  2004, \mn@doi [\apj] {10.1086/422547}, \href {https://ui.adsabs.harvard.edu/abs/2004ApJ...612..276S} {612, 276}

\bibitem[\protect\citeauthoryear{{Schober}, {Schleicher}, {Federrath}, {Klessen}  \& {Banerjee}}{{Schober} et~al.}{2012}]{Schoberetal2012}
{Schober} J.,  {Schleicher} D.,  {Federrath} C.,  {Klessen} R.,   {Banerjee} R.,  2012, \mn@doi [\pre] {10.1103/PhysRevE.85.026303}, \href {https://ui.adsabs.harvard.edu/abs/2012PhRvE..85b6303S} {85, 026303}

\bibitem[\protect\citeauthoryear{{Seta} \& {Federrath}}{{Seta} \& {Federrath}}{2020}]{Seta&Federrath2020}
{Seta} A.,  {Federrath} C.,  2020, \mn@doi [\mnras] {10.1093/mnras/staa2978}, \href {https://ui.adsabs.harvard.edu/abs/2020MNRAS.499.2076S} {499, 2076}

\bibitem[\protect\citeauthoryear{{Seta} \& {Federrath}}{{Seta} \& {Federrath}}{2021}]{Seta&Federrath2021b}
{Seta} A.,  {Federrath} C.,  2021, \mn@doi [Physical Review Fluids] {10.1103/PhysRevFluids.6.103701}, \href {https://ui.adsabs.harvard.edu/abs/2021PhRvF...6j3701S} {6, 103701}

\bibitem[\protect\citeauthoryear{{Seta} \& {Federrath}}{{Seta} \& {Federrath}}{2022}]{Seta&Federrath2022}
{Seta} A.,  {Federrath} C.,  2022, \mn@doi [\mnras] {10.1093/mnras/stac1400}, \href {https://ui.adsabs.harvard.edu/abs/2022MNRAS.514..957S} {514, 957}

\bibitem[\protect\citeauthoryear{{Seta}, {Bushby}, {Shukurov}  \& {Wood}}{{Seta} et~al.}{2020}]{Seta+2020}
{Seta} A.,  {Bushby} P.~J.,  {Shukurov} A.,   {Wood} T.~S.,  2020, \mn@doi [Physical Review Fluids] {10.1103/PhysRevFluids.5.043702}, \href {https://ui.adsabs.harvard.edu/abs/2020PhRvF...5d3702S} {5, 043702}

\bibitem[\protect\citeauthoryear{{Shalaby}, {Broderick}, {Chang}, {Pfrommer}, {Lamberts}  \& {Puchwein}}{{Shalaby} et~al.}{2017}]{SHARP2017}
{Shalaby} M.,  {Broderick} A.~E.,  {Chang} P.,  {Pfrommer} C.,  {Lamberts} A.,   {Puchwein} E.,  2017, \mn@doi [\apj] {10.3847/1538-4357/aa6d13}, \href {https://ui.adsabs.harvard.edu/abs/2017ApJ...841...52S} {841, 52}

\bibitem[\protect\citeauthoryear{{Simionescu} et~al.,}{{Simionescu} et~al.}{2019}]{Simionescu+2019}
{Simionescu} A.,  et~al., 2019, \mn@doi [\ssr] {10.1007/s11214-019-0590-1}, \href {https://ui.adsabs.harvard.edu/abs/2019SSRv..215...24S} {215, 24}

\bibitem[\protect\citeauthoryear{{Spitkovsky}}{{Spitkovsky}}{2005}]{TRISTAN-MP2005}
{Spitkovsky} A.,  2005, in {Bulik} T.,  {Rudak} B.,   {Madejski} G.,  eds,  American Institute of Physics Conference Series Vol. 801, Astrophysical Sources of High Energy Particles and Radiation. pp 345--350 (\mn@eprint {arXiv} {astro-ph/0603211}), \mn@doi{10.1063/1.2141897}

\bibitem[\protect\citeauthoryear{{Squire}, {Kunz}, {Arzamasskiy}, {Johnston}, {Quataert}  \& {Schekochihin}}{{Squire} et~al.}{2023a}]{Squire+2023a}
{Squire} J.,  {Kunz} M.~W.,  {Arzamasskiy} L.,  {Johnston} Z.,  {Quataert} E.,   {Schekochihin} A.~A.,  2023a, \mn@doi [Journal of Plasma Physics] {10.1017/S0022377823000727}, \href {https://ui.adsabs.harvard.edu/abs/2023JPlPh..89d9017S} {89, 905890417}

\bibitem[\protect\citeauthoryear{{Squire}, {Meyrand}  \& {Kunz}}{{Squire} et~al.}{2023b}]{Squire+2023b}
{Squire} J.,  {Meyrand} R.,   {Kunz} M.~W.,  2023b, \mn@doi [\apjl] {10.3847/2041-8213/ad0779}, \href {https://ui.adsabs.harvard.edu/abs/2023ApJ...957L..30S} {957, L30}

\bibitem[\protect\citeauthoryear{{St-Onge} \& {Kunz}}{{St-Onge} \& {Kunz}}{2018}]{St-Onge&Kunz2018}
{St-Onge} D.~A.,  {Kunz} M.~W.,  2018, \mn@doi [\apjl] {10.3847/2041-8213/aad638}, \href {https://ui.adsabs.harvard.edu/abs/2018ApJ...863L..25S} {863, L25}

\bibitem[\protect\citeauthoryear{{Stone}, {Gardiner}, {Teuben}, {Hawley}  \& {Simon}}{{Stone} et~al.}{2008}]{Athena2008}
{Stone} J.~M.,  {Gardiner} T.~A.,  {Teuben} P.,  {Hawley} J.~F.,   {Simon} J.~B.,  2008, \mn@doi [\apjs] {10.1086/588755}, \href {https://ui.adsabs.harvard.edu/abs/2008ApJS..178..137S} {178, 137}

\bibitem[\protect\citeauthoryear{{Stone}, {Tomida}, {White}  \& {Felker}}{{Stone} et~al.}{2020}]{Athena++2020}
{Stone} J.~M.,  {Tomida} K.,  {White} C.~J.,   {Felker} K.~G.,  2020, \mn@doi [\apjs] {10.3847/1538-4365/ab929b}, \href {https://ui.adsabs.harvard.edu/abs/2020ApJS..249....4S} {249, 4}

\bibitem[\protect\citeauthoryear{{Verscharen}, {Klein}  \& {Maruca}}{{Verscharen} et~al.}{2019}]{Verscharen+2019}
{Verscharen} D.,  {Klein} K.~G.,   {Maruca} B.~A.,  2019, \mn@doi [Living Reviews in Solar Physics] {10.1007/s41116-019-0021-0}, \href {https://ui.adsabs.harvard.edu/abs/2019LRSP...16....5V} {16, 5}

\bibitem[\protect\citeauthoryear{{Winske}, {Karimabadi}, {Le}, {Omidi}, {Roytershteyn}  \& {Stanier}}{{Winske} et~al.}{2022}]{Winske+2022}
{Winske} D.,  {Karimabadi} H.,  {Le} A.,  {Omidi} N.,  {Roytershteyn} V.,   {Stanier} A.,  2022, \mn@doi [arXiv e-prints] {10.48550/arXiv.2204.01676}, \href {https://ui.adsabs.harvard.edu/abs/2022arXiv220401676W} {p. arXiv:2204.01676}

\bibitem[\protect\citeauthoryear{{Yee}}{{Yee}}{1966}]{Yee1966}
{Yee} K.,  1966, \mn@doi [IEEE Transactions on Antennas and Propagation] {10.1109/TAP.1966.1138693}, \href {https://ui.adsabs.harvard.edu/abs/1966ITAP...14..302Y} {14, 302}

\bibitem[\protect\citeauthoryear{{Zenitani} \& {Umeda}}{{Zenitani} \& {Umeda}}{2018}]{Zenitani&Umeda2018}
{Zenitani} S.,  {Umeda} T.,  2018, \mn@doi [Physics of Plasmas] {10.1063/1.5051077}, \href {https://ui.adsabs.harvard.edu/abs/2018PhPl...25k2110Z} {25, 112110}

\bibitem[\protect\citeauthoryear{{Zinger}, {Dekel}, {Birnboim}, {Nagai}, {Lau}  \& {Kravtsov}}{{Zinger} et~al.}{2018}]{Zinger+2018}
{Zinger} E.,  {Dekel} A.,  {Birnboim} Y.,  {Nagai} D.,  {Lau} E.,   {Kravtsov} A.~V.,  2018, \mn@doi [\mnras] {10.1093/mnras/sty136}, \href {https://ui.adsabs.harvard.edu/abs/2018MNRAS.476...56Z} {476, 56}

\makeatother
\end{thebibliography}

\appendix
\section{Weight functions}
\label{app:weight_function}
The Cartesian components of $\vec{r}$ and $\vec{r}_l$ are $(x,y,z)$ and $(x_l,y_l,z_l)$, respectively. Using these, we define $\vec{h} = (h_x,h_y,h_z)$, where $h_x = {|x - x_l|}/{\dx}$, $h_y = {|y - y_l|}/{\dy}$ and $h_z = {|z - z_l|}/{\dz}$. $\dx$, $\dy$ and $\dz$ are the grid-cell size in the $x$, $y$ and $z$ directions, respectively.

The nearest-grid-point weight function, where the particle is assigned completely to its nearest grid cell, is defined as
\begin{equation}
    W_{\rm NGP} (\vec{r} - \vec{r}_l) = 
\begin{cases}
    1 ,& \text{if } |\vec{r}-\vec{r}_l| = \text{min}(|\vec{h}|) \\
    0,              & \text{otherwise}.
\end{cases}
\end{equation}
The distance between the particle and each grid-cell centre is calculated and the minimum value, $\text{min}(|\vec{h}|)$, gives us the grid-cell centre closest to the particle.

In the cloud-in-cell (CIC) interpolation kernel, a ``cloud" is created by each particle in the shape of a grid cell and the weight function can be described as 
\begin{equation}
    W_{\rm CIC} (\vec{r} - \vec{r}_l) = 
\begin{cases}
    (1-h_x) (1-h_y)(1-h_z) ,& \text{if } h_x,h_y,h_z < 1 \\
    0,              & \text{otherwise}.
\end{cases}
\end{equation}
The triangular-shaped-cloud weight function is defined co-ordinate wise as
\begin{equation}
    W_{\rm TSC} (\vec{r}_{i} - (\vec{r}_l)_i) = 
\begin{cases}
    1 - 2 h_i^{2} ,& \text{if } h_i \leq 0.5 \\
    2 (1 - h_i)^{2} ,& \text{if } 0.5 < h_i < 1 \\
    0,              & \text{otherwise}.
\end{cases}
\end{equation}
The total weight function, $W_{\rm TSC} (\vec{r} - \vec{r}_l)$, is then given by
\begin{equation}
    W_{\rm TSC} (\vec{r} - \vec{r}_l) = W_{\rm TSC} (\vec{x} - \vec{x}_l) W_{\rm TSC} (\vec{y} - \vec{y}_l)W_{\rm TSC} (\vec{z} - \vec{z}_l).
\end{equation}
In both CIC and TSC interpolation kernels, the size of the weight function is one grid cell ($\sim \dx$). One can also construct larger stencils for the weight function, however, the computational cost of the interpolation operation increases as the extent of the weight function increases.

\section{Cell-centered finite difference method}
\label{app:ccfd}
\begin{figure*}
    \centering
    \includegraphics[scale=1.0, trim={1.4cm 0.3cm 1.4cm 0.3cm},clip]{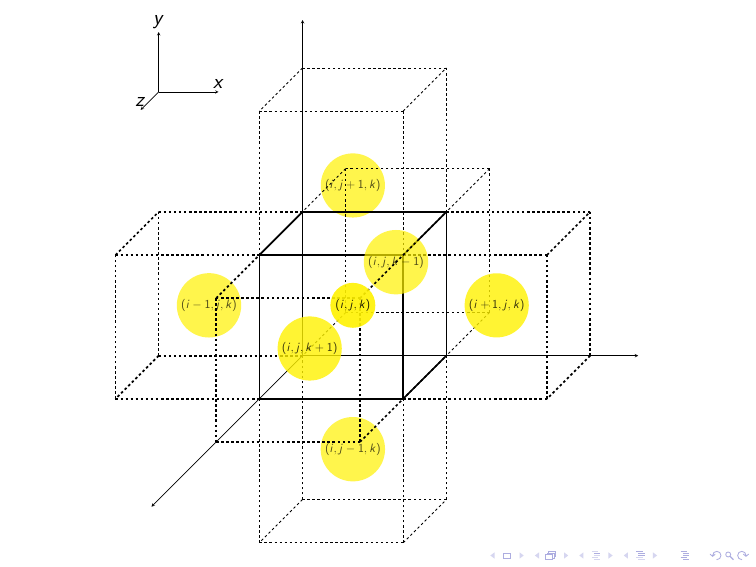}
    \caption{Cell-centered finite difference method to solve Faraday's law, \Eq{eqn:faradays_law}. The charge density and ion current are deposited on the grid-cell center, depicted by the yellow circle. The electric and magnetic fields are also calculated on the cell-centers. The magnetic field of the central grid cell is updated using the electric field of the neighbouring grid cells as described by \Eq{eqn:bx_CC} - \Eq{eqn:bz_CC}.}
    \label{fig:cell_centered}
\end{figure*}

We also have a cell-centered approach to update magnetic fields using Faraday's law, \Eq{eqn:faradays_law}. In this method, the electric and magnetic fields are defined and evolved on the cell-centers. This numerical stencil is depicted in \Fig{fig:cell_centered}. 

In this method, the magnetic field is updated in time in the following way
\begin{equation}
\begin{aligned}
\bx^{t+\dt}(i, j, k) = \bx^{t}(i, j, k) \\
+ \frac{\dt}{2\dz} \big[ \widetilde{E}^{t+\dt/2}_y(i, j, k+1) - \widetilde{E}^{t+\dt/2}_y(i, j, k-1) \big] \\
- \frac{\dt}{2\dy} \big[\widetilde{E}^{t+\dt/2}_z(i, j+1, k) - \widetilde{E}^{t+\dt/2}_z(i, j-1, k) \big],
\end{aligned}
\label{eqn:bx_CC}
\end{equation}
\begin{equation}
\begin{aligned}
\by^{t+\dt}(i, j, k) = \by^{t}(i, j, k) \\
+ \frac{\dt}{2\dx} \big[ \widetilde{E}^{t+\dt/2}_z(i+1, j, k) - \widetilde{E}^{t+\dt/2}_z(i-1, j, k)\big] \\
- \frac{\dt}{2\dz} \big[\widetilde{E}^{t+\dt/2}_x(i, j, k+1) - \widetilde{E}^{t+\dt/2}_x(i, j, k-1) \big],
\end{aligned}
\label{eqn:by_CC}
\end{equation}
\begin{equation}
\begin{aligned}
\bz^{t+\dt}(i, j, k) = \bz^{t}(i, j, k) \\
+ \frac{\dt}{2\dy} \big[ \widetilde{E}^{t+\dt/2}_x(i, j+1, k) - \widetilde{E}^{t+\dt/2}_x(i, j-1, k)\big] \\
- \frac{\dt}{2\dx} \big[\widetilde{E}^{t+\dt/2}_y(i+1, j, k) - \widetilde{E}^{t+\dt/2}_y(i-1, j, k) \big].
\end{aligned}
\label{eqn:bz_CC}
\end{equation}
The divergence of the magnetic field in the cell-centered finite difference method can be written as 
\begin{equation}
\begin{aligned}
    \nabla \cdot \B^{t+\dt}(i, j, k) = \frac{1}{2\dx} \big[ \bx^{t+\dt}(i+1, j, k) - \bx^{t+\dt}(i-1, j, k) \big] \\
    + \frac{1}{2\dy} \big[ \by^{t+\dt}(i, j+1, k) - \by^{t+\dt}(i, j-1, k) \big] \\
    + \frac{1}{2\dz} \big[ \bz^{t+\dt}(i, j, k+1) - \bz^{t+\dt}(i, j, k-1) \big].
\end{aligned}
\label{eqn:divB_CC}
\end{equation}
Using \Eq{eqn:bx_CC} - \Eq{eqn:bz_CC} in \Eq{eqn:divB_CC}, we obtain that $\nabla \cdot \B^{t+\dt}(i, j, k) = 0$. This is ensured by the construction of the cell-centered finite difference method similar to the constrained transport method (see \Sec{sec:CT}), albeit on a stencil twice the size. The above calculations show that as long as the construction of a numerical method ensures that the analytical expression, $\nabla \cdot \nabla \times \vec{C} = 0$, is satisfied for any vector $\vec{C}$, it can be used to ensure that magnetic fields are divergence-free.

\section{Hybrid precision - magnetic spectra and probability density functions} \label{app:precision_spectra_pdf}
\begin{figure*}
\begin{center}
\def\arraystretch{0}
\setlength{\tabcolsep}{0pt}
\begin{tabular}{ll}
\includegraphics[height=0.37\linewidth]{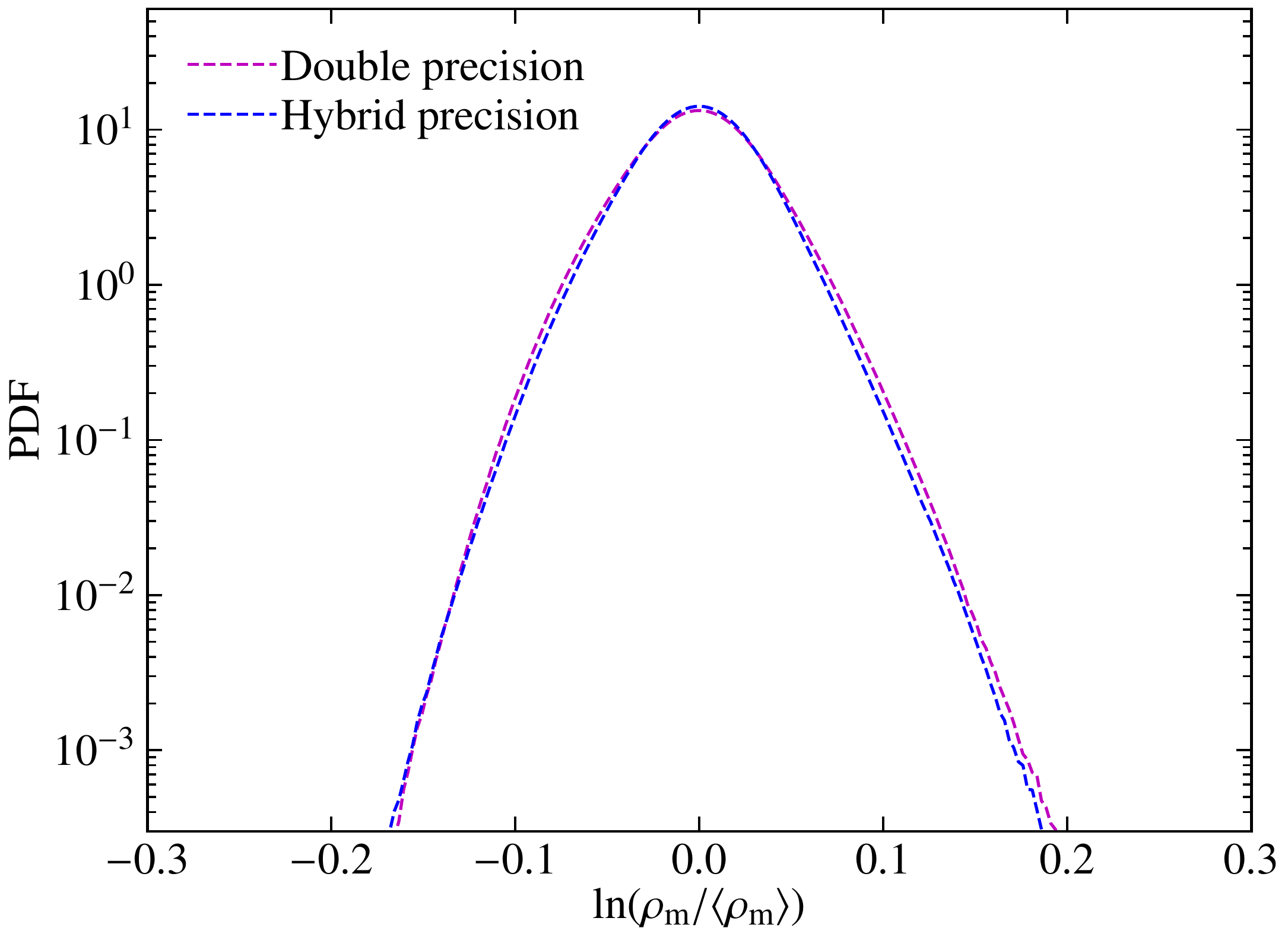} &
\includegraphics[height=0.37\linewidth]{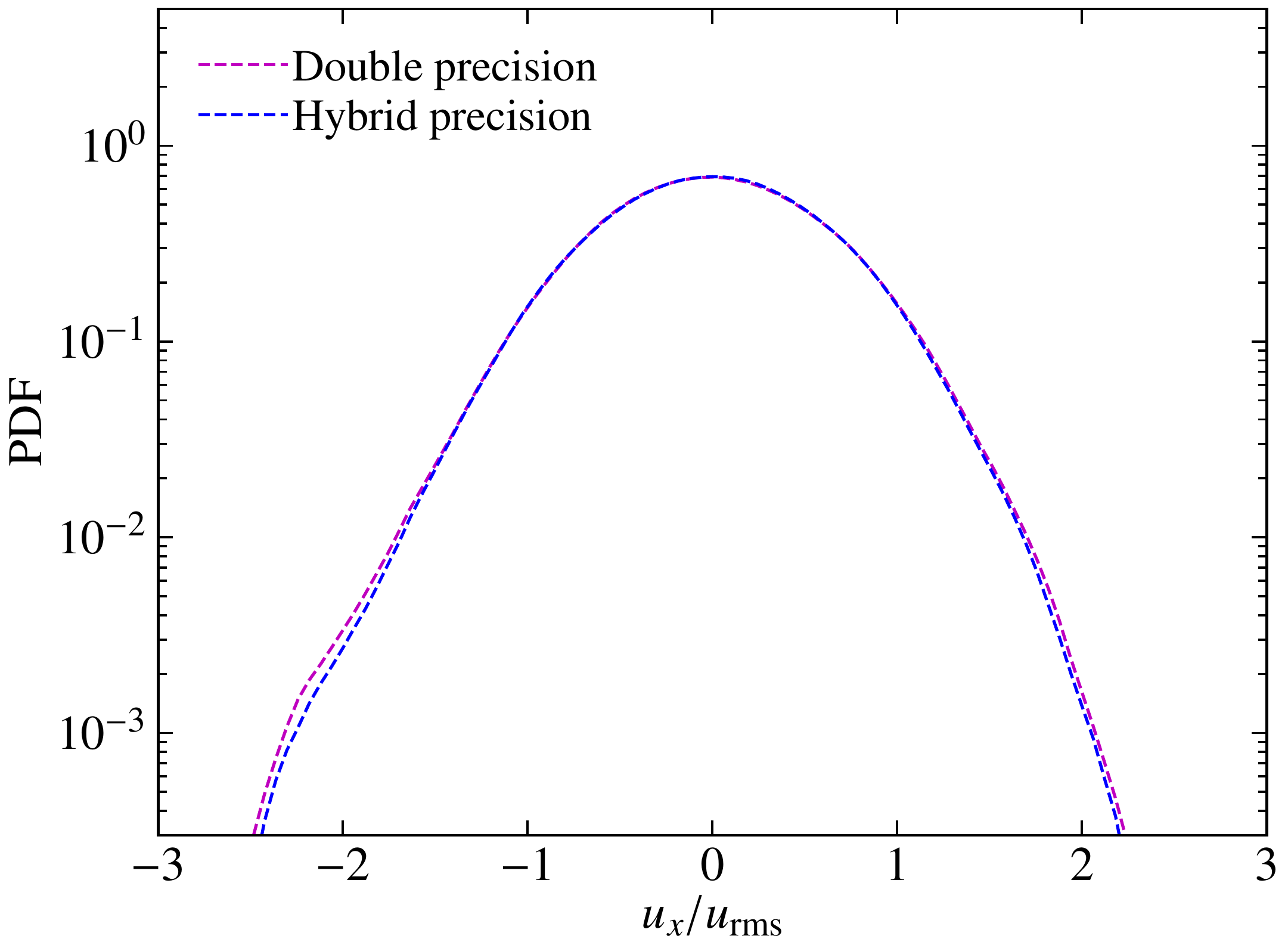} \\
\includegraphics[height=0.37\linewidth]{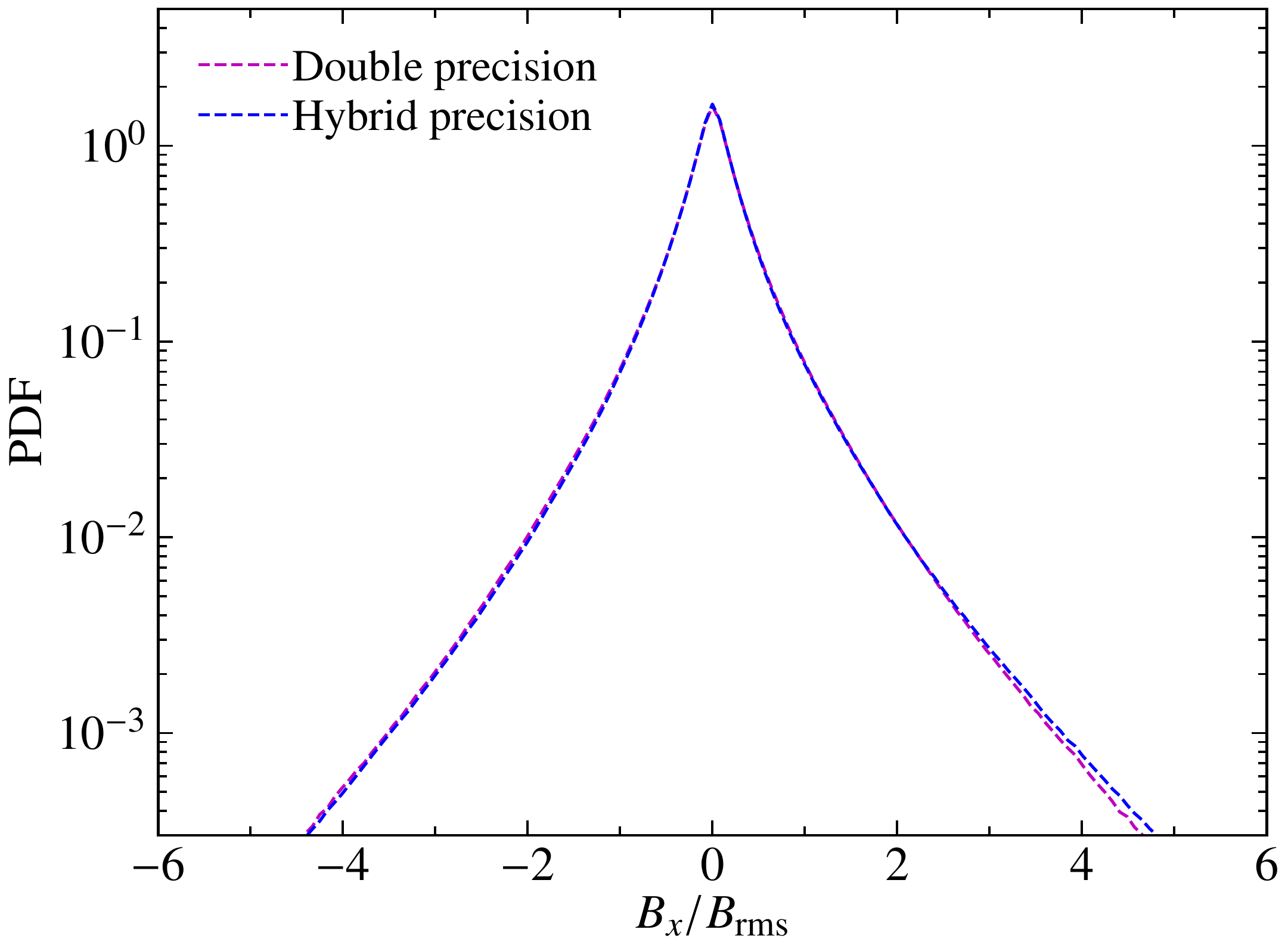} &
\includegraphics[height=0.375\linewidth]{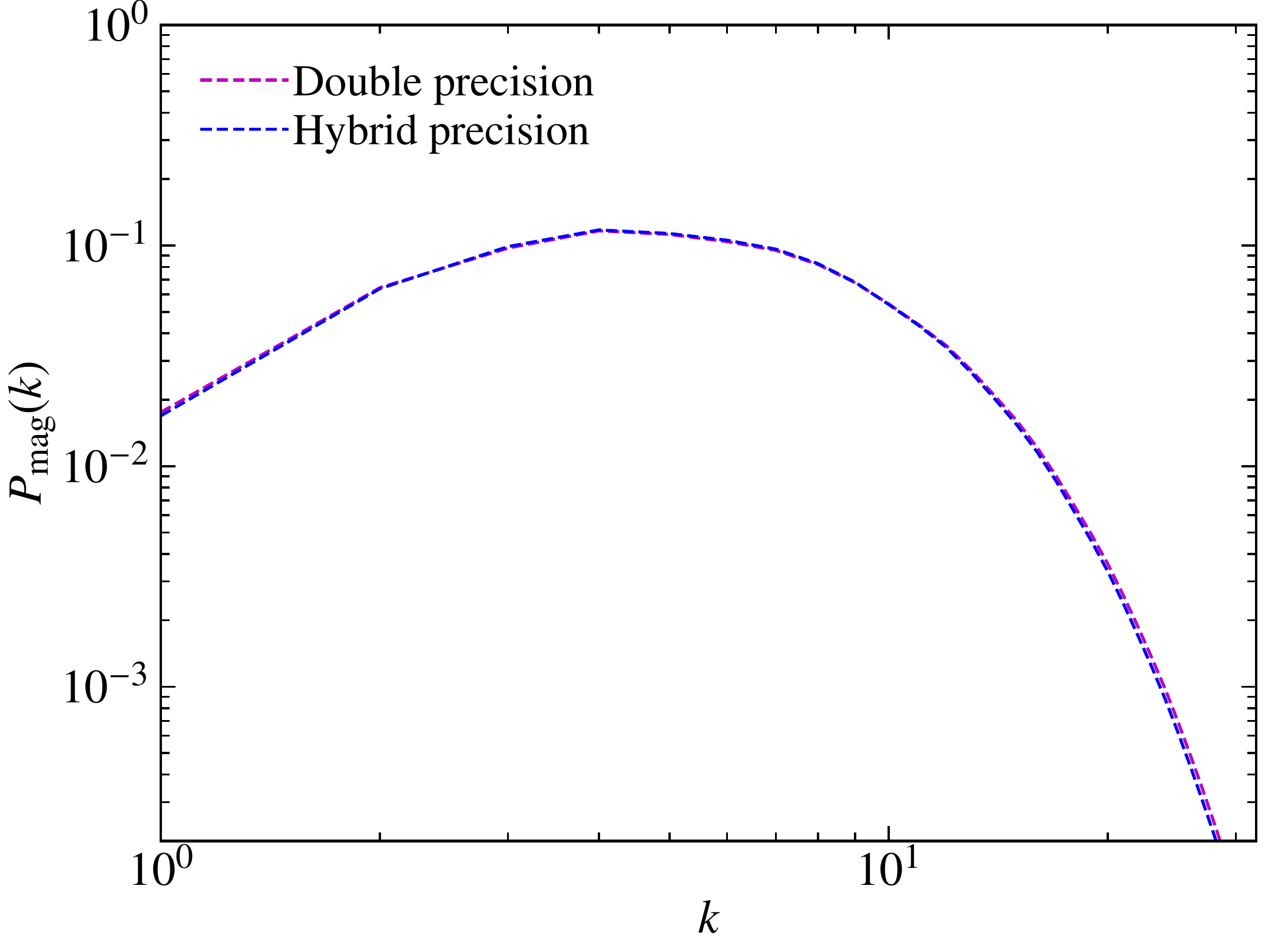}
\end{tabular}
\end{center}
\caption{Probability density functions (PDFs) of the natural logarithm of the mass density normalised to the mean mass density, ${\rm ln} (\rho_{\rm m}/\langle \rho_{\rm m} \rangle)$ (top left), the $x$-component of the velocity normalised to the root mean square velocity, $u_{x}/u_{\rm rms}$ (top right), the $x$-component of the magnetic field normalised to the root mean square magnetic field, $B_{x}/B_{\rm rms}$ (bottom left), and the power spectra of magnetic energy (bottom right), for the hybrid  vs.~double precision tests discussed in \Sec{sec:hybrid_precision}. The PDFs and power spectra are time-averaged in the kinematic regime of the turbulent dynamo ($\approx 3 - 15 \ted$). We find excellent agreement between the double and hybrid precision statistics.}
\label{fig:hp_pdf_spectra}
\end{figure*}

Probability density functions (PDFs) and power spectra are key quantities in the study of turbulence and magnetic field amplification \citep{Federrath2013,Federrath&Banerjee2015,Seta&Federrath2020,Seta&Federrath2021b}. In this section, we compare the PDFs and power spectra obtained in the hybrid vs.~double precision tests described in \Sec{sec:hybrid_precision}, with \Fig{fig:hp_pdf_spectra} showing the results. We find excellent agreement between the double and hybrid precision tests for all the examined quantities, further demonstrating that the numerical solutions with hybrid precision are in practice as accurate as those with double precision. 

\bsp
\label{lastpage}
\end{document}